\documentclass[twocolumn,english,aps,prb,amsmath,showpacs,amssymb]{revtex4}

\usepackage[T1]{fontenc}
\usepackage[latin9]{inputenc}
\usepackage{float}
\usepackage{bm}
\usepackage{amsmath}
\usepackage{amssymb}
\usepackage{graphicx}

\begin{document}

\title{Faraday effect in graphene enclosed in an optical cavity and the
equation of motion method for the study of magneto-optical transport
in solids}

\author{Aires Ferreira$^{1,2}$, J. Viana-Gomes$^{1}$, Yu. V. Bludov$^{1}$,
Vitor M. Pereira$^{2}$, N. M. R. Peres$^{1,2}$, A. H. Castro Neto$^{2,3}$}

\affiliation{$^{1}$ Department of Physics and Center of Physics, University of
Minho, P-4710-057, Braga, Portugal}

\affiliation{$^{2}$ Graphene Research Centre and Department of Physics, National
University of Singapore, 2 Science Drive 3, Singapore 117542}

\affiliation{$^{3}$ Department of Physics, Boston University, 590 Commonwealth
Avenue, Boston, Massachusetts 02215, USA}


\begin{abstract}
We show that by enclosing graphene in an optical cavity, giant Faraday
rotations in the infrared regime are generated and measurable Faraday
rotation angles in the visible range become possible. Explicit expressions
for the Hall steps of the Faraday rotation angle are given for relevant
regimes. In the context of this problem we develop an equation of
motion (EOM) method for calculation of the magneto-optical properties
of metals and semiconductors. It is shown that properly regularized
EOM solutions are fully equivalent to the Kubo formula. 
\end{abstract}

\maketitle

\section{Introduction\label{sec:Introduction}}

Electromagnetic radiation emitted by far stellar objects travels for
long periods of time through very diluted concentrations of interstellar
gases, traversing regions where week magnetic fields exist. In this
circumstance, the polarization of the electric field rotates due to
its interaction with the gases immersed in the magnetic field. Due
to the enormous traveling distances through such interstellar regions,
the degree of rotation of the polarization can be important. This
magnetic rotational effect turns out to be a problem in astrophysics,
since it modifies, in an unpredictable way, the polarization state
of the emitted radiation, introducing additional difficulties in the
interpretation of astronomical observations. In the electrodynamics
of metals and insulators the effect of polarization rotation induced
by a magnetic field was first discussed by Faraday\cite{M. Faraday 1846}
and, on Earth, has many different applications.

In magneto-optics, the effect coined optical Faraday rotation\cite{M. Faraday 1846}
refers to the rotation of the plane of polarization of light when
it transverses either a dielectric\cite{P. R. Berman 2010} or a metal,\cite{A. Stern 1964}
in the presence of a static magnetic field applied along the direction
of propagation of the electromagnetic wave. In addition to the rotation
of the plane of polarization, the polarization itself acquires a certain
degree of ellipticity. In dielectrics, the effect can be explained
using a model of harmonic oscillators coupled to light.\cite{P. R. Berman 2010}
In metals, the effect has its roots in the Hall effect.\cite{Palik 1970}

For a two-dimensional (2D) metal, such as graphene, in the Hall regime,
the conductivity becomes a tensor $\hat{\sigma}$, with finite (nonzero)
values for both diagonal and off-diagonal components. In magneto-optics,
the components of the tensor depend both on the frequency of the impinging
electromagnetic wave and on the cyclotron frequency of the electrons,
due to the magnetic field perpendicular to the plane of the metal.
The response of the electrons to the external magnetic field has two
regimes: (i) the semiclassical limit, of low fields and/or a high
electronic density; and (ii) the quantum Hall regime, of strong fields
and/or a low electronic density.

For interpretation of the optical Faraday rotation, in the semi-classical
regime, the Drude theory of metals suffices.\cite{A. Stern 1964}
In the case of graphene, it is possible to change its electronic density
either by use of a gate or by the adsorption of molecules.\cite{Novoselov 2005,Schedin 2007}
At high doping, graphene is in the semiclassical regime and Boltzmann
transport theory can be used to compute the Hall conductivity.\cite{Peres 2007}

In the absence of disorder and other relaxation mechanisms (such as
electron-phonon scattering), the conductivity of graphene (at zero
magnetic field) would be exclusively determined by interband transitions.
In the limit of no disorder, the optical conductivity of doped graphene,
in the infrared region of the spectrum and at zero magnetic field,
is given by\cite{Peres 2006,Gusynin PRB 2006,Carbotte2006_MWresponse,Ando_UnivCond,Falkovsky 2007,Falkovsky 2 2007,Peres IJMP 2008,Stauber_VisibleReg}
\begin{equation}
\sigma_{xx}=\sigma_{\textrm{g}}n_{F}(\hbar\omega-2E_{F})\,,\label{eq:universal_interband_conductivity}
\end{equation}
 where $\sigma_{\textrm{g}}=\pi e^{2}/(2h)$ is the so-called ac universal
conductivity of graphene.\cite{Peres 2006,PeresRMP,Nair 2008,Kuzmenko 2008}

When a magnetic field is applied perpendicularly to graphene's surface,
the system develops a finite Hall conductivity. In the quantum regime,
it was shown that the Faraday rotation angle $\theta_{F}$ is solely
determined by the fine structure constant $\alpha$, and presents
a step-like structure as the Fermi energy crosses different Landau
levels (LLs).\cite{Morimoto2009} The estimated Faraday rotation steps'
height in this case is of the order of $\theta_{F}\sim0.4^{\circ}$,\cite{Morimoto2009}
a magnitude that can be resolved experimentally.\cite{Shimano2008}
In the context of topological insulators, similar quantization rules
in certain thin-film geometries have been derived in Refs.~\onlinecite{Tse1}
and \onlinecite{Tse2}. We note in passing that, when the external
magnetic field is absent, a dynamic Hall effect can still be induced
by using circularly polarized light impinging on graphene at a finite
angle with the normal to the graphene surface.\cite{Karch}

On the theoretical side, the magneto-optical transport properties
of graphene have been investigated with the Green's function method\cite{Peres 2006,Gusynin PRB 2006},
and by means of numerical implementations of the Kubo formula, using
exact diagonalization\cite{Morimoto2009} and Chebyshev polynomial
expansions.\cite{Yuan_ResScatt} These approaches come with pros and
cons: numerical studies allow the exploration of general scenarios,
whereas Green's functions allows one to obtain analytic results, but
many times at the expense of lengthy calculations.

Motivated by the need for analytical flexible analytical tools, the
equation of motion (EOM) method employed in Ref.~\onlinecite{Peres
Excitons} is generalized to include the effect of a magnetic field.
As shown later, starting from a small set of EOMs, an adequate treatment
permits the derivation of response functions with correct analytical
properties (i.e., satisfying Kramers-Kronig causality relations). 

The present paper is divided into two main parts. In Sec.~\ref{Sec2 (Magneto-Optical Transport)}
we present the EOM method for calculation of the magneto-optical transport
in metals and semiconductors; to be concrete, the method is described
in the context of the properties of graphene. In Sec.~\ref{sec:Application:-the-Faraday}
we describe in detail the Faraday effect in graphene and propose an
experimental setup that is able to enhance the Faraday effect up to
the visible range. Section~\ref{sec:Application:-the-Faraday} relies
heavily on the results derived in Sec.~\ref{Sec2 (Magneto-Optical Transport)}.
Some technical details are given in the Appendixes. 

We have chosen to organize the subjects according to the following
interests of different readers: a reader having a primary interest
in the Faraday effect, and familiar with the details on the magneto-optical
properties of graphene, should be able to read Sec.~\ref{sec:Application:-the-Faraday}
with a bird's-eye reading of Sec.~\ref{Sec2 (Magneto-Optical Transport)}.
A reader interested in the Faraday effect in graphene but not well
acquainted with its magneto-optical properties may want to go through
Sec.~\ref{Sec2 (Magneto-Optical Transport)} first. Finally, reading
Sec.~\ref{Sec2 (Magneto-Optical Transport)} alone may appeal to
readers interested in applying the EOM method to another problem of
interest bearing no relation to graphene.

\section{Equation of Motion Method for Calculation of the Magneto-Optical
Conductivity \label{Sec2 (Magneto-Optical Transport)}}

Here, we develop the EOM approach to the calculation of the magneto-optical
properties of a semiconductor. To be concrete, the method is presented
in the context of the optical response of graphene. 

Electrons constrained to two dimensions are responsible for a variety
of quantum manifestations, a striking example being the integer quantum
Hall effect (IQHE). Measured in semiconductor 2D electron gases more
than 30 years ago\cite{vonKlitzing} and in the early days of graphene,
in both monolayer\cite{Novoselov 2005,PKim2005} and bilayer samples\cite{Novoselov 2006}
(very recently also in trilayer graphene\cite{IQHE_Trilayer}), the
static quantum Hall effect is a hallmark of elementary excitations
in electronic systems.\cite{Thouless_IQHE}

Its dynamical analog---the ac quantum Hall effect---can provide additional
information about charge carriers, such as the opening of gaps in
the spectrum.\cite{Carbotte2007} Recent advances in time-domain spectroscopy
in the Thz regime\cite{Shimano2008} have paved the way to measurement
of dynamical optical conductivities at impinging field energies closer
to the scale of interest. The goal is to reach cyclotronic energies,
usually $\mathcal{O}(10)$~meV in fields of 1-10 T, where strong
optical responses take place. The so-called \emph{optical} quantum
Hall conductivity of 2D electron gases shows a robust plateaux as
the Fermi energy is swept, although no quantization rule for the plateaux's
height exists.\cite{ShimanoThz} Due to its peculiar band structure,
graphene has been predicted to display a characteristic optical quantum
Hall effect which should be detectable via Faraday rotation measurements.\cite{Morimoto2009}
In the semiclassical regime, on the other hand, the Faraday rotation
of graphene was reported to be $\mathcal{O}(1)$ degrees in fields
of a few tesla,\cite{FaradayNaturePhys} a surprisingly high value
for a one-atom-thick electronic system.

\begin{figure}[ht]
\centering{}\includegraphics[clip,width=8cm]{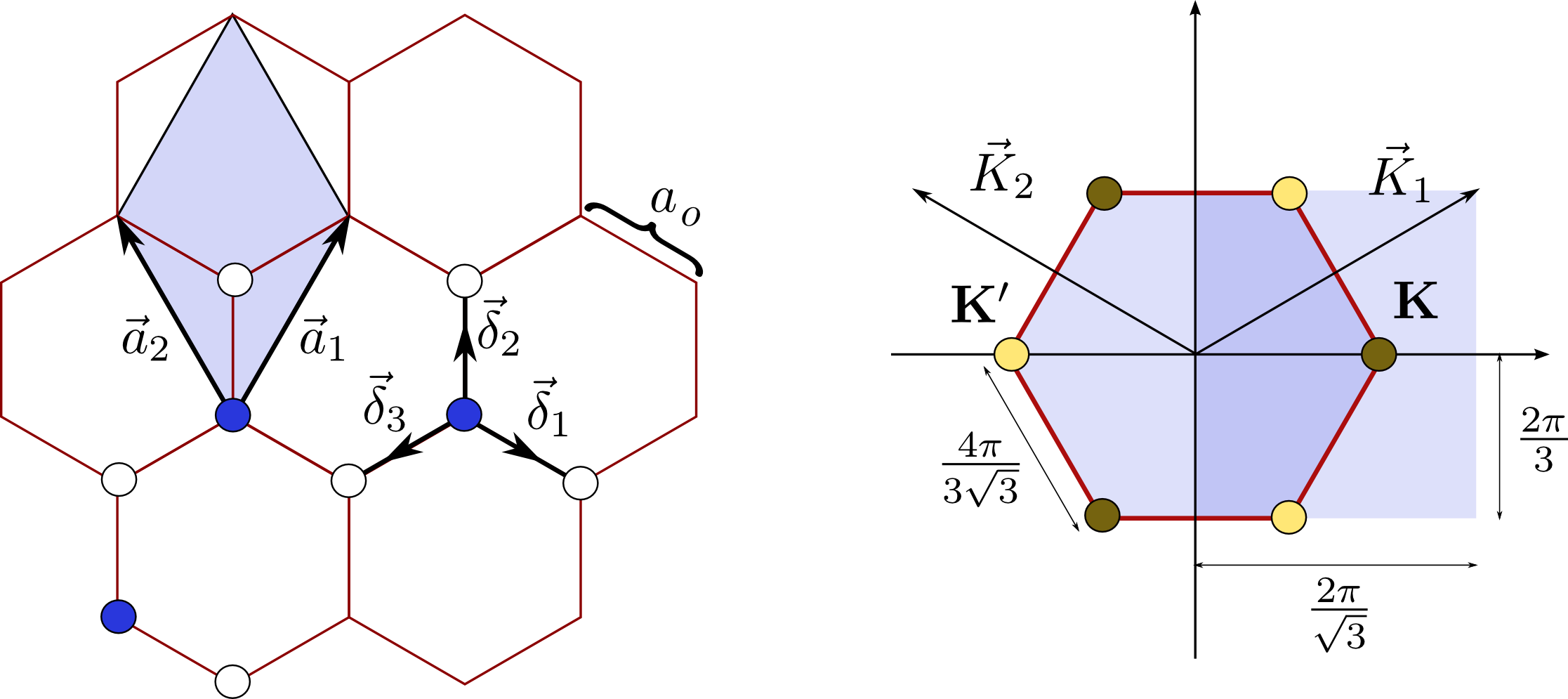}
\caption{\label{fig_monolayer_lattice}Lattice structure and Brillouin zone
of monolayer graphene. \textbf{Left:} Hexagonal lattice of graphene,
with the next nearest neighbor, $\bm{\delta}_{i}$, and the primitive,
$\bm{a}_{i}$, vectors depicted. The area of the primitive cell is
$A_{c}=3\sqrt{3}a_{0}^{2}/2\simeq5.1$~\AA{}$^{2}$, and $a_{0}\simeq1.4$~\AA{}.
\textbf{Right:} Brillouin zone of graphene, with the Dirac points
$\bm{K}$ and $\bm{K}'$ indicated. Close to these points, the dispersion
of graphene is conical and the density of states is proportional to
the absolute value of the energy.}
\end{figure}

\subsection{Graphene\label{sub:Graphene_Outline}}

The starting point of the present analysis is the low-energy continuum
description of single-layer graphene; having two (carbon) atoms per
unit cell and sixfold symmetry, its elementary excitations obey a
2D Dirac equation with linear electronic dispersion.\cite{CNeto RMP}
This section is meant to fix the notation. The Brillouin zone of graphene
has six corners, and among these, only two are inequivalent, the so-called
$\mathbf{K}$ and $\mathbf{K}^{\prime}$ Dirac points (see Fig.~\ref{fig_monolayer_lattice}).
At these points, the valence and conduction bands touch, with a linear
electronic spectrum up to energies of $\sim$2~eV.

We assume, in what follows, that the two Dirac points can be treated
independently, and introduce the valley degeneracy index, $g_{v}=2$,
when pertinent. This consideration is justified for typical experimental
conditions (i.e., low concentrations of scattering centers, finite
temperatures, etc.) and provides an accurate description of graphene's
electronic transport properties at finite densities.\cite{PeresRMP,Resonant Scatterers}

In accordance, we resort to the $2\times2$ Dirac Hamiltonian of graphene,
describing the physics of elementary excitations within the $\mathbf{K}$
valley, $H_{\mathbf{K}}=v_{F}\boldsymbol{\sigma}\cdot\mathbf{p}$,
where $v_{F}\simeq10^{6}$m/s is the Fermi velocity, $\boldsymbol{\sigma}=(\sigma_{x},\sigma_{y})$
{[}with $\sigma_{i}$ ($i=x,y$) denoting Pauli matrices{]}, and $\mathbf{p}$
is the momentum of the low-energy excitation (measured relative to
the $\mathbf{K}$ point).\cite{CNeto RMP} $H_{\mathbf{K}}$ has eigenvalues
given by 
\begin{equation}
E=\pm\hbar v_{F}|\mathbf{k}|\,,\label{eq:Eigenvalues_ZeroField}
\end{equation}
 {[}with $\mathbf{k}=(k_{x},k_{y})$ denoting a 2D wave vector{]},
and (normalized) wave functions given by
\begin{equation}
\psi_{\lambda,\mathbf{k}}(\mathbf{r})=\frac{1}{\sqrt{2A}}\left(\begin{array}{c}
1\\
\lambda e^{i\theta_{\mathbf{k}}}
\end{array}\right)e^{i\mathbf{k}\cdot\mathbf{r}}\,,\label{eq:Eigenstates_ZeroField}
\end{equation}
 where $A$ is the area of the graphene sample, $\lambda=+1$($-1$)
for electron(hole)-like excitation, and $\theta_{\mathbf{k}}=\arctan(k_{y}/k_{x})$.

The electromagnetic field can be incorporated via minimal coupling,
$\mathbf{p}\rightarrow\mathbf{p}+e\mathbf{A}_{\textrm{g}},$ where
$-e<0$ is the electron charge, and the vector potential $\mathbf{A}_{\textrm{g}}$
relates to the electromagnetic field according to the usual relations,
$\mathbf{B}=\boldsymbol{\nabla}\times\mathbf{A}_{\textrm{g}}$ and
$\mathbf{E}=-\partial\mathbf{A}_{\textrm{g}}/\partial t$.

Here, the vector potential contains the information about the impinging
electromagnetic radiation, and possible external static magnetic fields.
Assuming light linearly polarized along the $x$ axis, the radiation
term reads $\mathbf{A}=[A_{0}(\mathbf{r})e^{-i\omega t}+\textrm{c.c.}]\mathbf{e_{x}}$,
where $\omega$ stands for the frequency of the radiation field and
$A_{0}(\mathbf{r})$ describes its position dependence. For clarity
of exposition, we separate the light-matter interaction term from
the free Hamiltonian, 
\begin{equation}
H=H_{0}+ev_{F}\boldsymbol{\sigma}\cdot\mathbf{A}\,,\label{eq:Hamiltonian}
\end{equation}
 where $H_{0}\equiv H_{\mathbf{K}}+ev_{F}\boldsymbol{\sigma}\cdot\mathbf{A}_{B}$,
with $\mathbf{A}_{B}$ describing the static magnetic field.

A typical experimental scenario corresponds to a constant magnetic
field $B>0$ applied in the transverse direction with respect to the
graphene plane. In such case, LLs develop and the eigenenergies of
charge carriers become quantized according to\cite{McClure_LandaLev}
\begin{equation}
E_{n}=\textrm{sign}(n)\frac{\hbar v_{F}}{l_{B}}\sqrt{2|n|}\quad,n=0,\pm1,\pm2,...,\label{eq:LL}
\end{equation}
 with $l_{B}=\sqrt{\hbar/(eB)}$ denoting the magnetic length. Choosing
the gauge $\mathbf{A}_{B}=(0,Bx,0)$ results in the following set
of Landau eigenfunctions, 
\begin{equation}
\psi_{n,k_{y}}(\mathbf{r})=\frac{C_{n}}{\sqrt{L}}\left(\begin{array}{c}
\phi_{|n|-1}(x)\\
i\textrm{sign}(n)\phi_{|n|}(x)
\end{array}\right)e^{ik_{y}y}\,,\label{eq:Eigenstates_Field}
\end{equation}
 where $\phi_{n}(x)=e^{-\xi(x)^{2}/2}H_{n}(\xi(x))/\sqrt{n!2^{n}\sqrt{\pi}l_{B}}$,
$H_{n}(x)$ is the Hermite polynomial of degree $n\ge0$, $\phi_{-1}(x)=0$,
and $\xi(x)$ stands for the dimensionless center of the Landau orbit,
$\xi(x)=l_{B}k_{y}+x/l_{B}$. Here, $L$ is the linear dimension of
the system in the $y$ direction and $C_{n}$ is a normalization constant
that distinguishes the zero-energy level from the remaining levels,
$C_{n}=1$ for $n=0$ and $C_{n}=1/\sqrt{2}$ for $|n|\ge1$.

Having reviewed the basics of the graphene's electronic low-energy
theory, in what follows we present the EOM approach to the study of
magneto-optical transport.

\subsection{Theoretical methods\label{sub:Theoretical_Methods}}

In the context of electronic systems, the EOM was extensively used
in calculations of light polarization in semiconductor laser theory.\cite{Koch}
Recently, it has been used to study excitons in graphene in zero field.\cite{Peres Excitons}

The EOM approach avoids the calculation of current-current correlators
(i.e., Kubo formula), and, hence, provides a shortcut to determination
of the response of electronic systems to external perturbations. As
shown in detail in Appendix~\ref{Appendix_C}, with an appropriate
regularization procedure, the EOM solutions become fully equivalent
to the Kubo formula, and hence provide an accurate description of
transport in the linear response regime. Another advantage of the
present approach is that it allows for the calculation of non-linear
corrections to the conductivity.

At the heart of the EOM approach to calculation of the magneto-optical
conductivity is the Heisenberg equation for the electronic current
density, $\boldsymbol{J}(t)$, in the presence of an external electromagnetic
field, i.e., $d\boldsymbol{J}/dt=(i/\hbar)[H,\boldsymbol{J}]$, with
$H$ being the total Hamiltonian, Eq.~(\ref{eq:Hamiltonian}). Having
solved for the current density of the system in the presence of external
perturbation, in first order in the external field $\mathbf{A}$,
the optical conductivity follows from the constitutive electromagnetic
relation
\begin{equation}
\sigma_{ij}(\omega)=g_{s}g_{v}\times\frac{\tilde{J}_{i}(\omega)}{\tilde{E}_{j}(\omega)}\,,\label{eq:Ohm's Law}
\end{equation}
 where $\tilde{O}(\omega)$ relates to the average $O(t)$ {[}$O=J_{i},E_{j}${]}
according to $O(t)=\tilde{O}(\omega)e^{-i\omega t}+\textrm{c.c.}$,
with appropriate regularization implicit (Appendix~\ref{Appendix_C};
Sec.~\ref{sub:FiniteMagField}). Having graphene in the Dirac cone
approximation in mind, the latter equation contains the relevant degeneracies.
The spin contribution as a degeneracy factor, $g_{s}$, should be
valid for typical magnetic fields ($\lesssim$15 T) when the Zeeman
effect does not manifest.

The first step is to project the Heisenberg EOM for the current onto
the space of unperturbed single-particle states: we introduce the
field operator $\Psi_{\sigma}(\mathbf{r},t)=\sum_{\boldsymbol{\alpha}}\hat{c}_{\boldsymbol{\alpha},\sigma}(t)\psi_{\boldsymbol{\alpha}}(\mathbf{r})$
(and the respective Hermitian conjugate), where $\hat{c}_{\boldsymbol{\alpha},\sigma}$($\hat{c}_{\boldsymbol{\alpha},\sigma}^{\dagger}$)
is the annihilation (creation) operator obeying fermionic anticommutation
rules: $\{\hat{c}_{\boldsymbol{\alpha},\sigma},\hat{c}_{\boldsymbol{\alpha}^{\prime},\sigma^{\prime}}^{\dagger}\}=\delta_{\boldsymbol{\alpha}\boldsymbol{\alpha}^{\prime}}\delta_{\sigma,\sigma^{\prime}}$
and $\{\hat{c}_{\boldsymbol{\alpha},\sigma},\hat{c}_{\boldsymbol{\alpha}^{\prime},\sigma^{\prime}}\}=\{\hat{c}_{\boldsymbol{\alpha},\sigma}^{\dagger},\hat{c}_{\boldsymbol{\alpha}^{\prime},\sigma^{\prime}}^{\dagger}\}=0$.
The symbol $\boldsymbol{\alpha}=(\lambda,\mathbf{k},...)$ specifies
the single-particle state of the electron (or hole) and $\sigma=\pm1$
is the spin variable. The kets $|\boldsymbol{\alpha},\sigma\rangle\equiv\hat{c}_{\boldsymbol{\alpha},\sigma}^{\dagger}|0\rangle$
represent eigenstates of $H_{0}$, and, therefore, the position representation,
$\langle\mathbf{r}|\boldsymbol{\alpha},\sigma\rangle\equiv\psi_{\boldsymbol{\alpha},\sigma}(\mathbf{r})$,
equals Eq.~(\ref{eq:Eigenstates_ZeroField}) at zero magnetic field
or Eq.~(\ref{eq:Eigenstates_Field}) in the presence of a transverse
uniform magnetic field.

The second-quantized form of the full Hamiltonian and the current
density operator is given by 
\begin{align}
\hat{H}(t)= & \sum_{\sigma}\int d\mathbf{r}\Psi_{\sigma}^{\dagger}(\mathbf{r},t)H\Psi_{\sigma}(\mathbf{r},t)\,,\label{eq:H_2ndQ}\\
\hat{J}_{i}(t)= & \sum_{\sigma}\int d\mathbf{r}\Psi_{\sigma}^{\dagger}(\mathbf{r},t)j_{i}\Psi_{\sigma}(\mathbf{r},t)\,,\label{eq:J_2ndW}
\end{align}
 respectively, where 
\begin{equation}
\boldsymbol{j}=-\frac{ev_{F}}{A}\boldsymbol{\sigma}\,,\label{eq:current_op_graph}
\end{equation}
 is the current density of graphene in the continuum description.\cite{PeresRMP,Resonant Scatterers}
We omit the spin dependence of the operators hereafter for clarity
of exposition.

We now define the generic operator, 
\begin{equation}
\hat{P}_{\boldsymbol{\alpha\beta}}(t)\equiv\hat{c}_{\boldsymbol{\alpha}}^{\dagger}(t)\hat{c}_{\boldsymbol{\beta}}(t)\,,\label{eq:P_ab}
\end{equation}
 whose EOM reads 
\begin{equation}
\frac{d}{dt}\hat{P}_{\boldsymbol{\alpha\beta}}(t)=\frac{i}{\hbar}\sum_{\boldsymbol{\gamma},\boldsymbol{\delta}}h_{\boldsymbol{\gamma\delta}}\left[\hat{P}_{\boldsymbol{\gamma\delta}}(t),\,\hat{P}_{\boldsymbol{\alpha\beta}}(t)\right]\,,\label{eq:eq_of_motion}
\end{equation}
 where $h_{\boldsymbol{\gamma\delta}}=\langle\boldsymbol{\gamma}|\hat{H}|\boldsymbol{\delta}\rangle$
are the matrix elements of the full Hamiltonian {[}Eq.~(\ref{eq:Hamiltonian}){]}.
Solving for $\hat{P}_{\boldsymbol{\alpha\beta}}(t)$ gives directly
the current density according to, 
\begin{equation}
\hat{J}_{i}(t)=\sum_{\boldsymbol{\alpha},\boldsymbol{\beta}}\langle\boldsymbol{\alpha}|j_{i}|\boldsymbol{\beta}\rangle\hat{P}_{\boldsymbol{\alpha\beta}}(t),\label{eq:current density}
\end{equation}
 and hence the (yet non-regular) optical conductivity via Eq.~(\ref{eq:Ohm's Law}).
The regularization is the final step of the EOM approach needed for
obtaining a fully-consistent conductivity (in particular, obeying
Kramers-Kronig relations).\cite{Regularizaton_Procedure} The respective
technical procedure is given in Appendix~\ref{Appendix_C}.

In the following section, we solve Eq.~(\ref{eq:eq_of_motion}) explicitly
in the linear response regime (i.e., first order in the electric field)
for any pair of quantum states $\boldsymbol{\alpha}$,$\boldsymbol{\beta}$,
in the absence of a magnetic field. The case of finite (nonzero) magnetic
field intensity is left for Sec.~\ref{sub:FiniteMagField}.

\subsection{Graphene in a zero magnetic field\label{sub:ZeroMagField}}

The purpose of this section is to show the EOM method at work in the
context of a simple problem, which allows us to derive well-known
results. In the absence of magnetic fields, the macroscopic electronic
current follows the applied optical field, and thus only the longitudinal
conductivity is nonzero. From symmetry considerations, we also have
$\sigma_{xx}(\omega)=\sigma_{yy}(\omega)$. According to the statement
Eq.~(\ref{eq:current density}), the relevant set of EOMs to be solved
is determined by the non-zero matrix elements of the current density.

\begin{figure}
\begin{centering}
\includegraphics[clip,width=0.5\columnwidth]{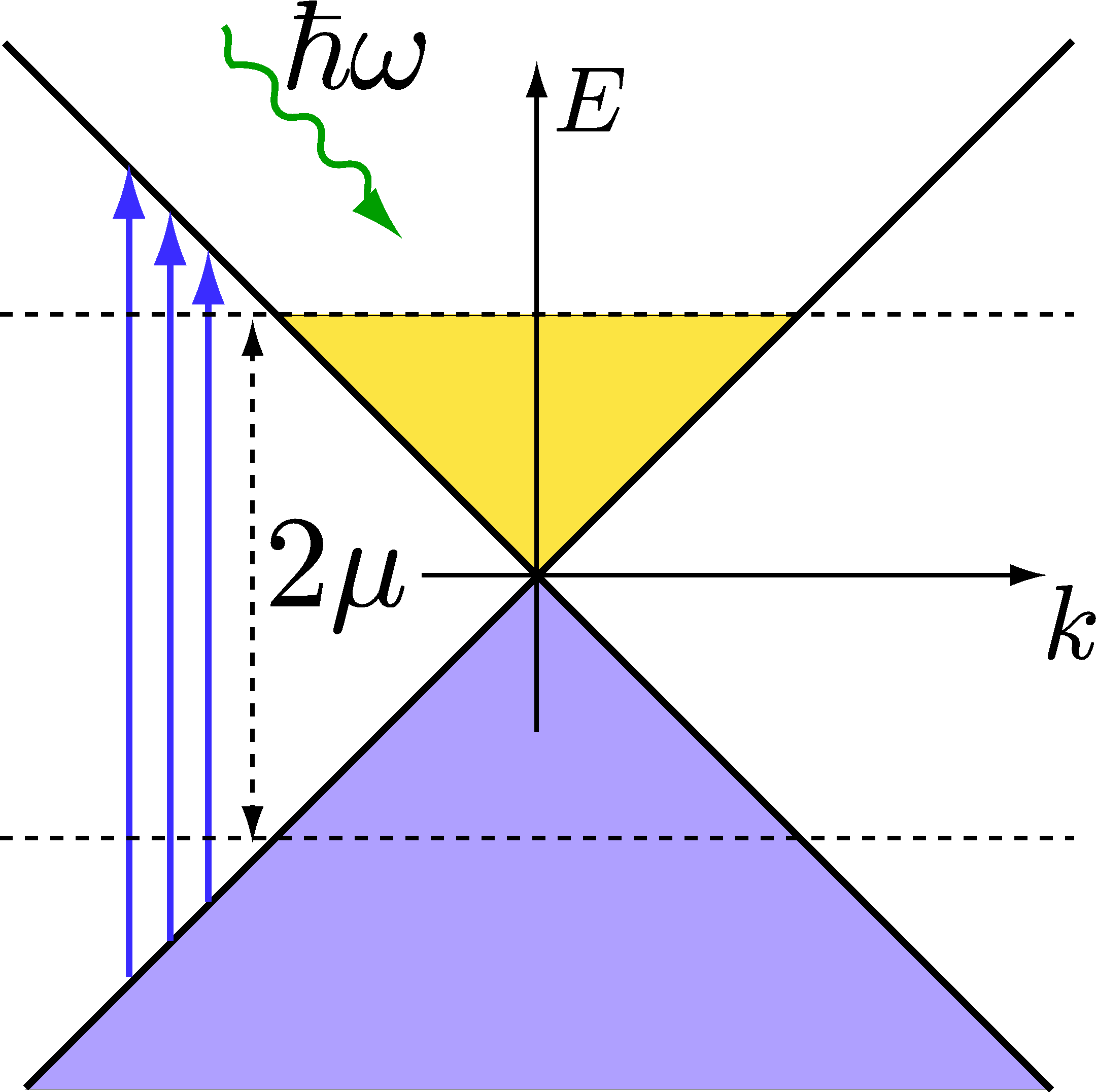} 
\par\end{centering}

\caption{\label{fig:Allowed-interband-transitions}Allowed interband transitions
(vertical arrows) in graphene; a photon of energy $\hbar\omega$ produces
an excitation from the lower to the upper Dirac, as long as $\hbar\omega>2\mu$.
The transitions conserve $\mathbf{k}$ and hence are said to be ``vertical.''
For $\hbar\omega\le2\mu,$ Pauli blocking forbids any (interband)
transition. In practice, due to disorder (impurities, etc.), the interband
conductivity can be non-zero even for $\hbar\omega\le2\mu$. }
\end{figure}

Defining $\langle\mathbf{k},\lambda|j_{x}|\mathbf{k}^{\prime},\lambda^{\prime}\rangle=-(ev_{F}/A)j_{\lambda,\lambda^{\prime},\mathbf{k},\mathbf{k}^{\prime}}^{x}$
and using the wave functions Eq.~(\ref{eq:Eigenstates_ZeroField}),
we easily find 
\begin{equation}
j_{\lambda,\lambda^{\prime},\mathbf{k},\mathbf{k}^{\prime}}^{x}=\frac{\delta_{\mathbf{k},\mathbf{k}^{\prime}}}{2}\left(\lambda^{\prime}e^{i\theta_{\mathbf{k}}}+\lambda e^{-i\theta_{\mathbf{k}}}\right)\,.\label{eq:matrix_elem_zerofield}
\end{equation}
 With this notation, the current density along the $x$ direction
reads, 
\begin{equation}
J_{x}(t)=-\frac{ev_{F}}{A}\sum_{\lambda,\lambda^{\prime},\mathbf{k}}j_{\lambda,\lambda^{\prime},\mathbf{k},\mathbf{k}}^{x}\langle\hat{c}_{\lambda,\mathbf{k}}^{\dagger}(t)\hat{c}_{\lambda^{\prime},\mathbf{k}}(t)\rangle\,.\label{eq:Px_av_s}
\end{equation}
 The non-null matrix elements in Eq.~(\ref{eq:matrix_elem_zerofield})
contributing to the conductivity correspond to transitions between
different bands conserving the momentum $\mathbf{k}$. These transitions
are said to be ``vertical,'' and, in addition, since they connect
states in different bands, they are refereed to as being interband-like
(see Fig.~\ref{fig:Allowed-interband-transitions}).

Taking the dipole approximation, $A(\mathbf{r})\rightarrow A_{0}$,
the Hamiltonian {[}Eq.~(\ref{eq:H_2ndQ}){]} reads
\begin{align}
\hat{H} & = & \sum_{\lambda,\mathbf{k}}E_{\lambda}(\mathbf{k})\hat{c}_{\lambda,\mathbf{k}}^{\dagger}\hat{c}_{\lambda,\mathbf{k}}+ev_{F}A_{0}\left(e^{-i\omega t}+\textrm{c.c.}\right)\times\nonumber \\
 &  & \times\sum_{\mathbf{k}}j_{c,v,\mathbf{k},\mathbf{k}}^{x}\hat{c}_{c,\mathbf{k},\sigma}^{\dagger}\hat{c}_{v,\mathbf{k},\sigma}+(c\longleftrightarrow v)\,.\label{eq:Ham_Zero_Field}
\end{align}
 In the latter equation, $E_{\lambda}(\mathbf{k})\equiv\lambda\hbar v_{F}k$
, and the subscripts $c$ ($v$) denote electrons (holes).

As described above, we need to compute the time evolution of the operator
$\hat{P}_{v,c,\mathbf{k}}(t)=\hat{c}_{v,\mathbf{k}}^{\dagger}(t)\hat{c}_{c,\mathbf{k}}(t)$.
Straightforward algebra yields
\begin{align}
\frac{d}{dt}\hat{P}_{v,c,\mathbf{k}} & =\frac{i}{\hbar}\left\{ \left[E_{v}(\mathbf{k})-E_{c}(\mathbf{k})\right]\hat{P}_{v,c,\mathbf{k}}\right.\nonumber \\
 & \left.+ev_{F}A_{0}\left(e^{-i\omega t}+\textrm{c.c.}\right)j_{c,v,\mathbf{k},\mathbf{k}}^{x}\left[\hat{n}_{c}(t)-\hat{n}_{v}(t)\right]\right\} \,,\label{eq:EqMotionPvc_ZeroField}
\end{align}
where we have defined the occupation operator for electrons (holes)
as $\hat{n}_{c(v)}(t)\equiv\hat{c}_{c(v),\mathbf{k}}^{\dagger}(t)\hat{c}_{c(v),\mathbf{k}}(t)$.
A similar equation holds for $\hat{P}_{c,v,\mathbf{k},\sigma}$ which
can be obtained by interchanging $c\longleftrightarrow v$.

To proceed, we take the average of Eq.~(\ref{eq:EqMotionPvc_ZeroField})
with respect to the unperturbed Hamiltonian, $H_{0}$, and approximate
$\langle\hat{n}_{c}(t)-\hat{n}_{v}(t)\rangle_{0}\simeq\langle\hat{n}_{c}-\hat{n}_{v}\rangle_{0}$.
Both procedures are consistent with an expansion of $\hat{J}_{x}(t)$
up to first order in the parameter $A_{0}$. The solution of the above
differential equation reads 
\begin{equation}
\langle\hat{P}_{v,c,\mathbf{k}}(t)\rangle_{0}=\tilde{P}_{v,c,\mathbf{k}}(\omega)e^{-i\omega t}+\tilde{P}_{v,c,\mathbf{k}}(-\omega)e^{i\omega t}\,,\label{eq:sol_P(t)}
\end{equation}
 with,
\begin{eqnarray}
\tilde{P}_{v,c,\mathbf{k}}(\omega) & = & ev_{F}A_{0}j_{c,v,\mathbf{k},\mathbf{k}}^{x}\frac{\langle\hat{n}_{c}\rangle_{0}-\langle\hat{n}_{v}\rangle_{0}}{E_{c}(\mathbf{k})-E_{v}(\mathbf{k})-\hbar\omega-i\Gamma}\,,\label{eq:Pvc_absorption}
\end{eqnarray}
 and we have introduced an imaginary energy $\Gamma$ by hand, so
to account for disorder phenomenologically. The remaining term $\tilde{P}_{v,c,\mathbf{k},\sigma}(\omega)$
can be obtained from the latter expression by making $\omega\rightarrow-\omega$
and $\Gamma\rightarrow-\Gamma$ . From Eq.~(\ref{eq:Px_av_s}), the
oscillator strength of the current density along the $x$ direction
$\tilde{J}_{x}(\omega)$ is seen to be given by 
\begin{equation}
\tilde{J}_{x}(\omega)=-\frac{ev_{F}}{A}\sum_{\mathbf{k}}[j_{v,c,\mathbf{k},\mathbf{k}}^{x}\tilde{P}_{v,c,\mathbf{k}}(\omega)+j_{c,v,\mathbf{k},\mathbf{k}}^{x}\tilde{P}_{c,v,\mathbf{k}}(\omega)]\,.\label{eq:Pol_Flux_zeroF}
\end{equation}
 The longitudinal optical conductivity, $\sigma_{xx}$, follows from
Eq.~(\ref{eq:Ohm's Law}),

\begin{align}
\sigma_{xx}^{\textrm{inter}}(\omega)= & g_{v}g_{s}\frac{e^{2}v_{F}^{2}}{i\omega}\int\frac{d^{2}\mathbf{k}}{4\pi^{2}}\left(\sin^{2}\theta_{\mathbf{k}}\right)\times\nonumber \\
 & \times\frac{n_{F}\left[E_{v}(\mathbf{k})\right]-n_{F}\left[E_{c}(\mathbf{k})\right]}{E_{c}(\mathbf{k})-E_{v}(\mathbf{k})-\hbar\omega-i\Gamma}+(c\leftrightarrow v)\,.\label{eq:Cond_Inter_Graph_ZeroField}
\end{align}
 where $n_{F}(E)=1/[e^{(E-\mu)/k_{B}T}+1]$ stands for the Fermi-Dirac
distribution ($\mu$ is the chemical potential). In deriving this
expression, we have used the relation $\tilde{E}_{x}(\omega)=i\omega A_{0}$.
Taking the clean limit $\Gamma\rightarrow0$ and considering $\omega>0$
and $T=0$, one obtains the well- known result 
\begin{eqnarray}
\textrm{Re}\,\,\sigma_{xx}^{\textrm{inter}}(\omega) & = & \frac{\pi e^{2}}{2h}\theta\left(\hbar\omega-2|\mu|\right)\,.\label{eq:Univ_Cond_InterBand}
\end{eqnarray}
 The latter result is the $T\rightarrow0$ limit of Eq.~(\ref{eq:universal_interband_conductivity}).
For photon energies higher than $2\mu$ (see Fig.~\ref{fig:Allowed-interband-transitions}),
the interband conductivity is essentially frequency independent (up
to energies of $\sim$2 eV) and equals 
\begin{equation}
\sigma_{\textrm{g}}=\frac{\pi e^{2}}{2h}\,,\label{eq:UnivCond}
\end{equation}
which is nothing other than the universal conductivity of graphene
mentioned in Sec.~\ref{sec:Introduction}. For $\mu=0$, and contrary
to ordinary semiconductors, there is no frequency threshold for interband
transitions: according to Eq.~(\ref{eq:Univ_Cond_InterBand}), some
interband transitions will always be available for a sufficiently
high photon frequency. As a consequence, Drude's description will
not suffice for a general description of the optical response of graphene.

In addition to the interband transitions discussed here, there is
an intraband contribution in graphene which can be appreciable for
$\mu\neq0$. This contribution comes from nonvertical processes (e.g.,
via collisions with phonons), not included in the Hamiltonian Eq.~(\ref{eq:Ham_Zero_Field}).
This contribution gives the Drude response and reads\cite{Stauber}
\begin{equation}
\textrm{Re}\,\sigma_{xx}^{\textrm{intra}}(\omega)=\frac{2e^{2}}{h}|\mu|\frac{\Gamma}{\hbar^{2}\omega^{2}+\Gamma^{2}}\,.\label{eq:Sigma_Intra_Zero.Field}
\end{equation}
 Interestingly enough, the latter result can be derived from a full
quantum mechanical calculation by considering a finite magnetic field
intensity and taking the limit $B\rightarrow0$ in the end.\cite{Gusynin PRB 2006}
This is because a magnetic field open gaps in the spectrum of a clean
system, allowing for intraband transitions (see Sec.~\ref{sub:FiniteMagField}).
A semiclassical calculation also leads to an equivalent result (Sec.~\ref{sub:SemiClassSolution}).

\subsection{Optical conductivity of graphene in a magnetic field\label{sub:FiniteMagField}}

In what follows, we show that the EOM method can be employed to study
the magneto-optical response of graphene along the same lines as in
Sec.~\ref{sub:ZeroMagField}. The presence of a transverse magnetic
field in the Hamiltonian develops LLs, and hence we must start from
the eigenstates given in Eq.~(\ref{eq:Eigenstates_Field}). The latter
defines the field operator, $\Psi(\mathbf{r},t)=\sum_{n,k_{y}}\hat{c}_{n,k_{y}}(t)\psi_{n,k_{y}}(\mathbf{r})$
(together with the respective Hermitian conjugate); the index $n$
labels the degenerate LL with energy given by Eq. (\ref{eq:LL}).
The field operator can be written as 
\begin{eqnarray}
\Psi(\mathbf{r},t) & = & \frac{1}{\sqrt{2L}}\sum_{n\neq0,k_{y}}\left(\begin{array}{c}
\phi_{|n|-1}(x)\\
i\textrm{sign}(n)\phi_{|n|}(x)
\end{array}\right)e^{ik_{y}y}\hat{c}_{n,k_{y}}\nonumber \\
 &  & +\frac{1}{\sqrt{L}}\sum_{k_{y}}\left(\begin{array}{c}
0\\
\phi_{0}(x)
\end{array}\right)e^{ik_{y}y}\hat{c}_{0,k_{y}}\,.\label{eq:Field_Operator_Field}
\end{eqnarray}
 This peculiar spinorial structure, with a single level being highlighted,
is on the basis of non-standard features in the magneto-optical conductivity
of graphene.\cite{AnamQHE_Sharapov05,Peres 2006,Carbotte2006_MWresponse,Carbotte2007}

\subsubsection{The longitudinal conductivity}

According to Eq.~(\ref{eq:Ohm's Law}), the calculation of the longitudinal
conductivity requires computation of the average value of the current
density operator along the $x$ direction,
\begin{equation}
J_{x}(t)=\sum_{n,n^{\prime}}\sum_{k_{y},k_{y}^{\prime}}\langle n,k_{y}|j_{x}|n^{\prime},k_{y}^{\prime}\rangle\langle\hat{c}_{n,k_{y}}^{\dagger}(t)\hat{c}_{n^{\prime},k_{y}^{\prime}}(t)\rangle\,.\label{eq:J_x_Landau}
\end{equation}
 Using the LL wavefunctions {[}Eq.~(\ref{eq:Eigenstates_Field}){]},
we easily find the non-zero matrix elements to be, 
\begin{align}
\langle0,k_{y}|j_{x}|\pm1,k_{y}^{\prime}\rangle= & -\frac{ev_{F}}{\sqrt{2}A}\delta_{k_{y},k_{y}^{\prime}}\,,\label{eq:dip_mat_1}\\
\langle n,k_{y}|j_{x}|n^{\prime},k_{y}^{\prime}\rangle= & -\frac{ev_{F}}{2A}i\left[\textrm{sign}(n^{\prime})\delta_{|n|-1,|n^{\prime}|}\right.\nonumber \\
 & \left.-\textrm{sign}(n)\delta_{|n|,|n^{\prime}|-1}\right]\delta_{k_{y},k_{y}^{\prime}}\,,\label{eq:dip_mat_2}
\end{align}
 where in the last line $n,n^{\prime}\neq0$. These statements show
that the optical transitions conserve $k_{y}$ and occur between levels
with indexes $n$ and $n^{\prime}$ satisfying $|n|-|n^{\prime}|=\pm1$.

Two sets of transitions are thus allowed: intraband transitions, occurring
within the same band, and, as in the absence of a magnetic field,
transitions connecting LLs in the valence and conduction bands, which
are interband-like. Transitions involving the zero-energy state $n=0$
can be considered either intraband- or interband-like, since the zero-energy
state is shared between electrons and holes. For the sake of simplicity
in defining the set of EOMs, throughout, we classify transitions involving
the zero-energy state as being interband.

In order to clearly distinguish among the possible types of transitions,
we define
\begin{equation}
\hat{c}_{n,k_{y}}\equiv\begin{cases}
c_{n} & \,\textrm{for }n>0\\
v_{|n|} & \,\textrm{for }n<0\\
a_{0} & \,\textrm{for }n=0
\end{cases}\,,\label{eq:redefinitions}
\end{equation}
with the Hermitian conjugates following identical redefinitions. Note
that with these definitions the subscript $n$ in the operators take
only positive integer values.

\emph{a. Interband transitions}---Using the field operator in the
presence of a magnetic field {[}Eq.~(\ref{eq:Field_Operator_Field}){]},
and keeping track of just the interband terms for the moment, the
full Hamiltonian takes the form
\begin{align}
\hat{H}= & \sum_{n\ge1}\left[E_{n}c_{n}^{\dagger}c_{n}+E_{-n}v_{n}^{\dagger}v_{n}\right]\nonumber \\
+ & \frac{ev_{F}A(t)}{\sqrt{2}}\left[c_{1}^{\dagger}a_{0}+v_{1}^{\dagger}a_{0}+\textrm{h.c.}\right]\nonumber \\
- & \frac{ev_{F}A(t)}{2}i\sum_{n\ge1}\left[\hat{P}_{n}^{(1)}+\hat{P}_{n}^{(2)}-\textrm{h.c.}\right]\,,\label{eq:Ham_Field}
\end{align}
 where $A(t)\equiv A_{0}(e^{-i\omega t}+\textrm{c.c.})$, and 
\begin{align}
\hat{P}_{n}^{(1)} & =c_{n}^{\dagger}v_{n+1}\,,\label{eq:Pn1}\\
\hat{P}_{n}^{(2)} & =c_{n+1}^{\dagger}v_{n}\,.\label{eq:Pn2}
\end{align}
 (Also, for clarity, we have omitted $k_{y}$ under all the summation
signs.) The first line in Eq.~(\ref{eq:Ham_Field}) describes massless
Dirac fermions in a transverse magnetic field and the remaining lines
contain the electronic transitions among different LLs induced by
the external electric field.

The interband current density along the $x$ direction can be recast
into the form

\begin{align}
\hat{J}_{x}(t) & =-\frac{1}{\sqrt{2}A}ev_{F}\left(c_{1}^{\dagger}a_{0}+v_{1}^{\dagger}a_{0}+\textrm{h.c.}\right)\nonumber \\
 & +\frac{1}{2A}ev_{F}\sum_{n\ge1}\left(i\hat{P}_{n}^{(1)}+i\hat{P}_{n}^{(2)}+\textrm{h.c.}\right)\,.\label{eq:Px_Inter_in_Field}
\end{align}
 From the form of the current we see that there are two basic sets
of EOMs to be solved: the first set refers to the time evolution of
operators involving the zero-energy state ($c_{1}^{\dagger}a_{0}$,
$v_{1}^{\dagger}a_{0}$, and Hermitian conjugates), while the other
set refers to higher energy LLs. Take, for instance, the operator
$\hat{P}_{n}^{(1)}$ belonging to the latter set; as in the case of
zero magnetic field (Sec.~\ref{sub:ZeroMagField}), the commutator
$[H,\hat{P}_{n}^{(1)}]$ gives rise to (i) occupation number operators
($v_{n+1}^{\dagger}v_{n+1}$ and $c_{n}^{\dagger}c_{n}$), and (ii)
a free evolution term, that is, the operator $\hat{P}_{n}^{(1)}$
itself. In addition, intraband terms with $|n|-|n^{\prime}|=\pm2$
show up, namely, $c_{n}^{\dagger}c_{n+2}$, $v_{n-1}^{\dagger}v_{n+1}$
and $a_{0}^{\dagger}v_{2}\delta_{n,1}$. These terms do not originate
real intraband transitions, since the respective current density matrix
elements are null.

We are now in the position to write the prototype EOMs governing the
interaction of Landau quasiparticles with an external oscillating
electric field, 
\begin{align}
\frac{\hbar}{i}\frac{d}{dt}\hat{P}_{n}^{(1)} & =\left[E_{n}-E_{-(n+1)}\right]\hat{P}_{n}^{(1)}-\frac{i}{2}ev_{F}A(t)\times\nonumber \\
 & \times[v_{n+1}^{\dagger}v_{n+1}-c_{n}^{\dagger}c_{n}]\,,\label{eq:EOM_Pn1}\\
\frac{\hbar}{i}\frac{d}{dt}\hat{P}_{c} & =E_{1}\hat{P}_{c}+\frac{1}{\sqrt{2}}ev_{F}A(t)[a_{0}^{\dagger}a_{0}-c_{1}^{\dagger}c_{1}]\,,\label{eq:EOM_Q1}
\end{align}
 where we have omitted the time dependence of the operators and defined
$\hat{P}_{c(v)}=c(v){}_{1}^{\dagger}a_{0}$. The remaining operators
obey similar equations. {[}The EOM for $\hat{P}_{n}^{(2)}$ is obtained
making $\hat{P}_{n}^{(1)}\rightarrow\hat{P}_{n}^{(2)}$and interchanging
$n$ with $n+1$ on the right-hand side of Eq.~(\ref{eq:EOM_Pn1}).
As for $\hat{P}_{v}$, we let $\hat{P}_{c}\rightarrow\hat{P}_{v}$,
$E_{1}\rightarrow E_{-1}$, and $c_{1}(c_{1}^{\dagger})\rightarrow v_{1}(v_{1}^{\dagger})$
in Eq.~(\ref{eq:EOM_Q1}).{]}

\begin{figure}
\begin{centering}
\includegraphics[clip,width=0.7\columnwidth]{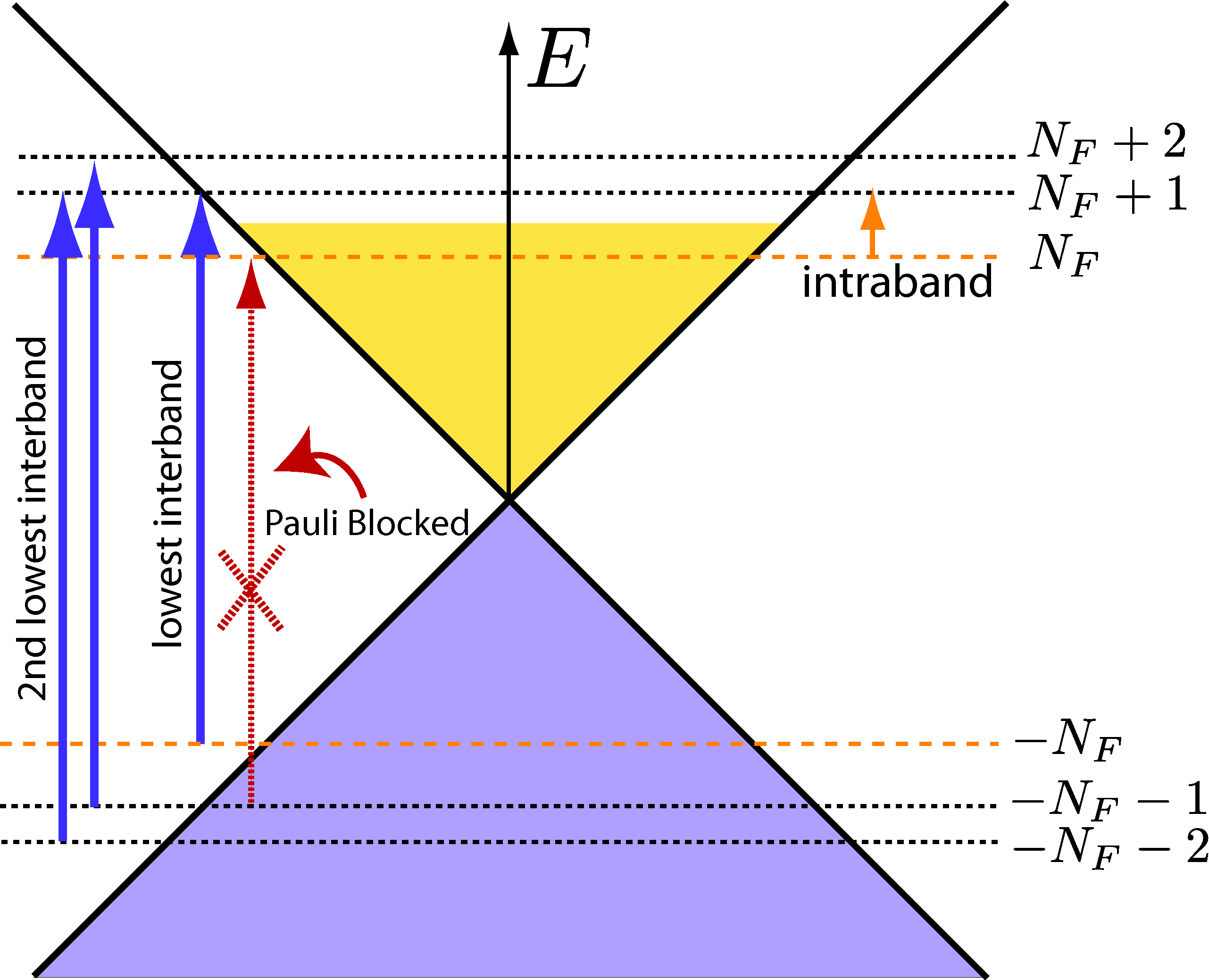} 
\par\end{centering}

\caption{\label{fig:Int_Transitions_SXX_Field}Schematic of electronic transitions
contributing to $\sigma_{xx}(\omega)$ of doped graphene in a magnetic
field. In this example, $E_{F}\ge E_{1}$, and thus the last occupied
LL, $n=N_{F}\ge1$, belongs to the conduction band. Two types of transitions
take place: (i) interband transitions, connecting LLs from the lower
cone (valence band) with LLs in the upper cone (conduction band),
and (ii) intraband transitions within the upper cone. Intraband transitions
are limited to adjacent LLs: $N_{F}\rightarrow N_{F}+1$. The figure
shows the following interband transitions: (a) the pair $-N_{F}\rightarrow N_{F}+1$
and $-N_{F}-1\rightarrow N_{F}$, whose energy difference is $E_{N+1}+E_{N}$
(the lowest interband energy; note that transitions $-N_{F}-1\rightarrow N_{F}$
are forbidden because $n=N_{F}$ is occupied); and (b) the pair $-N_{F}-1\rightarrow N_{F}+2$
and $-N_{F}-2\rightarrow N_{F}+1$. The respective energy difference
is $E_{N+1}+E_{N+2}$ (the second lowest interband energy difference),
and in this case both transitions take place. Transitions with higher
energy differences are not represented. }
\end{figure}

To solve the above set of differential equations to first order in
$A_{0}$, we proceed as in Sec.~\ref{sub:ZeroMagField}. Taking the
average value $\langle...\rangle_{0}$ of each EOM with respect to
the unperturbed Hamiltonian, $H_{0}$, the solution for each operator
$O$ can be written as $\langle O(t)\rangle_{0}=\tilde{O}(\omega)e^{-i\omega t}+\tilde{O}(-\omega)e^{i\omega t}$,
where the oscillator strengths read 
\begin{align}
\tilde{P}_{n}^{(1)}(\omega) & =-\frac{i}{2}ev_{F}A_{0}\frac{\langle v_{n+1}^{\dagger}v_{n+1}\rangle_{0}-\langle c_{n}^{\dagger}c_{n}\rangle_{0}}{E_{-(n+1)}-E_{n}-\hbar\omega-i\Gamma}\,,\label{eq:Sol_Pn1}\\
\tilde{P}_{c}(\omega) & =\frac{1}{\sqrt{2}}ev_{F}A_{0}\frac{\langle a_{0}^{\dagger}a_{0}\rangle_{0}-\langle c_{1}^{\dagger}c_{1}\rangle_{0}}{-E_{1}-\hbar\omega-i\Gamma}\,,\label{eq:Sol_Q1}
\end{align}
 and where, as in Sec.~\ref{sub:ZeroMagField}, we have added a imaginary
energy $\Gamma$ to account for level broadening. The solutions for
$\tilde{P}_{n}^{(2)}(\omega)$ and $\tilde{P}_{v}(\omega)$ can be
obtained from the latter expressions as described below Eq.~(\ref{eq:EOM_Q1}).

Combining these results and Eq.~(\ref{eq:Px_Inter_in_Field}), we
easily find 
\begin{eqnarray}
\tilde{J}_{x}(\omega) & = & \frac{1}{2A}ev_{F}\sum_{k_{y}}\left\{ \sum_{n\ge1}\left[i\tilde{P}_{n}^{1}(\omega)+i\tilde{P}_{n}^{2}(\omega)\right]\right.\nonumber \\
 &  & \left.-\sqrt{2}\left[\tilde{P}_{c}(\omega)+\tilde{P}_{v}(\omega)\right]+"\textrm{c.c. term}"\right\} \,,\label{eq:Jx(omega)Jx(omega)}
\end{eqnarray}
 where the summation over $k_{y}$ has been restored. This summation
yields the degeneracy of the LLs $\sum_{k_{y}}=A/(2\pi l_{B}^{2})$.
The last term in the above equation (i.e., the c.c. term) is obtained
taking the complex conjugate and making $\omega\rightarrow-\omega$
of all the previous terms.

The final expression for the longitudinal (interband) conductivity
is derived in two steps: (i) dividing the Eq.~(\ref{eq:Jx(omega)Jx(omega)})
by $\tilde{E}_{x}(\omega)$ {[}Eq.~(\ref{eq:Ohm's Law}){]}, and
(ii) undertaking appropriate regularization to remove the divergent
factor $1/\omega$,

\begin{align}
\sigma_{xx}^{\textrm{inter}}(\omega) & =\frac{e^{2}v_{F}^{2}\hbar}{2\pi l_{B}^{2}}i\sum_{n=0}^{N_{\textrm{c}}}(1+\delta_{n,0})\sum_{\alpha=\pm1}\alpha\nonumber \\
 & \times\left[\frac{1}{E_{-(n+1)}-E_{n}}\times\frac{n_{F}[E_{-(n+1)}]-n_{F}[E_{n}]}{E_{-(n+1)}-E_{n}-\alpha(\hbar\omega+i\Gamma)}\right.\nonumber \\
 & \left.+(n\leftrightarrow n+1)\right]\,.\label{eq:Sigma_xx_inter_reg}
\end{align}
 The above expression is analytic in the upper-half plane and finite
at $\omega=0$, thus obeying Kramers-Kronig causality relations. (We
refer to Appendix~\ref{Appendix_C} for the derivation and physical
grounds of the regularization procedure.) Note that, as usual when
dealing with low-energy theories, a cutoff energy $E_{\textrm{cut}}$
of the order of the bandwidth must be considered for consistency;
we take $n\le N_{\mathrm{c}}$, with $N_{\textrm{c}}=\textrm{int}[(E_{\textrm{cut}}/E_{1})^{2}]$,
where $\textrm{int}[...]$ denotes the integer part. $N_{\textrm{c}}$
varies roughly as $10^{4}B^{-1}$ with $B$ in teslas. Within the
physical relevant range for $E_{\textrm{cut}}$, these summations
converge quite rapidly; the figures in the present work have $E_{\textrm{cut}}\approx t\simeq$2.7~eV.

\emph{b. Intraband transitions}---The intraband interaction Hamiltonian
reads 
\begin{equation}
\hat{H}_{\textrm{int}}^{\textrm{intra}}=\frac{i}{2}ev_{F}A(t)\sum_{n\ge1}\left[v_{n}^{\dagger}v_{n+1}-c_{n}^{\dagger}c_{n+1}-\textrm{h.c.}\right]\,,\label{eq:H_intra}
\end{equation}
 and the zero-energy operators ($a_{0}$ and $a_{0}^{\dagger}$) are
absent given our classification of intraband transitions {[}see Eq.~(\ref{eq:redefinitions})
and the following text{]}. The calculation follows identical steps
to the interband conductivity and, hence, is not repeated. The final
expression for the (regular) intraband diagonal conductivity reads,
\begin{align}
\sigma_{xx}^{\textrm{intra}}(\omega) & =\frac{e^{2}v_{F}^{2}\hbar}{2\pi l_{B}^{2}}i\sum_{\alpha=\pm1}\alpha\sum_{n=1}^{N_{c}}\left[\frac{1}{E_{n+1}-E_{n}}\times\right.\nonumber \\
 & \times\frac{n_{F}[E_{n+1}]-n_{F}[E_{n}]}{E_{n+1}-E_{n}-\alpha\left(\hbar\omega+i\Gamma\right)}\nonumber \\
 & \left.+(E_{n}\rightarrow-E_{n}\wedge E_{n+1}\rightarrow-E_{n+1})\right]\,.\label{eq:Sigma_xx_intra_reg}
\end{align}

The full longitudinal conductivity $\sigma_{xx}(\omega)$ is given
by adding its interband and intraband counterparts, that is, Eqs.~(\ref{eq:Sigma_xx_inter_reg})
and (\ref{eq:Sigma_xx_intra_reg}), respectively; straightforward
algebra yields 
\begin{equation}
\sigma_{xx}(\omega)=\frac{e^{2}}{h}\sum_{n\neq m=-N_{c}}^{N_{c}}\frac{\Lambda_{nm}^{xx}}{iE_{nm}}\frac{n_{F}(E_{n})-n_{F}(E_{m})}{\hbar\omega+E_{nm}+i\Gamma}\,,\label{eq:sigma_xx_final}
\end{equation}
 with $E_{nm}=E_{n}-E_{m}$, and where we have defined the longitudinal
matrix elements 
\begin{equation}
\Lambda_{nm}^{xx}=\frac{\hbar^{2}v_{F}^{2}}{l_{B}^{2}}(1+\delta_{m,0}+\delta_{n,0})\delta_{|m|-|n|,\pm1}\,.\label{eq:eta_mn_xx}
\end{equation}

Equation~(\ref{eq:sigma_xx_final}) is the main result of the present
section. It coincides with Eq.~(7) in Ref.~\onlinecite{CarbotteIJMP2007}
obtained via a Green's function calculation in the bubble approximation
and, also, with a Kubo formula calculation within the Dirac cone approximation
(see Appendix~\ref{Appendix_C}). We note in passing that, on top
of the interband and intraband contributions discussed here, there
is a correction arising from phonon-electron coupling. At low temperatures
and zero field, this correction is expected to be small.\cite{Stauber}
At a high magnetic field, though, a recent calculation shows that
phonon energy peaks split the LLs nearby,\cite{Carbotte_Phonons}
which can lead to a measurable signature in magneto-optical experiments.

\subsubsection{The general properties of $\sigma_{xx}(\omega)$}

In what follows, we overview the main features of graphene's longitudinal
magneto-optical conductivity, an essential step to understanding the
Faraday rotation in graphene (Sec.~\ref{sec:Application:-the-Faraday}).

\begin{figure}
\begin{centering}
\includegraphics[clip,width=0.9\columnwidth]{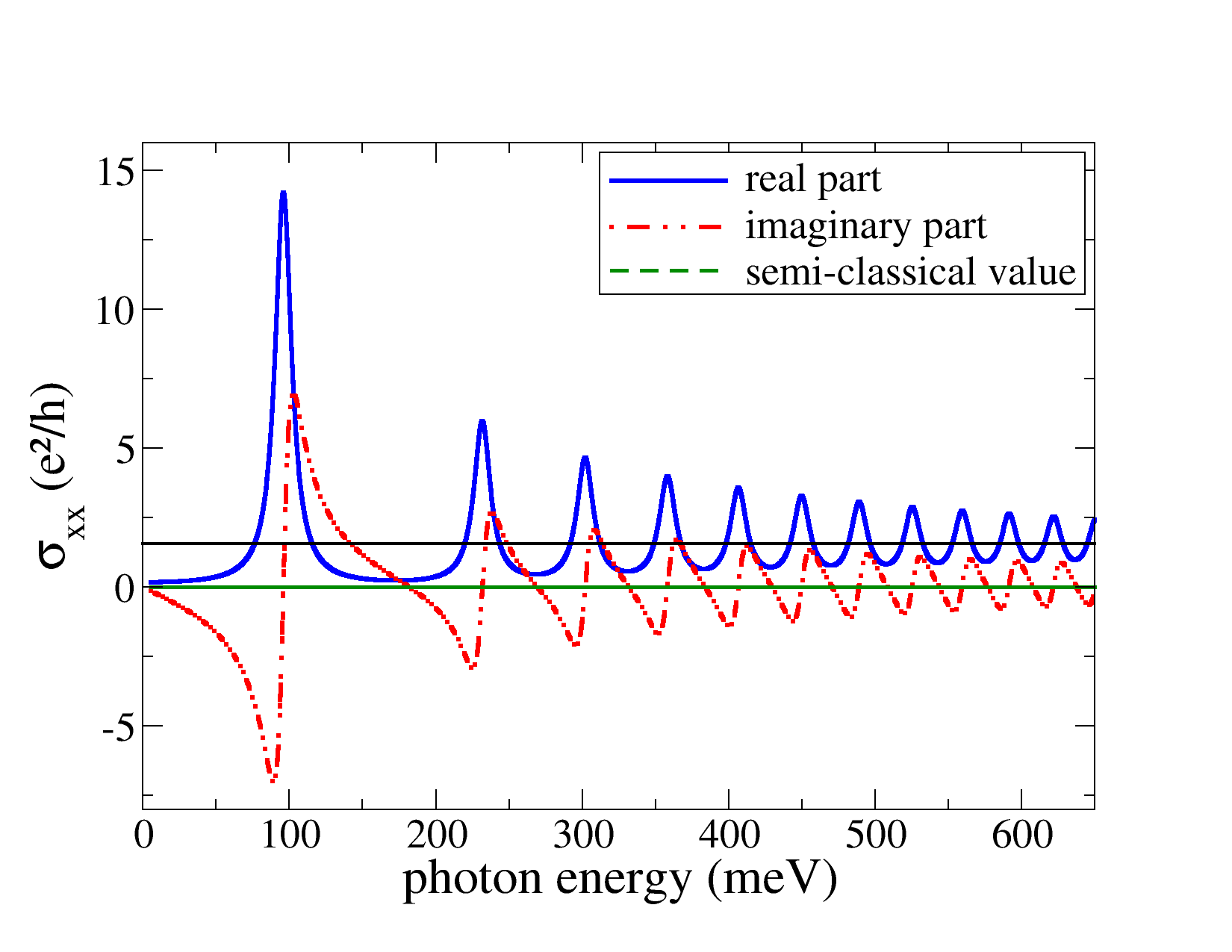}
\quad{} 
\par\end{centering}

\caption{\label{fig:Cond_xx_vs_freq}Longitudinal magneto-optical conductivity
as a function of the photon energy for a field of $7$~T, zero chemical
potential, $T=17$~K, and $\Gamma=6.8$~meV ($\sim79$~K). The
horizontal dashed-dot (black) line marks the graphene's universal
ac-conductivity background {[}Eq.~(\ref{eq:universal_interband_conductivity}){]}.}
\end{figure}

\emph{a. Low electronic density}---At a low electronic density, more
precisely, for $|E_{F}|<E_{1}$, no intraband transitions can take
place. Because the LL energy scale in graphene is relatively high
(e.g., $E_{1}\simeq36$~meV for a field of 1~T), the magneto-optical
conductivity is fully driven by interband transitions even close to
room temperature.

Figure~\ref{fig:Cond_xx_vs_freq} shows a plot of Eq.~(\ref{eq:sigma_xx_final})
for zero Fermi energy and a magnetic field of 7~T: a sequence of
absorption peaks, corresponding to the maximum of the real part of
each term in Eq.~(\ref{eq:Sigma_xx_inter_reg}), $\hbar\omega\simeq E_{1},\, E_{2}-E_{-1},\, E_{3}-E_{-2}$,
etc., is clearly observed {[}see Eq.~(\ref{eq:interband_peaks})
and text thereafter{]}. The conductivity never vanishes, even though
the concentration of carriers is low ($E_{F}\rightarrow0$), a genuine
signature of graphene's LL structure.\cite{Novoselov 2005}

The contributions from different interband transitions {[}Eq.~(\ref{eq:Sigma_xx_inter_reg}){]}
partially overlap at a high frequency, with the effect that the real
part of $\sigma_{xx}(\omega)$ displays the so-called Shubnikov\textendash{}de
Haas oscillations around the universal ac optical conductivity of
graphene, $\sigma_{\textrm{g}}$ (the imaginary part, in turn, oscillates
around 0).\cite{Peres 2006,Gusynin PRB 2006,Carbotte2006_MWresponse,Ando_UnivCond,Falkovsky 2007,Falkovsky 2 2007,Peres IJMP 2008,Stauber_VisibleReg}
The semiclassical conductivity is null, on the other hand, thus failing
to describe the magneto-transport in neutral graphene.

\emph{b. High electronic density}---Away from charge neutrality, more
precisely, for $|E_{F}|>E_{1}$, the picture is more involved; intraband
transitions can now occur, while some interband transitions will be
blocked. We take $T=0$ and, without loss of generality, assume that
$E_{F}>0$ (similar conclusions hold for holes); direct inspection
of Eq.~(\ref{eq:Sigma_xx_intra_reg}) shows that a single type of
intraband transition is allowed, whose contribution to the optical
conductivity reads
\begin{equation}
\sigma_{xx}^{\textrm{intra}}(\omega)=\frac{e^{2}}{h}\frac{2i\hbar v_{F}^{2}}{\Delta\omega_{N_{F}}l_{B}^{2}}\frac{\hbar\omega+i\Gamma}{\left(\hbar\omega+i\Gamma\right)^{2}-\hbar^{2}\Delta\omega_{N_{F}}^{2}}.\label{eq:s_xx_intra_T=00003D00003D0}
\end{equation}
 In the above formula, 
\begin{equation}
\Delta{}_{N_{F}}\equiv\hbar\Delta\omega_{N_{F}}=E_{N_{F}+1}-E_{N_{F}}\,,\label{eq:intraband_gap}
\end{equation}
denotes the intraband gap, with $N_{F}$ being the index for the last
occupied LL.

\begin{figure}
\begin{centering}
\includegraphics[clip,width=0.9\columnwidth]{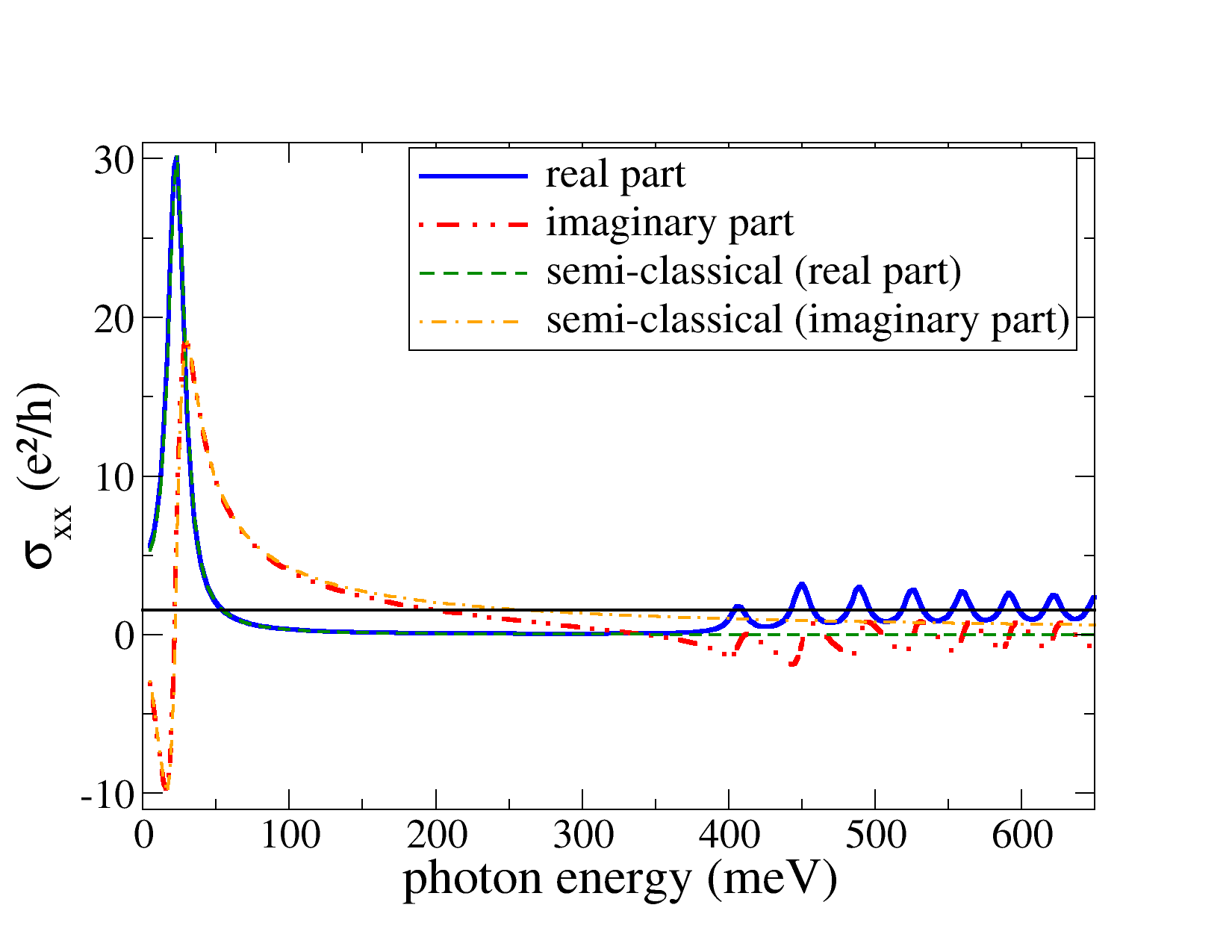} 
\par\end{centering}

\caption{\label{fig:Cond_xx_vs_freq_2}The longitudinal conductivity as function
of the photon energy for $E_{F}=0.2$~eV. Other parameters as in
Fig.~\ref{fig:Cond_xx_vs_freq}. The solid horizontal (black) line
shows graphene's universal ac-conductivity background {[}Eq.~(\ref{eq:universal_interband_conductivity}){]}.}
\end{figure}

Let us first consider the limiting case when the energy gap $\Delta{}_{N_{F}}$
is larger than the level broadening, $\Delta{}_{N_{F}}\gtrsim\Gamma$.
The latter typically happens at high magnetic fields and not too high
Fermi energies; in this limit, the real part of Eq.~(\ref{eq:s_xx_intra_T=00003D00003D0})
displays a maximum at $\omega\simeq\Delta\omega_{N_{F}}$, with an
intensity falling off as $B/\Delta\omega_{N_{F}}$, 
\begin{equation}
\textrm{Re}\,\sigma_{xx}^{\textrm{intra}}(\Delta\omega_{N_{F}})\simeq\left(\frac{2eBv_{F}^{2}}{\pi\Gamma\Delta\omega_{N_{F}}}\right)\times\sigma_{\textrm{g}}\,.\label{eq:intraband_peak}
\end{equation}
 The intraband magneto-peak, Eq.~(\ref{eq:intraband_peak}), is the
lowest frequency peak in the absorption spectrum of graphene with
$E_{F}>E_{1}$; its magnitude increases with increasing Fermi energy
and/or magnetic field intensity. An example of an intraband absorption
line occurring at $\omega\simeq\Delta\omega{}_{N_{F}}$ is shown in
Fig.~\ref{fig:Cond_xx_vs_freq_2}. In that case, the parameters correspond
to $\Delta{}_{N_{F}}=22.6$~meV and $\Gamma=6.8$~meV, and hence
$\Delta{}_{N_{F}}\gtrsim\Gamma$. Some points are worth mention: (i)
the intraband contribution to the conductivity {[}Eq.~(\ref{eq:s_xx_intra_T=00003D00003D0}){]}
dominates at low photon frequencies; and (ii) the curve for $\textrm{Re}\,\sigma_{xx}(\omega)$
shows that the remaining absorption peaks are found in the higher
frequency part of the spectrum, above the threshold for interband
transitions, $\hbar\omega\ge E_{N_{F}}+E_{N_{F}+1}$. (Note that,
at a low magnetic field and/or high Fermi energy, the level spacing
between adjacent LLs is so reduced that $E_{N_{F}}\simeq E_{N_{F}+1}\simeq E_{F}$,
and thus one recovers the condition found earlier, namely, $\hbar\omega>2E_{F}$.)
Such interband peaks cause Shubnikov\textendash{}de Haas oscillations
despite the finite electronic density.

For a general relation between the broadening and the energy gap $\Delta_{N_{F}}$,
the maximum for the intraband peak occurs at
\begin{equation}
\omega_{\textrm{peak}}^{\textrm{intra}}=\textrm{Re}\,\sqrt{2\Delta\omega_{N_{F}}\sqrt{\Delta\omega_{N_{F}}^{2}+\Gamma^{2}/\hbar^{2}}-\Delta\omega_{N_{F}}^{2}-\Gamma^{2}/\hbar^{2}}\,.\label{eq:intraband_peak_general}
\end{equation}
When $\Delta\omega_{N_{F}}\le\Gamma/(\sqrt{3}\hbar)$ (typically the
case for a very high Fermi energy and/or low magnetic field), the
intraband conductivity is maximal at null frequency, with an intensity
given by Eq.~(\ref{eq:intraband_peak}) multiplied by a factor of
2.

The regime $\Delta_{N_{F}}\lesssim\Gamma$ is illustrated in the bottom
panel in Fig.~\ref{fig:Fig_Theta_Exp}. Two magnetic fields are considered,
at a fixed Fermi energy, $E_{F}=0.3$~eV, with $\textrm{Re}\:\sigma_{xx}(\omega)$
being represented by the solid lines. When $B=7$T (left-hand panel),
although a considerable number of levels are occupied ($N_{F}=9$),
one has $\Delta_{N_{F}}\simeq1.4\Gamma$, which, according to Eq.~(\ref{eq:intraband_peak_general}),
corresponds to a maximum of the longitudinal conductivity at $\omega\simeq\Delta\omega_{N_{F}}$.
This is indeed confirmed by the numerical calculation shown there.
Decreasing the magnetic field down to $B=$3~T (right-hand panel),
reduces $\Delta_{N_{F}}$ (recall that the LL energy varies as $l_{B}^{-1}\sim\sqrt{B}$),
which in turn increases the number of occupied levels to $N_{F}=22$.
As a consequence, $\Delta_{N_{F}}\simeq0.67\Gamma$, and the maximum
of the intraband peak is seen to be shifted to zero frequency, again
in accordance with Eq.~(\ref{eq:intraband_peak_general}).

Given the intrinsic large cyclotron gap of graphene, $E_{1}$, the
intraband contribution {[}Eq.~(\ref{eq:s_xx_intra_T=00003D00003D0}){]}
controls the magneto-optical response of this material in the microwave
region up to terahertz (THz) frequencies in samples with a finite
electronic density ($E_{F}>E_{1}$).

The interband contribution, on the other hand, is important both in
samples with a low electronic density, $E_{F}<E_{1}$, where it determines
the full magneto-optical response (discarding the effect of phonons
as discussed above), and in samples with arbitrary carriers concentrations,
for photon energies above the threshold for interband transitions,
$\hbar\omega=E_{N_{F}}+E_{N_{F}+1}$ (typically within the near-infrared
region).

The positions of each interband peak can be obtained from Eq.~(\ref{eq:intraband_peak_general}),
with $\Delta\omega_{N_{F}}$ replaced by 
\begin{equation}
\Delta\Omega_{n}=(E_{n+1}+E_{n})/\hbar\,,\label{interband_gap}
\end{equation}
 with the constraint $n\ge N_{F}$. At finite electronic densities,
$N_{F}\ge1$, typically one has $\hbar\Delta\Omega_{n}\gtrsim\Gamma$,
and thus we arrive at the following useful approximation 
\begin{equation}
\omega_{\textrm{peak}}^{\textrm{inter}\,(n)}\simeq\Delta\Omega_{n}\quad,n\ge N_{F}\,.\label{eq:interband_peaks}
\end{equation}
 For not too small fields, $B\gtrsim0.1$~T, the cyclotron gap $E_{1}\simeq36\times\sqrt{B}$~meV$\cdot$T$^{-1/2}$
is larger than the LL broadening, and thus, in practice, the latter
statement can be generalized to include the case of $N_{F}=0$.

For general parameters, the intensity of each interband peak is no
longer given by a simple expression, because many interband transitions
can contribute to the spectral weight close to each of the resonances
$\omega\simeq\Delta\Omega_{n}$. As a result, as $\omega$ varies,
the real part of $\sigma_{xx}(\omega)$ oscillates around a constant
value of about $\sigma_{\textrm{g}}$. Examples are shown in Fig.~\ref{fig:Cond_xx_vs_freq}
for $E_{F}=0$ and in Fig.~\ref{fig:Cond_xx_vs_freq_2} for $E_{F}=0.2$~eV.
In the first case, we have $N_{F}=0$ and therefore all the observed
peaks are interband-like. The second case has $N_{F}=4$ and therefore
one intraband peak is observed, corresponding to transitions $n=4\rightarrow n=5$,
at low photon energy, whereas the interband peaks appear at energies
$\hbar\omega\gtrsim2E_{F}=0.4$~eV.

We finally remark that, as long as not too low magnetic fields are
considered ($B\lesssim0.1$~T), the above considerations are valid
even close to room temperature (e.g., for $B=1$~T, the first LL
corresponds to a thermal energy of $420$~K).

\subsubsection{The Hall conductivity}

The Hall optical conductivity of graphene, $\sigma_{xy}(\omega)$,
follows directly from Eq.~(\ref{eq:Ohm's Law}); choosing $i=y$,
$j=x$, we obtain 
\begin{equation}
\sigma_{xy}(\omega)=-g_{s}g_{v}\times\frac{\tilde{J}_{y}(\omega)}{\tilde{E}_{x}(\omega)}\,,\label{eq:Hall_Cond}
\end{equation}
 where we have invoked graphene's sixfold crystallographic symmetry
to write $\sigma_{xy}(\omega)=-\sigma_{yx}(\omega)$. The central
quantity to be computed this time is the average value of the current
density operator along the $y$ direction; using Eqs.~(\ref{eq:Eigenstates_Field})
and (\ref{eq:Field_Operator_Field}), we get 
\begin{equation}
J_{y}(t)=-ev_{F}\sum_{n,n^{\prime}}\langle n,k_{y}|j_{y}|n^{\prime},k_{y}\rangle\langle\hat{c}_{n,k_{y}}^{\dagger}(t)\hat{c}_{n^{\prime},k_{y}^{\prime}}(t)\rangle\,.\label{eq:J_y_Landau}
\end{equation}
 The non-zero matrix elements read 
\begin{align}
\langle0,k_{y}|j_{y}|\pm1,k_{y}\rangle= & -i\frac{ev_{F}}{\sqrt{2}A}\,,\label{eq:dip_mat_y_1}\\
\langle n,k_{y}|j_{y}|n^{\prime},k_{y}\rangle= & -\frac{ev_{F}}{2A}\left[\textrm{sign}(n^{\prime})\delta_{|n|-1,|n^{\prime}|}\right.\nonumber \\
 & \left.+\textrm{sign}(n)\delta_{|n|,|n^{\prime}|-1}\right]\,,\label{eq:dip_mat_y_2}
\end{align}
 (plus respective complex conjugates) where, in the last line, $n,n^{\prime}\neq0$.
Omitting the summation over $k_{y}$, the total current density reads
\begin{align}
\hat{J}_{y}(t) & =\frac{i}{\sqrt{2}A}ev_{F}\left(c_{1}^{\dagger}a_{0}+v_{1}^{\dagger}a_{0}-\textrm{h.c.}\right)\nonumber \\
 & -\frac{1}{2A}ev_{F}\sum_{n\ge1}\left(\hat{P}_{n}^{(1)}-\hat{P}_{n}^{(2)}+\textrm{h.c.}\right)\nonumber \\
 & -\frac{1}{2A}ev_{F}\sum_{n\ge1}\left(c_{n}^{\dagger}c_{n+1}-v_{n}^{\dagger}v_{n+1}+\textrm{h.c.}\right).\label{eq:Py_total_field}
\end{align}
 The EOMs resemble those derived for the longitudinal conductivity
{[}Eqs.~(\ref{eq:EOM_Pn1})-(\ref{eq:EOM_Q1}){]}, the reason being
that the current matrix elements in the $x$ and $y$ directions are
the same except for phase factors {[}compare Eqs.~(\ref{eq:dip_mat_1})
and (\ref{eq:dip_mat_2}) with Eqs.~(\ref{eq:dip_mat_y_1}) and (\ref{eq:dip_mat_y_2}){]}.
The final formula (after regularization) yields, 
\begin{equation}
\sigma_{xy}^{\textrm{reg}}(\omega)=\frac{e^{2}}{h}\sum_{n\neq m=-N_{c}}^{N_{c}}\frac{\Lambda_{nm}^{xy}}{iE_{nm}}\frac{n_{F}(E_{n})-n_{F}(E_{m})}{\hbar\omega+E_{nm}+i\Gamma}\,,\label{eq:sigma_xy_final}
\end{equation}
 with matrix elements $\Lambda_{mn}^{xy}$ related to $\Lambda_{nm}^{xx}$
{[}Eq.~(\ref{eq:sigma_xx_final}){]} according to, 
\begin{equation}
\Lambda_{nm}^{xy}=i\Lambda_{nm}^{xx}(\delta_{|m|,|n|-1}-\delta_{|m|-1,|n|})\,.\label{eq:Eta_mn_xy}
\end{equation}
 Likewise $\sigma_{xx}(\omega)$, the result for the Hall conductivity
based on the EOM method coincides with the result obtained using Green
functions calculations.\cite{CarbotteIJMP2007}

\begin{figure}
\begin{centering}
\includegraphics[clip,width=0.7\columnwidth]{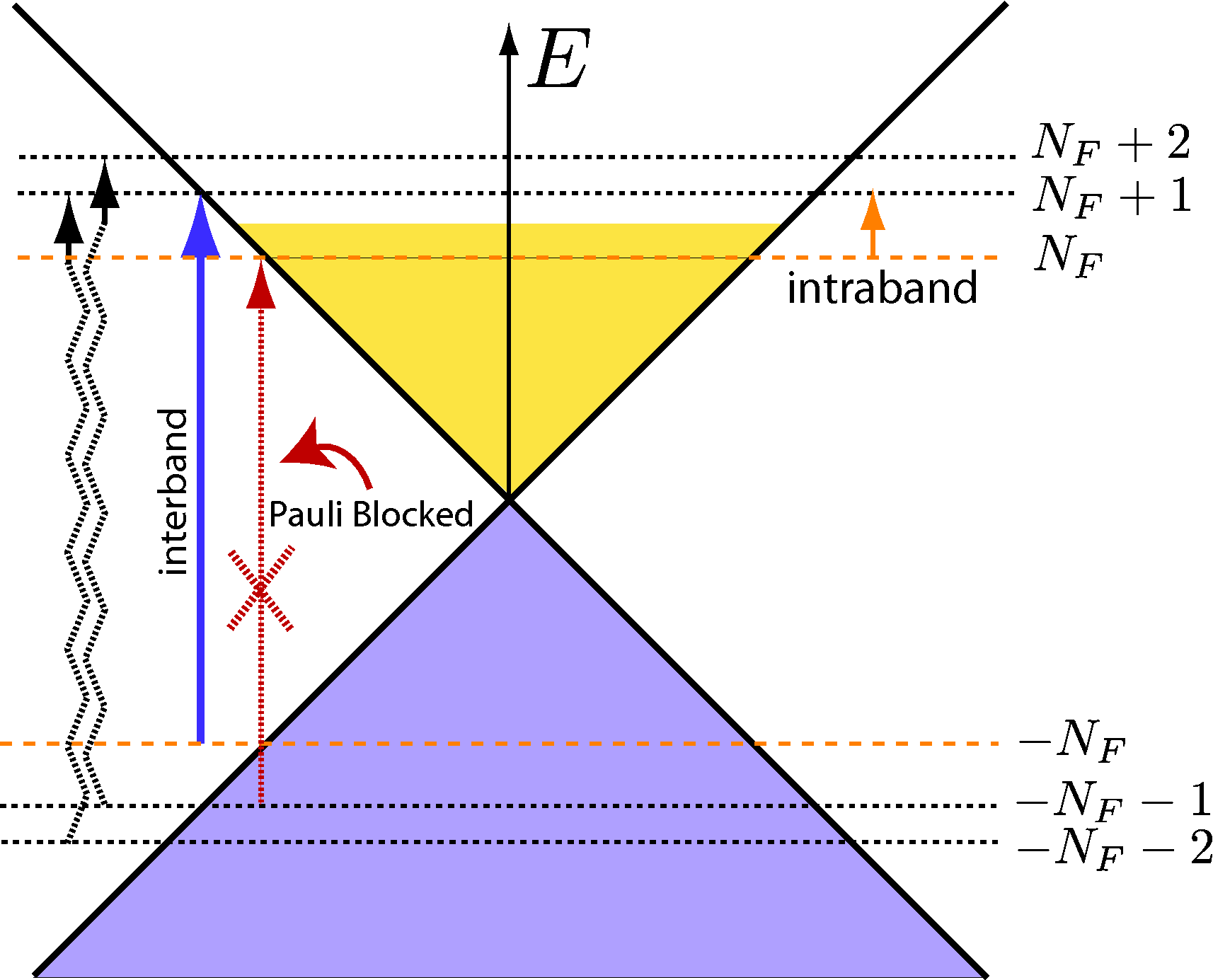} 
\par\end{centering}

\caption{\label{fig:Transitions_Hall}Schematic of electronic transitions contributing
to the Hall conductivity of doped graphene in a magnetic field. Contrary
to the longitudinal conductivity (Fig.~\ref{fig:Int_Transitions_SXX_Field}),
symmetry implies that only interband transitions involving the smallest
energy difference, $\hbar\Delta\Omega_{N_{F}}=E_{N_{F}}+E_{N_{F}+1}$,
contribute to $\sigma_{xy}$. The remaining interband transitions
($\Delta\Omega_{n}$, with $n>N_{F}$) come in pairs whose contribution
to the Hall current mutually cancel as explained in the text: an example
of a pair of interband transitions that cancel is shown in zig-zag
arrows. Note: The schematic picture is strictly adequate for $N_{F}\ge1$;
the case of $N_{F}=0$ admits a single type of electronic transition,
namely, $n=0\rightarrow n=1$. }
\end{figure}

Symmetry considerations imply that only two terms contribute in general
for the zero-temperature Hall conductivity, and hence the formula
Eq.~(\ref{eq:sigma_xy_final}) can be considerably simplified. The
first term is the intraband contribution and reads,
\begin{equation}
\sigma_{xy}^{\textrm{intra}}(\omega)=\frac{e^{2}}{h}\frac{2\hbar^{2}v_{F}^{2}}{l_{B}^{2}}\frac{1-\delta_{N_{F},0}}{\left(\hbar\omega+i\Gamma\right)^{2}-\hbar^{2}\Delta\omega_{N_{F}}^{2}}\,,\label{eq:sigma_xy_intra}
\end{equation}
 and the second is interband-like, connecting electronic states with
$n=-N_{F}$ and $n=N_{F}+1$, and reads,
\begin{equation}
\sigma_{xy}^{\textrm{inter}}(\omega)=\frac{e^{2}}{h}\frac{2\hbar^{2}v_{F}^{2}}{l_{B}^{2}}\frac{1+\delta_{N_{F},0}}{\left(\hbar\omega+i\Gamma\right)^{2}-\hbar^{2}\Delta\Omega_{N_{F}}^{2}}\,.\label{eq:sigma_xy_inter}
\end{equation}
 A single interband transition play a role in setting the Hall conductivity,
even for zero Fermi energy. This is at odds with the situation for
$\sigma_{xx}(\omega)$, where many non equivalent interband transitions
contribute to the optical spectral weight. To understand this peculiar
feature of $\sigma_{xy}(\omega)$, let us consider the second lowest
interband resonant energy, namely, $\Delta E_{2}=E_{N_{F}+2}-E_{-N_{F}-1}$:
there are two distinct sorts of interband transitions $n\rightarrow m$
involving such energy difference, namely, the pair $n_{1}=-N_{F}-2\,\wedge\, m_{1}=N_{F}+1$
and $n_{2}=-N_{F}-1\,\wedge\, m_{2}=N_{F}+2$, whose Hall matrix elements
read, $\Lambda_{n_{1}m_{1}}^{xy}=i\Lambda_{n_{1}m_{1}}^{xx}$ and
$\Lambda_{n_{2}m_{2}}^{xy}=-i\Lambda_{n_{2}m_{2}}^{xx}$, respectively.
When substituting into Eq.~(\ref{eq:sigma_xy_final}), these contributions
cancel each other at $T=0$ because $\Lambda_{n_{2}m_{2}}^{xx}=\Lambda_{n_{1}m_{1}}^{xx}$.
The same argument applies to all transitions involving an energy difference
larger than the interband gap, $\hbar\Delta\Omega_{N_{F}}$. The only
exception is indeed the interband transition $-N_{F}\rightarrow N_{F}+1$
because, contrary to interband transitions involving larger energy
differences, it cannot be canceled by the other member of the pair,
$n=-N_{F}-1\wedge m=N_{F}$, since the latter is forbidden via Pauli
blockade; a schematic picture is given in Fig.~\ref{fig:Transitions_Hall}.

The extremum points of the real part of the Hall conductivity occurs
at zero frequency, $\omega=0$, and 
\begin{align}
\omega_{\pm}^{\textrm{intra}}\simeq & \Delta\omega_{N_{F}}\pm\Gamma/\hbar\,,\label{eq:omega_intra_pm}\\
\omega_{\pm}^{\textrm{inter}}\simeq & \Delta\Omega_{N_{F}}\pm\Gamma/\hbar\,,\label{eq:omega_inter_pm}
\end{align}
 where we have considered $\Gamma/\hbar\lesssim\Delta\omega_{N_{F}}$
{[}see Eq.~(\ref{eq:intraband_peak}) and text therein{]} and made
use of $\Gamma/\hbar\ll\Delta\Omega_{N_{F}}$. The latter consideration
is true in virtually all situations except for graphene at a low electronic
density and small magnetic field $B$. Within the same accuracy, the
Hall conductivity at $\omega=0$ reads

\begin{equation}
\textrm{Re}\,\sigma_{xy}(0)\simeq-\left(\frac{1-\delta_{N_{F},0}}{\Delta\omega_{N_{F}}^{2}}+\frac{1+\delta_{N_{F},0}}{\Delta\Omega_{N_{F}}^{2}}\right)\left(\frac{4eBv_{F}^{2}}{\hbar\pi}\right)\times\sigma_{\textrm{g}}\,,\label{eq:interband_Hall_extrema_w.eq.0}
\end{equation}
 whereas at the point $\omega_{\pm}^{\textrm{intra}}$ it is given
by
\begin{equation}
\textrm{Re}\,\sigma_{xy}(\omega_{\pm}^{\textrm{intra}})\simeq F_{\Delta\omega_{N_{F}}}^{\pm}\left(\frac{1-\delta_{N_{F},0}}{\Delta\omega_{N_{F}}}\right)\left(\frac{eBv_{F}^{2}}{\pi\Gamma}\right)\times\sigma_{\textrm{g}}\,,\label{eq:interband_Hall_extrema_intra}
\end{equation}
and for $\omega_{\pm}^{\textrm{inter}}$ it reads
\begin{equation}
\textrm{Re}\,\sigma_{xy}(\omega_{\pm}^{\textrm{inter}})\simeq F_{\Delta\Omega_{N_{F}}}^{\pm}\left(\frac{1+\delta_{N_{F},0}}{\Delta\Omega_{N_{F}}}\right)\left(\frac{eBv_{F}^{2}}{\pi\Gamma}\right)\times\sigma_{\textrm{g}}\,,\label{eq:interband_Hall_extrema_inter}
\end{equation}
where we have defined $F_{\omega}^{\pm}=\pm\hbar\omega/(\hbar\omega\pm\Gamma)$.
The intensity of the Hall peaks dependence on the magnetic field intensity
$B$ is the same as for the longitudinal (intraband) peaks {[}Eq.~(\ref{eq:intraband_peak}){]},
i.e., as $\sim\sqrt{B}$. Also, similarly to $\sigma_{xx}(\omega)$,
in doped graphene with $N_{F}>1$, the interband peak is very low
compared with the intraband Hall peak for $\Delta\omega_{N_{F}}\ll\Delta\Omega_{N_{F}}$.
We, finally, remark that the anomaly associated with the zero energy
LL is present in all the latter expressions via the factor $1+\delta_{N_{F},0}$.

\begin{figure}
\begin{centering}
\includegraphics[clip,width=0.95\columnwidth]{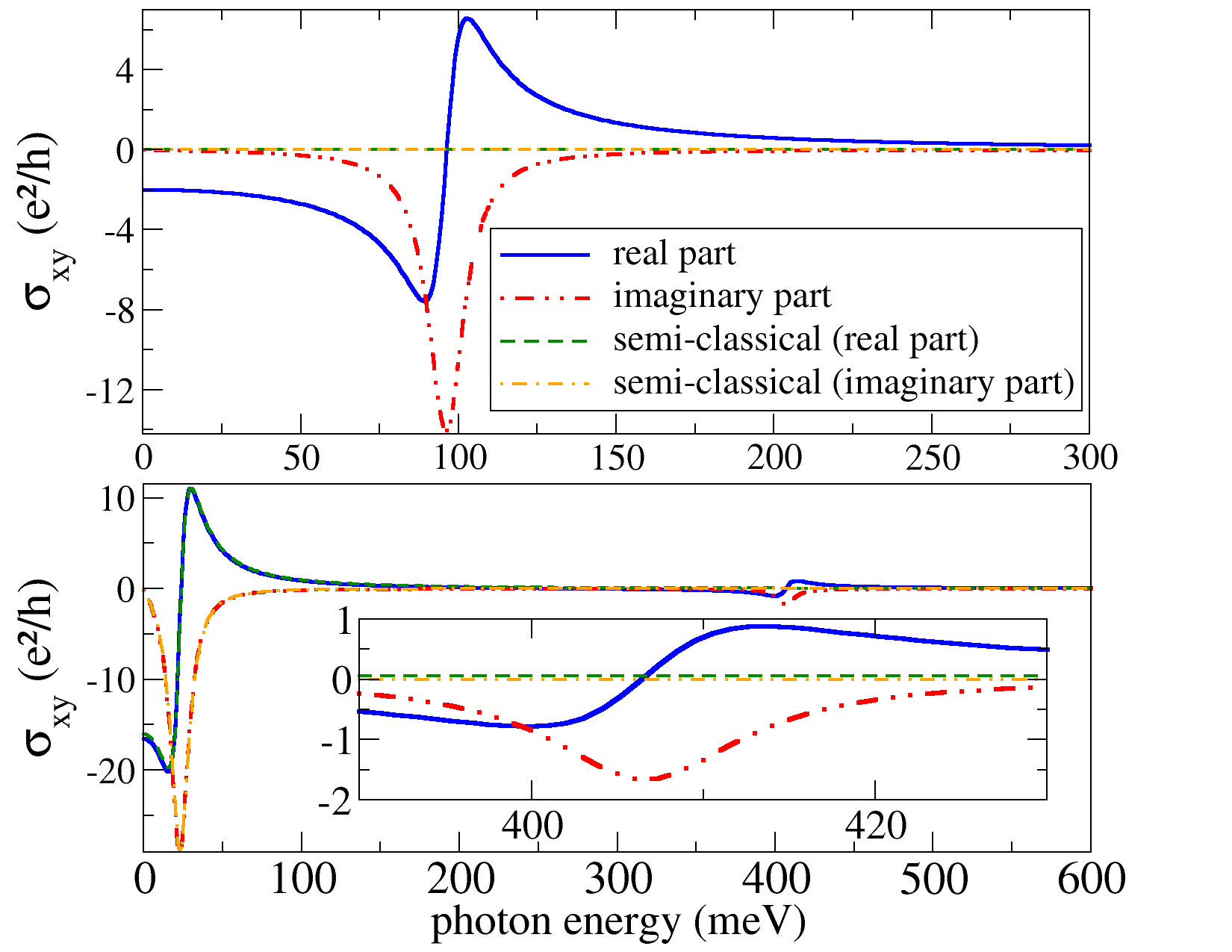} 
\par\end{centering}

\caption{\label{fig:Cond_xy_vs_freq_a}Hall conductivity as a function of photon
energy for $E_{F}=0$ (top) and $E_{F}=0.2$~eV (bottom). In both
plots $T=0$ (other parameters as in Fig.~\ref{fig:Cond_xx_vs_freq}).
At zero Fermi energy (top), $\sigma_{xy}(\omega)$ originates in a
single type of interband transition, centered at $\hbar\omega\approx E_{1}\simeq96$~meV,
and therefore cannot be described by a semi-classical treatment {[}Eq.~(\ref{eq:sigma_xy_inter}){]}.
When $E_{F}=0.2$~eV (bottom), the first four LLs are fulfilled,
which results in a classical intraband contribution {[}Eq.~(\ref{eq:sigma_xy_intra}){]},
centered at $\hbar\omega\simeq E_{5}-E_{4}\thickapprox23$~meV, and
a single interband transition {[}Eq.~(\ref{eq:sigma_xy_inter}){]}
centered at $\hbar\omega\simeq E_{5}+E_{4}\thickapprox0.4$~eV~$\simeq2E_{F}$.
The latter is shown in the inset. }
\end{figure}

Figure~\ref{fig:Cond_xy_vs_freq_a} shows the Hall conductivity of
graphene at a high magnetic field ($B=7$~T) for $N_{F}=0$ (top)
and $N_{F}=4$ (bottom), corresponding to neutral and highly doped
graphene samples, respectively. The main characteristics of $\textrm{Re}\,\sigma_{xy}(\omega)$
can be explained using Eqs.~(\ref{eq:interband_Hall_extrema_w.eq.0})-(\ref{eq:interband_Hall_extrema_inter}).
In particular, for doped graphene, the spectral weight concentrates
around two well-separated parts of the spectrum: (i) an intraband-dominated
region ($n=4\rightarrow n=5$ ), at low photon energies, with a maximum
(minimum) intensity occurring at $\hbar\omega_{+}\simeq\Delta_{4}+\Gamma\simeq30$~meV
($\hbar\omega_{-}\simeq\Delta_{4}-\Gamma\simeq16$~meV ) {[}intensity
equal to $\simeq10e^{2}/h$ ($\simeq-20e^{2}/h$), in accordance with
Eq.~(\ref{eq:interband_Hall_extrema_intra}){]}, and (ii) an interband-dominated
region $n=-4\rightarrow n=5$, at high photon energies, with a maximum
(minimum) intensity occurring at $\hbar\Omega_{+}\simeq\hbar\Delta\Omega_{4}+\Gamma\simeq413$~meV
($\hbar\Omega_{-}\simeq\hbar\Delta\Omega_{4}-\Gamma\simeq400$~meV)
{[}intensity equal to $\simeq0.81e^{2}/h$ ($\simeq-0.85e^{2}/h$),
in accordance with Eq.~(\ref{eq:interband_Hall_extrema_inter}){]}.

\begin{figure}
\begin{centering}
\includegraphics[clip,width=1\columnwidth]{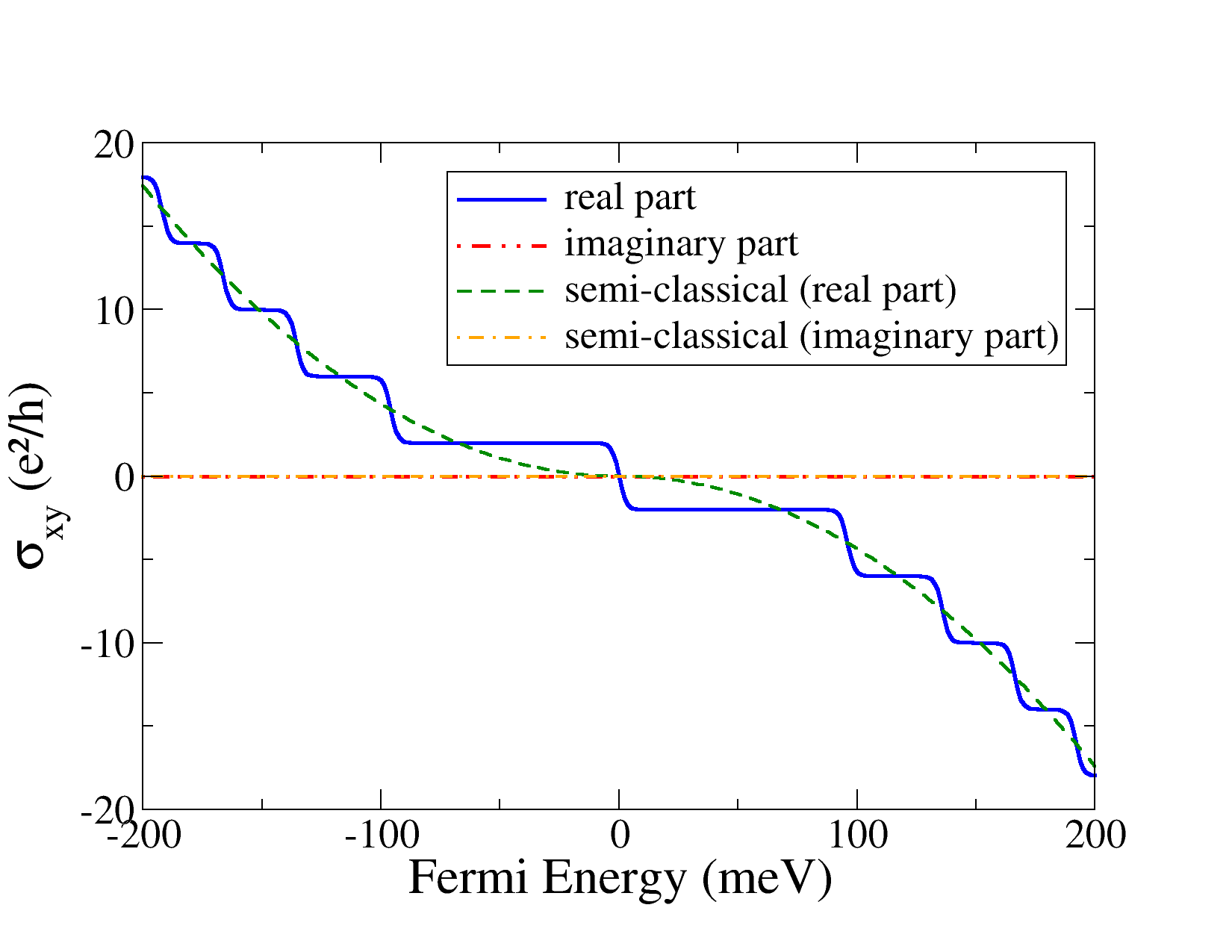} 
\par\end{centering}

\caption{\label{fig:cond_xy_vs_mu}The dc Hall conductivity as a function of
the Fermi energy. The parameters are $T=17$~K and $\Gamma=0.68$~meV.
The plateaux show Hall quantization values according to the theoretical
prediction for massless Dirac fermions {[}Eq.~(\ref{eq:Hall_quantification}){]}. }
\end{figure}

\emph{Dependence on the Fermi energy}---The variation of conductivity
with the Fermi energy reveals other peculiar feature of 2D systems:
Hall quantization.\cite{vonKlitzing,Thouless_IQHE} Figure~\ref{fig:cond_xy_vs_mu}
shows the formation of plateau in the static (or dc) Hall conductivity,
$\sigma_{xy}(0)$, a direct evidence for discrete energy levels. In
conventional 2D electron gases, the widths of such plateau are constant
(the LLs energy scales as $n$), whereas in graphene the plateau's
width decreases with increasing Fermi energy (the LLs energy scales
as $\sqrt{n}$). As for the steps heights, they are equidistant in
graphene, $\Delta\sigma_{xy}(0)=4e^{2}/h$, even when crossing $E_{F}=0$,
whereas in conventional 2D systems the step from the first electron
LL ($n=1$) and the first hole LL ($n=-1$) is twice the value of
the remaining steps (a manifestation of the zero-energy LL graphene
anomaly).

The Hall conductivity quantization rule for graphene can be readily
obtained by adding the intraband and interband Hall conductivities,
\begin{equation}
\sigma_{xy}(0)=-\frac{4e^{2}}{h}\left(N_{F}+\frac{1}{2}\right)\,,\label{eq:Hall_quantification}
\end{equation}
 where we have used $\Gamma\ll E_{1}$ in order to simplify the denominators
of Eqs.~(\ref{eq:sigma_xy_intra}) and (\ref{eq:sigma_xy_inter}).
Despite the filling factor, $\nu=4N_{F}+2$, being an integer number,
there is no complete correspondence with the conventional 2D IQHE,
for which $\sigma_{xy}=-4e^{2}N_{F}/h$; an extra $1/2$ factor due
to the contribution of the zero-energy state, shared by both electrons
and holes, shows up, which must be taken separately, making $\nu$
always even---this is known as the \emph{anomalous }IQHE and is a
hallmark of chiral massless fermions. The anomalous IQHE was predicted
theoretically\cite{AnamQHE_Sharapov05,Peres 2006} and measured\cite{Novoselov 2005,PKim2005}
in the early days of graphene.

\subsection{The semiclassical solution\label{sub:SemiClassSolution}}

Within the semiclassical approach, the spinorial nature of the electrons'
wave function is immaterial. On the other hand, the massless nature
of the spectrum invalidates a straightforward Drude-like approach
\cite{Ziman,Mermin} to the calculation of the transport coefficients,
and Boltzmann transport theory is required, since in this formulation
the central quantity to be computed is the deviation of the momentum
distribution function from the equilibrium Fermi distribution.

In the semiclassical regime (that is, a high electronic density and/or
low magnetic fields), the physics of the Hall effect can be explained
in terms of Boltzmann's theory of transport, where the electric current
is given, in the case of graphene, by:
\begin{equation}
\boldsymbol{J}=\frac{e^{2}}{h}\int d\mathbf{k}g(B,\mathbf{k},\omega)\boldsymbol{v}_{\mathbf{k}}\,,\label{eq:semi-classical_current}
\end{equation}
 with spin and valley degeneracies included, and where $g(B,\mathbf{k},\omega)\equiv g_{\mathbf{k}}$,
is the deviation of the carriers' (electrons or holes) distribution
function from the equilibrium Fermi distribution, $f_{0}(\epsilon)$,
$e$ is the charge of the carrier, the static magnetic field \textbf{$B$}
is considered to be perpendicular to graphene's surface, $\omega$
is the frequency of the electromagnetic field, and the carrier's velocity
reads $\boldsymbol{v}_{\mathbf{k}}=(v_{x},v_{y})=v_{F}(\cos\theta,\sin\theta)$.
In the presence of both an electric and a magnetic field, the distribution
$g_{\mathbf{k}}$ is the solution of the equation\cite{Ziman} 
\begin{equation}
-e\mathbf{E}\cdot\boldsymbol{v}_{\mathbf{k}}\frac{\partial f_{0}}{\partial\epsilon}=\frac{g_{\mathbf{k}}}{\tau_{\mathbf{k}}}+\frac{\partial g_{\mathbf{k}}}{\partial t}+\frac{e}{\hbar}\left(\boldsymbol{v}_{\mathbf{k}}\times\mathbf{B}\right)\cdot\nabla_{\mathbf{k}}g_{\mathbf{k}}\,,\label{eq:boltzmann_eq}
\end{equation}
 where we have employed the the standard relaxation approximation,\cite{Ziman}
i.e., 
\begin{equation}
\left.\frac{\partial f_{\mathbf{k}}}{\partial t}\right|_{\textrm{scatt}}=-\frac{g_{\mathbf{k}}}{\tau_{\mathbf{k}}}\,,\label{eq:relaxation_approx}
\end{equation}
where $\tau_{\mathbf{k}}$ is the relaxation scattering time, $\mathbf{E}=(E_{0,x},E_{0,y})$
is the electric field, and $\nabla_{\mathbf{k}}$ is the gradient
operator with respect to the momentum $\mathbf{k}$. Writing $g_{\mathbf{k}}$
as,
\begin{equation}
g_{\mathbf{k}}=e^{-i\omega t}\mathbf{k}\cdot\mathbf{A}\,,\label{eq:g_k}
\end{equation}
 and noting that $(\boldsymbol{v}_{\mathbf{k}}\times\mathbf{B})\cdot\nabla_{\mathbf{k}}g_{\mathbf{k}}=\boldsymbol{v}_{\mathbf{k}}\cdot(\mathbf{B}\times\nabla_{\mathbf{k}}g_{\mathbf{k}})$,
Eq.~(\ref{eq:boltzmann_eq}) can be solved exactly, where the vector
\textbf{$\mathbf{A}$} needs to be determined. Solving Eq.~(\ref{eq:boltzmann_eq}),
the components of the vector $A=(A_{x},A_{y})$ are obtained in the
form
\begin{align}
A_{x} & =\frac{(1-i\omega\tau_{\mathbf{k}})E_{x}-\tau_{\mathbf{k}}\omega_{c}E_{y}}{(1-i\omega\tau_{\mathbf{k}})^{2}+\omega_{c}^{2}\tau_{\mathbf{k}}^{2}}\,,\label{eq:vec_A_x}\\
A_{y} & =\frac{(1-i\omega\tau_{\mathbf{k}})E_{y}+\tau_{\mathbf{k}}\omega_{c}E_{x}}{(1-i\omega\tau_{\mathbf{k}})^{2}+\omega_{c}^{2}\tau_{\mathbf{k}}^{2}}\,,\label{eq:vec_A_y}
\end{align}
 where 
\begin{equation}
\omega_{c}=ev_{F}^{2}B/|E_{F}|\,,\label{eq:cyclotron_freq}
\end{equation}
 is the graphene's cyclotron frequency, and $E_{x(y)}$ is defined
as
\begin{equation}
E_{x(y)}=-eE_{0,x(y)}v_{x(y)}\frac{\partial f_{0}}{\partial\epsilon}\,.\label{eq:def_Ex/y}
\end{equation}
 Introducing $g_{\mathbf{k}}$ in Eq.~(\ref{eq:semi-classical_current}),
and assuming $T=0$, we obtain the components of the conductivity
tensor, which read
\begin{align}
\sigma_{xx} & =\frac{e^{2}}{h}\frac{2|E_{F}|\tau_{k_{F}}}{\hbar}\frac{1-i\omega\tau_{k_{F}}}{(1-i\omega\tau_{k_{F}})^{2}+\omega_{c}^{2}\tau_{k_{F}}^{2}}\,,\label{eq:sigma_xx_semiclass}\\
\sigma_{xy} & =-\frac{e^{2}}{h}\frac{2E_{F}\tau_{k_{F}}}{\hbar}\frac{\omega_{c}\tau_{k_{F}}}{(1-i\omega\tau_{k_{F}})^{2}+\omega_{c}^{2}\tau_{k_{F}}^{2}}\,,\label{eq:sigma_xy_semiclass}
\end{align}
Note that setting $\omega_{c}=0$ in Eq.~(\ref{eq:sigma_xx_semiclass})
leads to the semi-classical longitudinal conductivity at zero field
mentioned in Sec.~\ref{sub:ZeroMagField}.

\emph{Validity of the semiclassical calculation}---The results presented
so far demonstrate the reliability of the Boltzmann approach in regions
of the spectrum where the optical weight is mostly due to intraband
transitions. This is borne out in Fig.~\ref{fig:Cond_xx_vs_freq_2}
{[}Fig.~\ref{fig:Cond_xy_vs_freq_a} (bottom){]}, where $\sigma_{xx}(\omega)$
{[}$\sigma_{xy}(\omega)${]} is plotted as a function of $\hbar\omega$,
for $B=7$~T and $E_{F}=0.2$~eV: the agreement between the real
part (imaginary part) of the quantum calculation shown by the solid
(blue) line {[}dashed-double-dotted (red) line{]} and the semi-classical
calculation shown in dashed (green) curve {[}dashed-dotted (orange)
line{]} in these figures is confined to energies $\hbar\omega\lesssim2E_{F}$.
For high photon frequencies---more precisely, above the threshold
for interband transitions, $\hbar\omega\simeq2E_{F}$---the conductivity
cannot be described by Boltzmann's transport theory.

The fine agreement observed at low photon energies is not accidental
and ceases to occur only for a very low Fermi energy. To see why,
we note that Eqs.~(\ref{eq:s_xx_intra_T=00003D00003D0}) and (\ref{eq:sigma_xy_intra})
(intraband conductivity) and Eqs.~(\ref{eq:sigma_xx_semiclass})
and (\ref{eq:sigma_xy_semiclass}) (semiclassical conductivity) coincide,
upon identification of the intraband energy gap $\Delta_{N_{F}}$,
with the cyclotron energy $\hbar\omega_{c}$. This identification
is justified when a sufficient number of LLs are filled. In fact,
expressing the Fermi energy as $E_{F}=(\hbar v_{F}/l_{B})\sqrt{2N^{\star}}$,
we obtain $\Delta_{N_{F}}\rightarrow\hbar\omega_{c}$ provided that
\begin{equation}
\sqrt{N_{F}+1}-\sqrt{N_{F}}\rightarrow\frac{1}{2\sqrt{N^{\star}}}\,.\label{eq:quantum-classical_trans}
\end{equation}
Noting that $N_{F}=\textrm{int}[N^{\star}]$, we then see that the
latter limit is achieved when $N^{\star}\gg1$, as anticipated.

\begin{figure}
\begin{centering}
\includegraphics[clip,width=0.95\columnwidth]{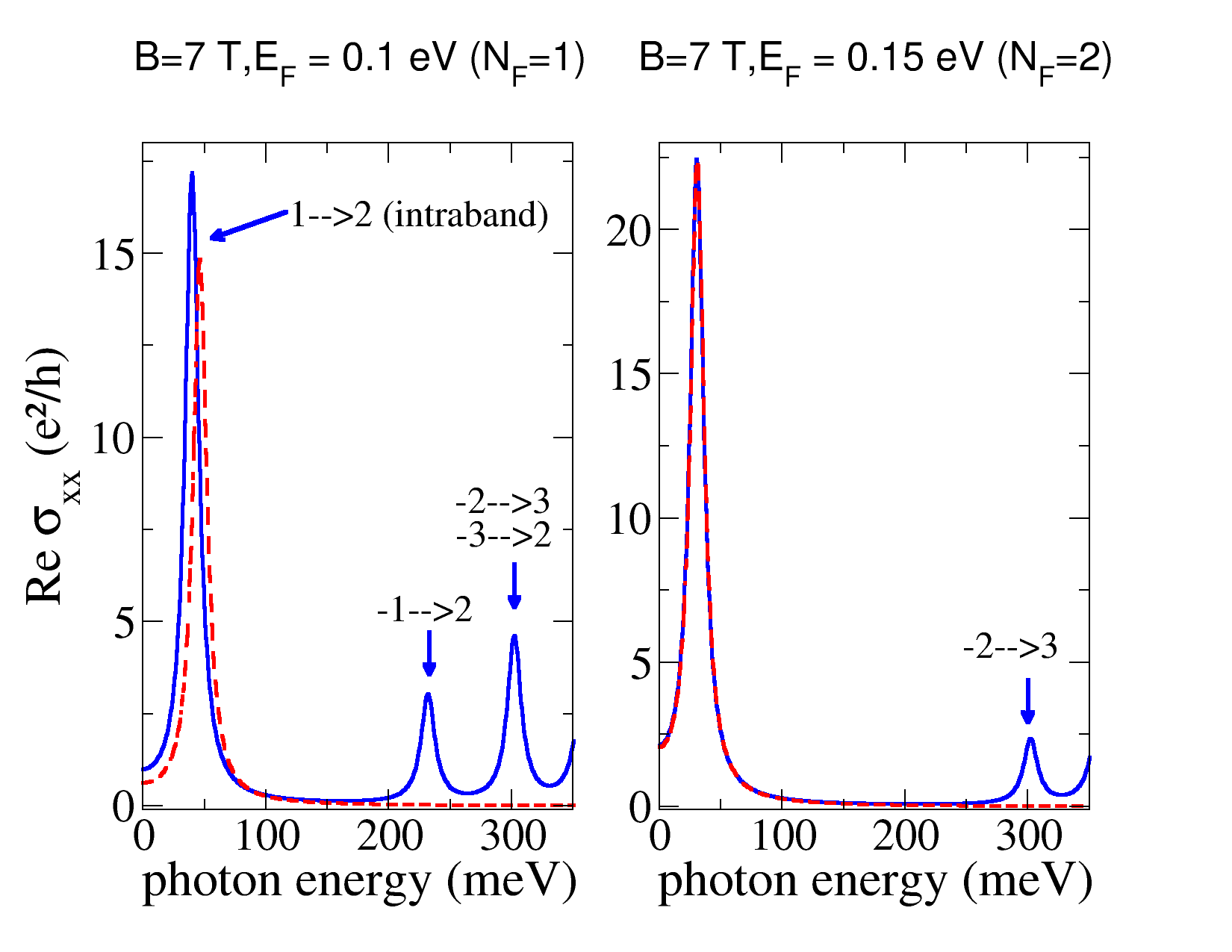} 
\par\end{centering}

\caption{\label{fig:CondXX_Re_2_cases}The real part of the longitudinal conductivity
is plotted as a function of the photon energy for $E_{F}=0.1$~eV
(left) and $E_{F}=0.15$~eV (right). In these plots, $B=7$~T and
$\Gamma=6.8$~meV. The dashed (red) represents the semiclassical
result {[}Eq.~(\ref{eq:sigma_xx_semiclass}){]} and the solid (blue)
line represents the EOM quantum solution {[}Eq.~(\ref{eq:sigma_xx_final}){]}.
Note that, in the right panel, there is no interband peak $n=-1\rightarrow m=2$,
at $E_{F}\approx230$~meV, and the peak at $\hbar\omega\approx300$~meV
loses half of its intensity because the $n=-3\rightarrow m=2$ transitions
get blocked when the Fermi energy crosses the LL with $n=2$.}
\end{figure}

For the parameters in Fig.~\ref{fig:Cond_xx_vs_freq_2} (see also
Fig.~\ref{fig:Cond_xy_vs_freq_a}, bottom), even though only a few
LLs are fulfilled, i.e., $N_{F}=4$, the values of $\Delta_{N_{F}}$
and $\hbar\omega_{c}$ are quite similar, $\Delta_{N_{F}}=0.0226$~eV
and $\hbar\omega_{c}=0.0230$~eV, explaining the consistence between
the two theories in describing the intraband electronic transport.
In practice, only for a very low Fermi energy and/or a very high magnetic
field, such that $N_{F}=0$, does the semi-classical calculation fail
to describe the conductivity in the whole optical spectrum, since
all transitions are interband-like in this case. Remarkably, already
for a single occupied LL, $N_{F}=1$, the semiclassical calculation
provides a reasonable description of the optical conductivity, as
long as one remains inside the portion of the spectrum where the interband
processes have little or no weight, that is, $\hbar\omega\lesssim E_{1}+E_{2}$
(see Fig.~\ref{fig:CondXX_Re_2_cases}). We note again, however,
that the intraband region extends for a large range of frequencies
given the large intrinsic cyclotron gap of graphene.

In summary, the validity of the semiclassical calculation is bound
to photon energies below the interband threshold, $\hbar\omega\lesssim E_{N_{F}}+E_{N_{F}+1}$,
and for not a too low Fermi energy, $N_{F}\gtrsim1$. For the parameters
used in Figs.~\ref{fig:Cond_xx_vs_freq_2}, \ref{fig:Cond_xy_vs_freq_a}
(bottom), \ref{fig:CondXX_Re_2_cases} and \ref{fig:Faraday_Various},
we list in Table I the corresponding values of $\Delta_{N_{F}}/\hbar\omega_{c}$.
These figures have $N_{F}>1$ and hence the semi-classical conductivity
agrees well within the far-infrared part of the spectrum. For completeness,
Fig.~\ref{fig:CondXX_Re_2_cases} shows the real part of $\sigma_{xx}(\omega)$
for $N_{F}=1$ (left) and $N_{F}=2$ (right). The former has $\Delta_{N_{F}}/\hbar\omega_{c}\simeq0.86$
and hence the semi-classical calculation is only partially accurate.
In particular, it underestimates the maximum intensity for intraband
light absorption. The right panel in Fig.~\ref{fig:CondXX_Re_2_cases},
with $N_{F}=2$, has $\Delta_{N_{F}}/\hbar\omega_{c}\simeq0.99$,
which explains the excellent agreement between the two curves in the
intraband region, $\hbar\omega\lesssim0.3$ eV.

\begin{table}[H]
 \centering{}%
\begin{tabular}{|c|c|c|c|}
\hline 
$B$ (T)  & $E_{F}$ (eV)  & $N_{F}$  & $\Delta_{N_{F}}/\hbar\omega_{c}$\tabularnewline
\hline 
\hline 
1  & 0.30  & 68  & 0.9990\tabularnewline
\hline 
2  & 0.30  & 34  & 0.9954\tabularnewline
\hline 
5  & 0.30  & 13  & 1.0066\tabularnewline
\hline 
7  & 0.20  & 4  & 0.9837\tabularnewline
\hline 
7  & 0.15  & 2  & 0.9933\tabularnewline
\hline 
7  & 0.10  & 1  & 0.8629\tabularnewline
\hline 
\end{tabular}\caption{\label{tab:TableI}Values of several relevant quantities related to
the numerical simulations given in Figs.~\ref{fig:Cond_xx_vs_freq_2},\ref{fig:Cond_xy_vs_freq_a},\ref{fig:CondXX_Re_2_cases}
, and \ref{fig:Faraday_Various}. The agreement between the semiclassical
calculation and the quantum intraband expression comes from the similarity
between $\Delta_{N_{F}}$ and $\hbar\omega_{c}$.}
\end{table}

Having presented the calculation method for the magneto-optical properties
of graphene based on the EOM method, we we now turn to study of the
Faraday effect.

\section{The Faraday effect in graphene\label{sec:Application:-the-Faraday}}

We discuss the transmission of electromagnetic radiation between two
dielectric media separated by graphene. The scattering geometry is
given in Fig.~\ref{fig:FaradayRot_Graph}, where the transverse magnetic
mode is chosen as a particular example. Since we are interested in
a normal incidence, there is no distinction between the transverse
magnetic and the transverse electric modes.

The present section is organized as follows: in Sec.~\ref{sub:Faraday_Rot},
we derive general expressions for transmission, ellipticity, and Faraday
rotation angle. These quantities depend on the frequency of the impinging
light, $\omega$, magnitude of the (transverse) magnetic field, $B$,
scattering mechanisms (i.e., level broadening, $\Gamma$), temperature
$T$, and Fermi energy, $E_{F}$, via the magneto-optical conductivity
tensor of graphene derived in Sec.~\ref{Sec2 (Magneto-Optical Transport)}.

Our theoretical results are tested against experimental data measured
recently by Crassee \emph{et al.} using graphene samples with a high
electronic density\emph{.\cite{FaradayNaturePhys}} The limit of a
low electronic density is studied in Sec.~\ref{sub:Low_electronic_dens},
where the Faraday rotation angle is shown to display quantum jumps
as a function of the Fermi energy.

Finally, in Sec.~\ref{sub:Enhancement-of-Faraday}, an experimental
setup is proposed that is able to greatly enhance the Faraday rotation
angle in the entire optical spectrum.

\subsection{Faraday rotation in graphene\label{sub:Faraday_Rot}}

\begin{figure}
\begin{centering}
\includegraphics[clip,width=0.9\columnwidth]{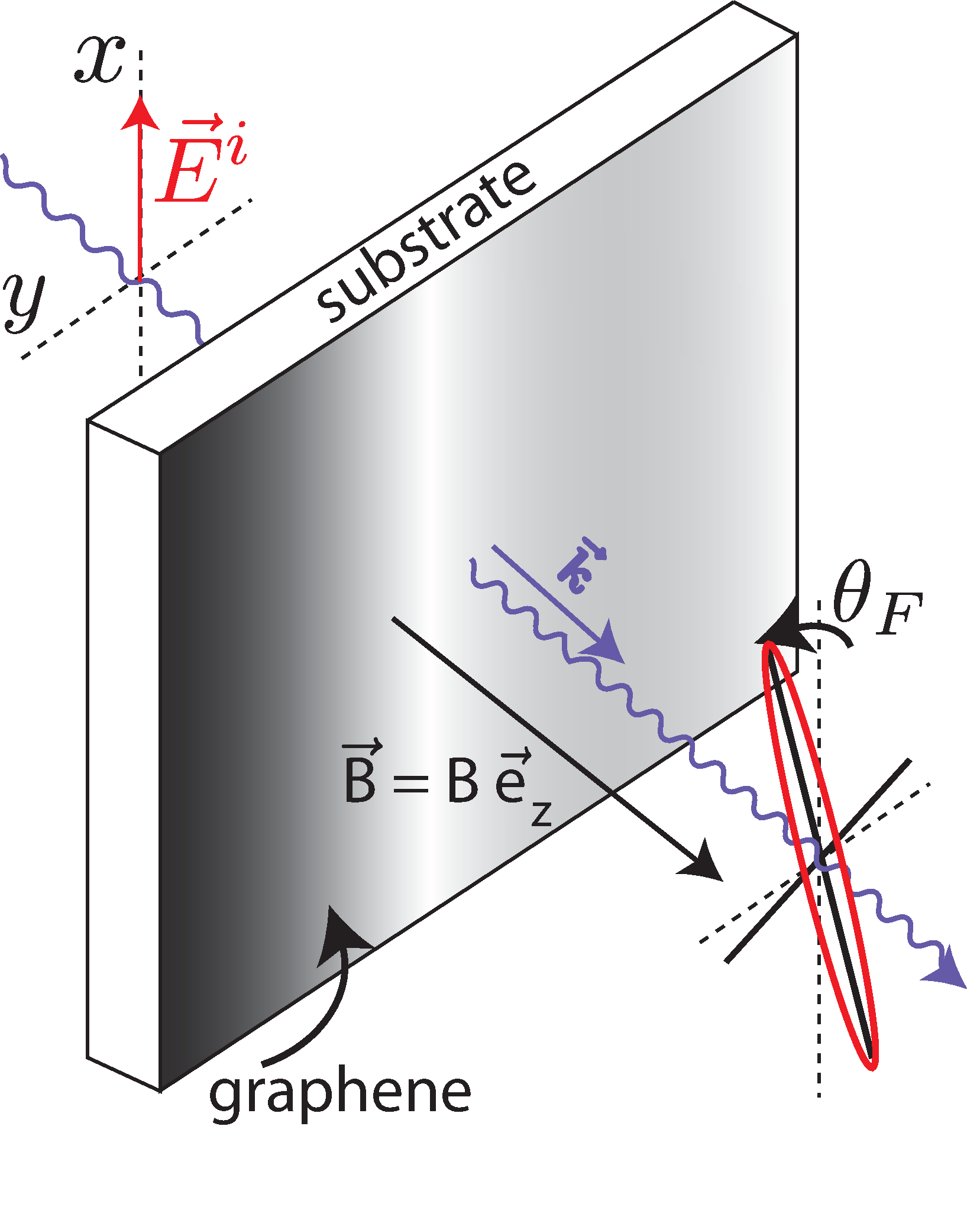} 
\par\end{centering}

\caption{\label{fig:FaradayRot_Graph}Schematic of the Faraday effect: an electromagnetic
wave polarized in the $xy$ plane (transverse magnetic mode) and traveling
in the positive $z$ direction passes through a graphene film subjected
to a transverse magnetic field $B$. In this case, graphene is adhered
to a substrate (typically SiO$_{2}$), but the experiment can also
be made with suspended graphene. The transmitted field sees its plane
of polarization rotated by an angle $\theta_{F}$ and acquires a certain
degree of ellipticity. }
\end{figure}

We now solve the problem posed in Fig.~\ref{fig:FaradayRot_Graph},
considering only a single graphene sheet separating two dielectrics.
In what follows, we assume that graphene is deposited on top of a
lossless dielectric medium (i.e., fully transparent to impinging light),
of relative permittivity $\epsilon_{r}$. The generalization of the
problem to the case of a lossy dielectric poses no difficulties, except
for the introduction of a complex index of refraction associated with
the dielectric medium. We further assume that the incoming electromagnetic
field is linearly polarized along the $x$ axis and propagates along
the $z$ direction, as shown in the diagram in Fig.~\ref{fig:FaradayRot_Graph};
that is,
\begin{equation}
\boldsymbol{E}^{i}=\mathbf{e}_{\mathbf{x}}E_{x}^{i}e^{i(qz-\omega t)}\,,\label{eq:E_i}
\end{equation}
 such that $q=\sqrt{\epsilon_{r}}\omega/c$. 

Due to the optical Faraday rotation of the plane of polarization of
the electric field, both the reflected, $\boldsymbol{E}^{r}$, and
the transmitted, $\boldsymbol{E}^{t}$, fields acquire a finite component
along the $y$-direction; that is,
\begin{align}
\boldsymbol{E}^{r}= & (E_{x}^{r},E_{y}^{r})e^{-i(qz-\omega t)}\,,\label{eq:E_r}\\
\boldsymbol{E}^{t}= & (E_{x}^{t},E_{y}^{t})e^{i(kz-\omega t)}\,,\label{eq:E_t}
\end{align}
 where $k=\omega/c$. For this problem, Maxwell's equation for the
electric field reads (in MKS units)
\begin{equation}
\frac{\partial^{2}E_{i}}{\partial z^{2}}+i\omega\mu_{0}\delta(z)\sum_{j=x,y}\sigma_{ij}E_{j}+\omega^{2}\epsilon_{r}\mu E_{i}=0\,,\label{eq:maxwell_eq}
\end{equation}
 where $E_{i}$ is the $i$-component of the electric field (we have
$i=x,y$), $\mu_{0}$ is the vacuum permeability, and $\sigma_{ij}$
are the components of the magneto-optical tensor of graphene (see
Sec.~\ref{Sec2 (Magneto-Optical Transport)}). The boundary conditions
at the substrate-graphene-air interface are the continuity of the
tangential components of the electric field at the surface of graphene
($z=0$),
\begin{equation}
\left(E_{x}^{i},0\right)+\left(E_{x}^{r},E_{y}^{r}\right)=\left(E_{x}^{t},E_{y}^{t}\right)\,,\label{eq:continuity_field}
\end{equation}
 and (the derivatives are evaluated at $z=0$)
\begin{equation}
\frac{\partial E_{l}^{t}}{\partial z}-\frac{\partial E_{l}^{i}}{\partial z}-\frac{\partial E_{l}^{r}}{\partial z}=-i\omega\mu_{0}\sum_{j=x,y}\sigma_{lj}E_{j}^{t}\,,\label{eq:continuity_derivative}
\end{equation}
 where the last condition was derived from integrating Eq.~(\ref{eq:maxwell_eq})
in the interval $z\in[0^{-},0^{+}]$ and $l=x,y$. Calculation of
the transmitted intensities becomes easier to perform if we rewrite
the boundary conditions in terms of circularly polarized waves:
\begin{align}
-2qE_{x}^{i}+(k+q)E_{\pm}^{t} & =-\mu\omega\sigma_{\mp}E_{\pm}^{t}\,,\label{eq:circular1}
\end{align}
 where $E_{\pm}=E_{x}\pm iE_{y}$ and $\sigma_{\pm}=\sigma_{xx}\pm i\sigma_{xy}$,
for in this representation the two circular polarizations decouple
from each other. From Eq.~(\ref{eq:circular1}) the transmission
amplitudes follow in the form:
\begin{equation}
t_{\pm}\equiv\frac{E_{\pm}^{t}}{E_{x}^{i}}=\frac{2\sqrt{\epsilon_{r}}}{1+\sqrt{\epsilon_{r}}+c\mu_{0}\sigma_{\mp}}=|t_{\pm}|e^{i\theta_{\pm}}\,.\label{eq:t_p_m}
\end{equation}

\begin{figure}
\begin{centering}
\includegraphics[clip,width=0.9\columnwidth]{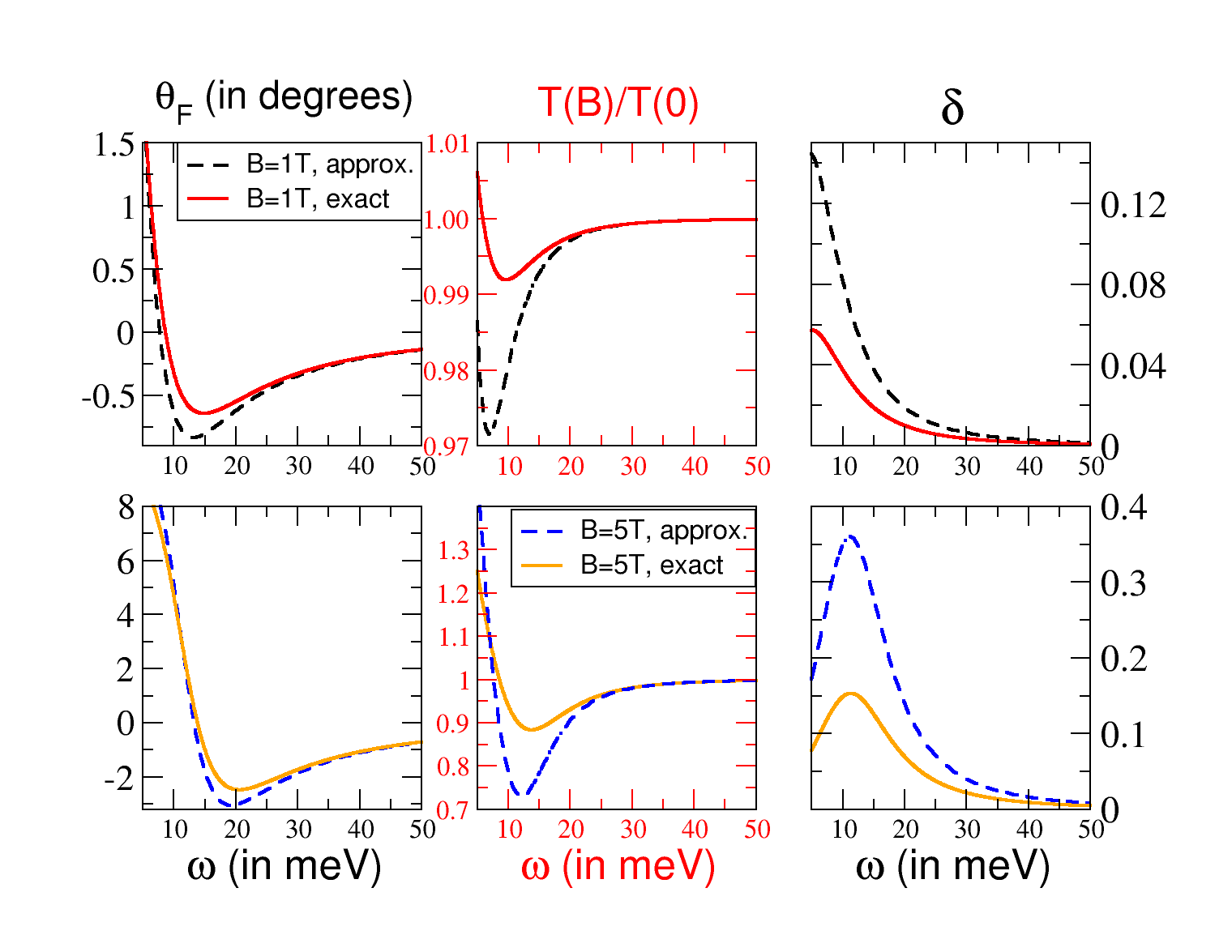} 
\par\end{centering}

\bigskip{}

\begin{centering}
\includegraphics[clip,width=0.9\columnwidth]{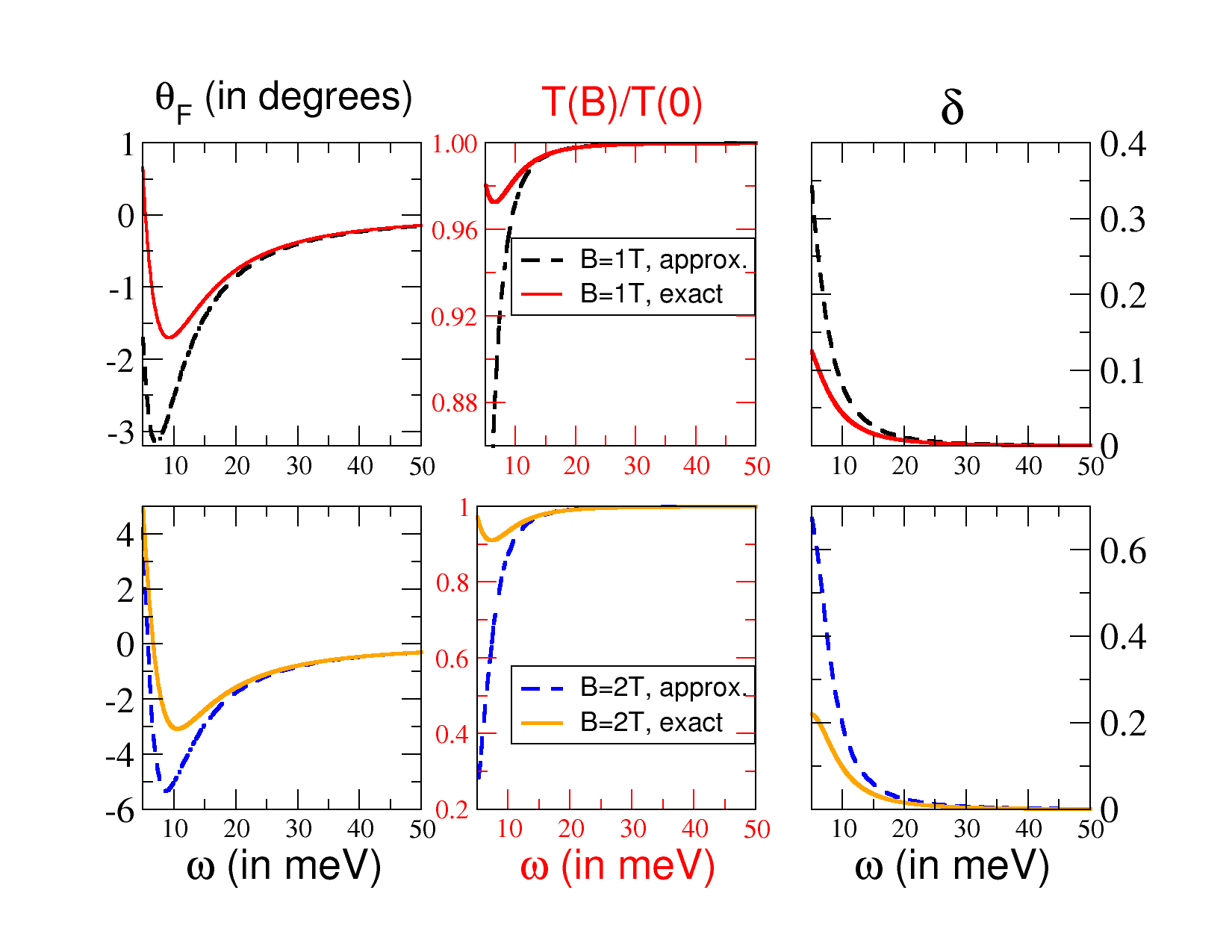} 
\par\end{centering}

\caption{\label{fig:Faraday_Various}Faraday rotation angle (in degrees), normalized
transmittance, and ellipticity of electromagnetic radiation passing
through graphene subjected to a perpendicular magnetic field. The
graphene sample is assumed to have a finite electronic density, $E_{F}=0.3$~eV,
and to be on top of SiO$_{2}$ ($\epsilon_{r}=3.9$). Top six panels:
Simulation of $\theta_{F}$, $T(B)/T(0)$, and $\delta$, considering
a broadening of $\Gamma=7$~meV. Bottom six panels: Simulation of
the same quantities as above for $\Gamma=3.7$~meV. In all panels,
dashed lines correspond to approximate calculations, as given by Eqs.~(\ref{eq:theta_F_approx})
and (\ref{eq:delta_approx}), and $T=17$~K. }
\end{figure}

The transmittance can be written as,
\begin{equation}
T(B)=\frac{1}{2\sqrt{\epsilon_{r}}}\left(|t_{+}|^{2}+|t_{-}|^{2}\right)\,,\label{eq:transmittance}
\end{equation}
where the factor $1/2$ comes from the proper normalization of circularly
polarized waves (omitted in the definition above, for simplicity of
writing) and the factor $1/\sqrt{\epsilon_{r}}$ is due to flux conservation.
Faraday's rotation angle $\theta_{F}$, and the ellipticity $\delta$
are given by\cite{Chiu,Wallace,Vassilevich}
\begin{align}
\theta_{F} & =\frac{1}{2}\left(\theta_{+}-\theta_{-}\right)\,,\label{eq:theta_faraday-1}\\
\delta & =\frac{|t_{+}|-|t_{-}|}{|t_{+}|+|t_{-}|}\,,\label{eq:delta}
\end{align}
 respectively. From Eq.~(\ref{eq:t_p_m}), $\theta_{F}$ is given
in terms of the conductivity $\sigma_{\pm}$, since
\begin{equation}
\theta_{\pm}=-\arctan\frac{\mu c\sigma_{\mp}^{\prime\prime}}{1+\sqrt{\epsilon_{r}}+c\mu\sigma_{\mp}^{\prime}}\,,\label{eq:tpm_comp}
\end{equation}
 where $\sigma_{\pm}=\sigma_{\pm}^{\prime}+i\sigma_{\pm}^{\prime\prime}$,
and $\sigma_{\pm}^{\prime}$ and $\sigma_{\pm}^{\prime\prime}$ are
the real and imaginary parts of $\sigma_{\pm}$, respectively. Explicitly,
we have
\begin{equation}
\sigma_{\pm}=\left(\sigma_{xx}^{\prime}\mp\sigma_{xy}^{\prime\prime}\right)+i\left(\sigma_{xx}^{\prime\prime}\pm\sigma_{xy}^{\prime}\right)\,,\label{eq:sigma_pm}
\end{equation}
 from which follows the approximate expression
\begin{equation}
\theta_{F}\approx-\frac{c\mu_{0}}{1+\sqrt{\epsilon_{r}}}\sigma_{xy}^{\prime}\,,\label{eq:theta_F_approx}
\end{equation}
 where we have assumed that $\theta_{F}\lesssim1$ and that $1+\sqrt{\epsilon_{r}}\gg c\mu_{0}\sigma_{\mp}^{\prime}$.
The latter assumption is the more stringent of the two. For comparison,
in the numerical studies shown in Fig.~\ref{fig:Faraday_Various},
we represent both the exact and the approximate results for $\theta_{F}$,
$\delta$, and $T$. This allows us to check the validity of the approximate
results. Discarding terms of the order of $(c\mu_{0}\sigma_{\mp})^{2}$
in Eq.~(\ref{eq:transmittance}), we obtain an approximate expression
for the total transmitted light in the form
\begin{equation}
T(B)\approx\frac{4\sqrt{\epsilon_{r}}}{(1+\sqrt{\epsilon_{r}})^{2}}\left(1-\frac{2c\mu_{0}}{1+\sqrt{\epsilon_{r}}}\sigma_{xx}^{\prime}\right)\,.\label{eq:T_approx}
\end{equation}
Within the same degree of approximation used to derive Eq.~(\ref{eq:theta_F_approx}),
the ellipticity is given by
\begin{equation}
\delta\approx-\frac{2c\mu}{1+\sqrt{\epsilon_{r}}}\sigma_{xy}^{\prime\prime}\,.\label{eq:delta_approx}
\end{equation}

The validity of these approximations depends on the photon frequency,
as can be seen in Fig.~\ref{fig:Faraday_Various}. In what follows,
the exact expression is used in all numerical studies.

In our simulations of the Faraday effect, we assume broadenings of
the order of $10$~meV. Our assumption is consistent with the values
found in pump-probe experiments performed in exitaxial and exfoliated
graphene samples\cite{pump_probe_epitaxial,pump_probe_exfoliated},
and in infrared spectroscopy studies of the Drude conductivity of
graphene.\cite{Horgn2011}

\subsection{Fit to experimental data in the high-density regime\label{sub:Fit_Exp_high_doping}}

\begin{figure}
\begin{centering}
\includegraphics[clip,width=0.9\columnwidth]{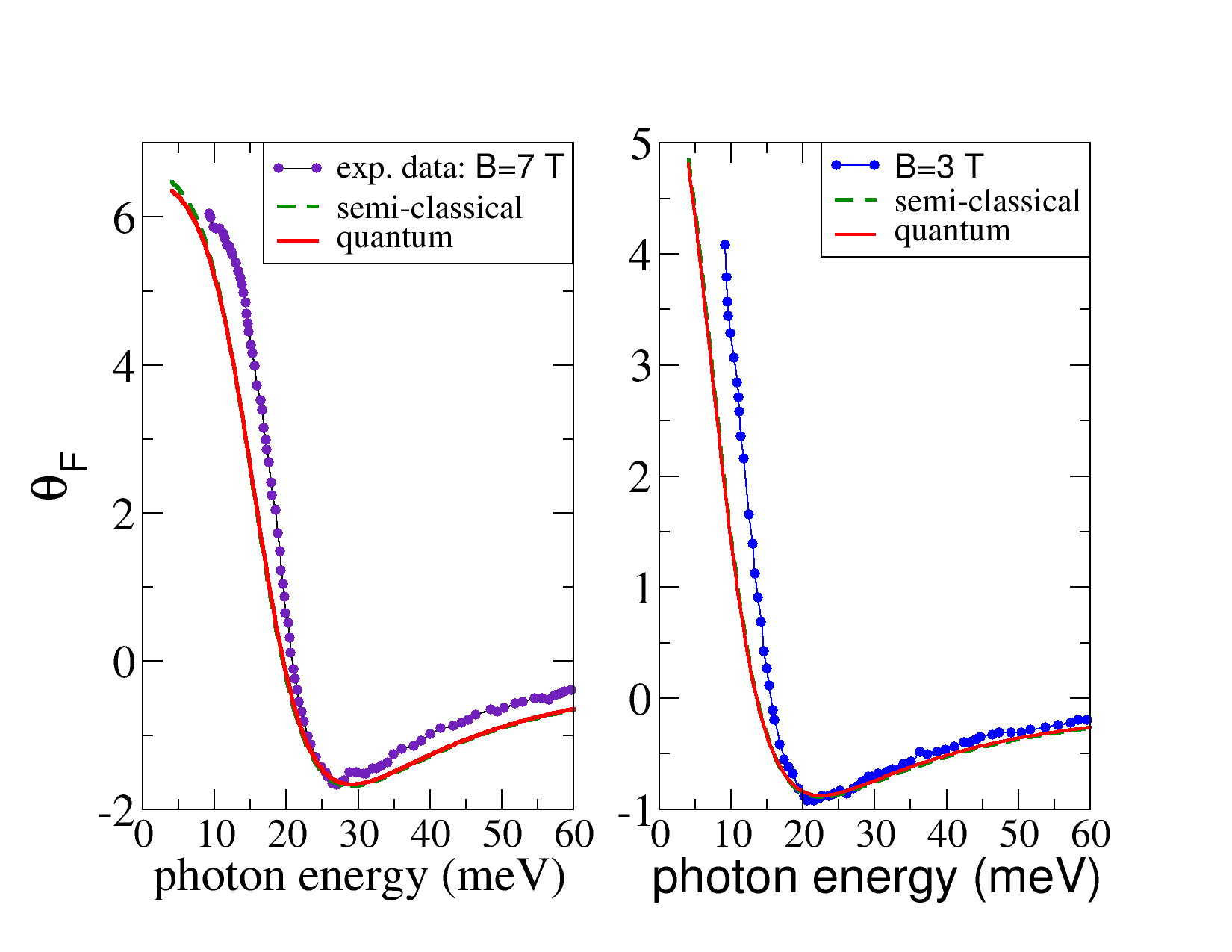} 
\par\end{centering}

\begin{centering}
\includegraphics[clip,width=0.9\columnwidth]{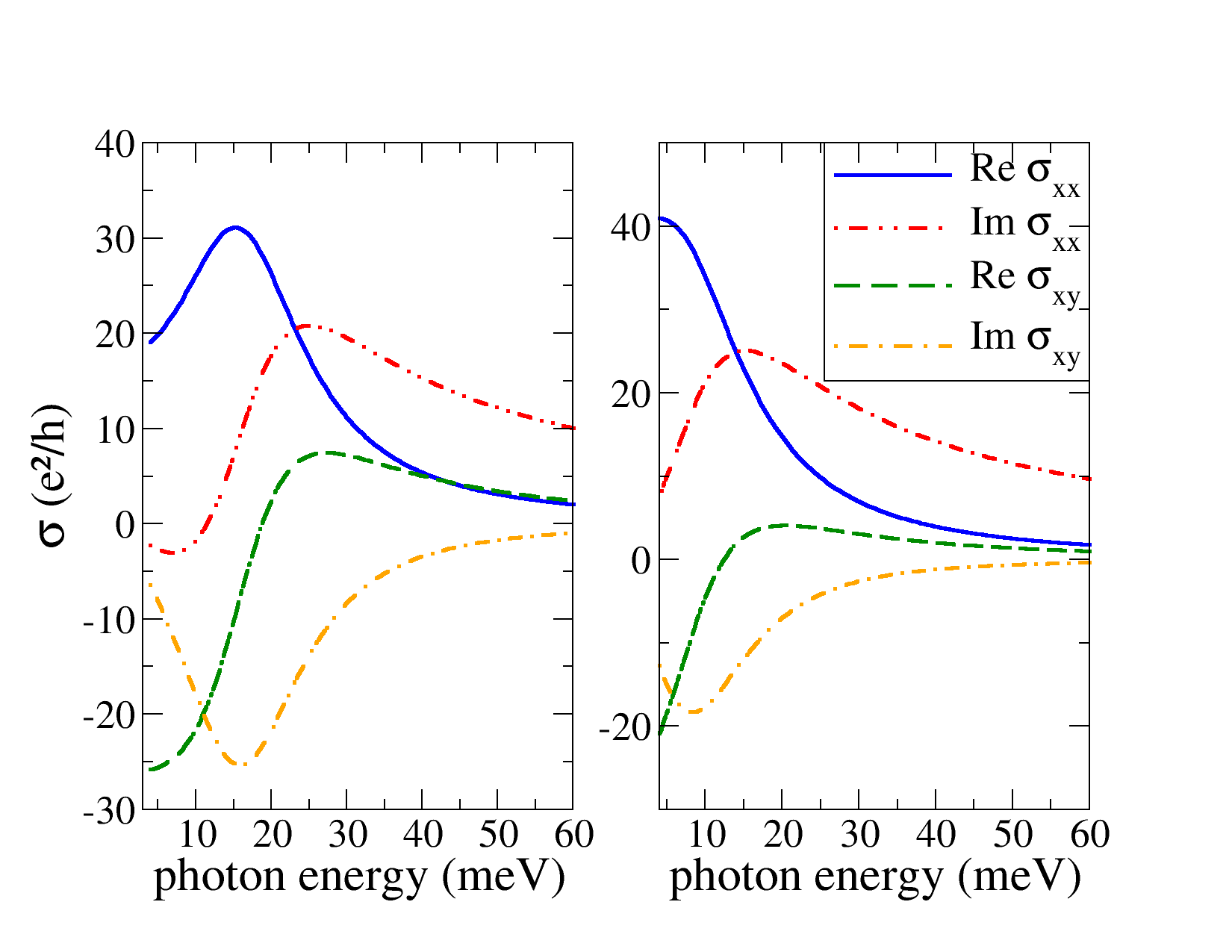} 
\par\end{centering}

\centering{}\caption{\label{fig:Fig_Theta_Exp}Faraday effect in doped graphene. \textbf{Top}:
The Faraday rotation angle (in degrees) when graphene is grown on
silicon carbide. Fit to the experimental values of $\theta_{F}$,
at a magnetic field of $B=7$~T (left) and $B=3$~T (right), using
the semiclassical approach {[}dashed (green) line{]} and the full
quantum calculation {[}solid (red) line{]}. Parameters are $E_{F}=0.3$~eV,
$\Gamma=10.5$~meV, $T=6$~K, and $\epsilon_{r}=4.4$. \textbf{Bottom}:
Theoretical optical conductivity {[}Eqs.~(\ref{eq:sigma_xx_final})
and (\ref{eq:sigma_xy_final}){]} for the same parameters used to
fit the experimental data: $B=7$~T (left) and $B=3$~T (right).}
\end{figure}

Figure~\ref{fig:Fig_Theta_Exp} shows fits for two sets of experimental
data for $\theta_{F}$,\cite{FaradayNaturePhys} measured when electromagnetic
radiation passes through graphene epitaxially grown on silicon carbide
(data taken at a temperature of 6~K). According to the experiments
by Crassee \emph{et al.,\cite{FaradayNaturePhys}} it was possible
to produce a single graphene sheet grown on the Si-terminated surface
of 6H-SiC (the sample underwent H passivation of the Si dangling bonds,
resulting in quasi-free-standing single-layer graphene). Two sets
of experimental data are shown in Fig.~\ref{fig:Fig_Theta_Exp} (top),
corresponding to two magnetic field intensities, $B=7$~T and $B=3$~T.

In the course of the experiments it was found that the bare substrate
did not reveal any Faraday effect, and therefore the measured rotation
angle is intrinsic to graphene. This statement is confirmed by the
model developed in Sec.~\ref{sub:Faraday_Rot}. ARPES measurements
on the sample used indicated a Fermi energy of the order of $E_{F}\simeq0.34\pm0.01$~eV.

In order to fit the data we have used $E_{F}=0.3$~eV. We do not
expect a perfect fit because we are considering a lossless dielectric.
Nevertheless, the fit is fairly accurate, given the simplicity of
the model. Moreover, the value of $\epsilon_{r}$ was set to 4.4 which
is not the relative permittivity of SiC and must be understood as
an effective number, given that the experimental data were taken with
epitaxially grown graphene. Although the calculation in Sec.~\ref{sub:FiniteMagField}
does not include this fact explicitly, the fits are satisfactory,
for they reproduce the main features of the experimental data: a decrease
in $\theta_{F}$ with the photon energy until a minimum is reached
for $\hbar\omega\approx26$~meV~(20~meV) when the magnetic field
intensity is 7~T~(3~T).

Comparing the top and bottom panels in Fig.~\ref{fig:Fig_Theta_Exp},
it can be seen that the minimum (maximum) of the Faraday rotation
angle coincides roughly with the maximum (minimum) of $\sigma_{xy}^{\prime}$.
The latter fact agrees well with what could be concluded from the
approximated result stated in Eq.~(\ref{eq:theta_F_approx}). In
order to interpret the variation of the Faraday rotation angle with
the photon energy, it is sufficient to use the simplified results
derived in Sec.~\ref{sub:FiniteMagField} for $T=0$, namely, Eqs.~(\ref{eq:sigma_xy_intra})-(\ref{eq:interband_Hall_extrema_inter}).
(This is clearly justified given the low temperature in the experiment
in Ref.~\onlinecite{FaradayNaturePhys}; the respective thermal
energy corresponds to about $0.01$ times the level spacing $\Delta_{1}=E_{1}-E_{0}$
{[}see Eq.~(\ref{eq:intraband_gap}) for the definition of $\Delta_{n}${]}
for both intensities of magnetic field.)

For a magnetic field of 7~T~(3~T), intraband transitions $n=9\rightarrow n=10$
($n=22\rightarrow n=23$) control the variation of $\theta_{F}$,
from positive up to negative values, as the photon energy varies.
Here, the index $n$ denotes LLs with energy given by $E_{n}=\textrm{sign}(n)\sqrt{2|n|}\hbar v_{F}/l_{B}$
{[}see Eq.~(\ref{eq:LL}) and text thereafter{]}. The remaining transitions
contributing to the Hall conductivity are interband-like and occur
at much higher photon energies $\hbar\omega\simeq2E_{F}$, and thus
it does not influence the Faraday rotation in the range of energy
plotted in Fig.~\ref{fig:Fig_Theta_Exp}.

In this example, intraband transitions involve a very small difference
in energy, even when the magnetic field is 7~T. The value of the
intraband gap {[}Eq.~(\ref{eq:intraband_gap}){]} is $\Delta_{N_{F}}\simeq16$~meV
($\Delta_{N_{F}}\simeq7$~meV) for $B=7$~T~($B=3$~T), which
is comparable to $\Gamma$ (here $N_{F}$ denotes the last occupied
LL for a given Fermi energy). The exact calculation shows that the
extrema points of the real part of the intraband Hall conductivity
{[}Eq.~(\ref{eq:sigma_xy_intra}){]} occur at $\omega=0$, and,
\begin{align}
\omega_{\pm}^{\textrm{intra}} & =\frac{1}{\hbar}\textrm{Re}\,\sqrt{\Delta{}_{N_{F}}^{2}+\Gamma^{2}\pm2\Gamma\sqrt{\Delta{}_{N_{F}}^{2}+\Gamma^{2}}}\,.\label{eq:omega_intra_hall_exact}
\end{align}
Substituting the values given in the caption to Fig.~\ref{fig:Fig_Theta_Exp}
into the latter formula, we obtain $\omega_{+}^{\textrm{intra}}\simeq27$~meV
($\omega_{+}^{\textrm{intra}}\simeq20$~meV) for a field intensity
of 7~T~(3~T). As mentioned above, these are the points where the
Faraday rotation reaches its minimum value. Increasing further the
photon energy, $\hbar\omega>\hbar\omega_{+}^{\textrm{intra}}$, the
Faraday rotation increases toward zero, essentially because at large
$\omega$, below the interband threshold, the Hall conductivity becomes
very low (Fig.~\ref{fig:Fig_Theta_Exp}) and no distinction arises
between $\sigma_{-}$ and $\sigma_{+}$, and thus $t_{+}\approx t_{-}$.
Increasing the photon energy up to $\hbar\omega\sim2E_{F}$, the interband
transition comes into play and drives the Faraday rotation. Interband
transitions are important in samples with low electronic densities,
as explained in the following section.

The curves for $\theta_{F}$, computed either from the semiclassical
expressions for the conductivity {[}Eqs.~(\ref{eq:sigma_xx_semiclass})
and (\ref{eq:sigma_xy_semiclass}){]} or via the EOM expressions {[}Eqs.~(\ref{eq:sigma_xx_final})
and (\ref{eq:sigma_xy_final}){]} are almost indistinguishable (see
Fig.~\ref{fig:Fig_Theta_Exp}, top), in the range of photon energies
considered, except for $\hbar\omega\approx10$~meV, where a very
small deviation is observed when the intensity of the magnetic field
is 7~T.

The agreement between the quantum and the semiclassical solutions
is explained by the similarity of the intraband gap $\Delta_{N_{F}}$
and the cyclotron energy $\hbar\omega_{c}$ {[}see Eq.~(\ref{eq:cyclotron_freq}){]}.
The values for these quantities are $\Delta_{N_{F}}\simeq6.62$($15.6$)~meV
and $\hbar\omega_{c}\simeq6.58$($15.4$)~meV for a field of $3$($7$)~T.
The agreement between the methods breaks down near the interband threshold,
$\hbar\omega\simeq2E_{F}\simeq0.6$~eV, where the quantum contribution
arising from the interband transition cannot be neglected.

\begin{figure}
\begin{centering}
\includegraphics[clip,width=1\columnwidth]{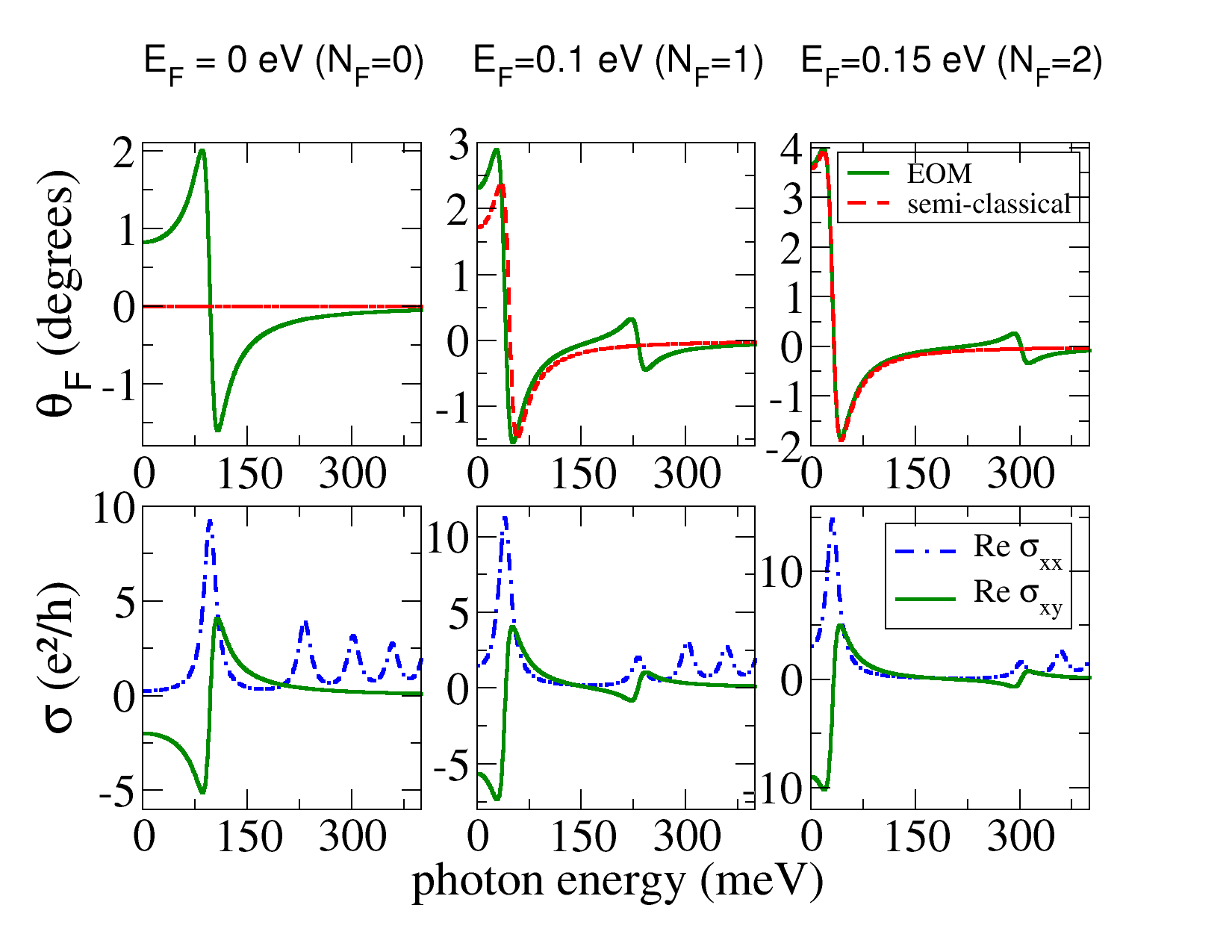} 
\par\end{centering}

\caption{\label{fig:Farad.Quantum.Reg}Low electronic density limit. \textbf{Top}:
Faraday rotation angle (given in degrees) for free-standing graphene
($\epsilon_{r}=1$) for different LL occupations: from left to right,
$N_{F}=0,1,$ and 2. The magnetic field intensity is $B=7$~T, $\Gamma=10.5$~meV
and $T=0$. Adding a dielectric substrate to graphene decreases the
maximum amount of Faraday rotation that is achievable, without introducing
major qualitative changes {[}see Eq.~(\ref{eq:theta_F_approx}){]}.
\textbf{Bottom}: Real part of the quantum conductivity tensor for
the Fermi energies considered in the top panel.}
\end{figure}

\subsection{Quantum jumps in the Faraday rotation: the low electronic density
limit \label{sub:Low_electronic_dens}}

When low Fermi energies are considered, energy quantization becomes
important (see Secs.~\ref{sub:FiniteMagField} and \ref{sub:SemiClassSolution}).
The limiting case occurs for $0\le E_{F}<\sqrt{2}\hbar v_{F}/l_{B}$,
i.e.,~$N_{F}=0$. In this case, at $T=0$, LLs with $n\ge1$ are
all empty, and a single type of transition contributes to the Hall
conductivity, $n=0\rightarrow n=1$. Since this transition is interband-like,
it cannot be explained within the semiclassical treatment (Secs.~\ref{sub:FiniteMagField}
and \ref{sub:SemiClassSolution}). This situation is illustrated in
Fig.~\ref{fig:Farad.Quantum.Reg} (bottom): when $N_{F}=0$, the
real part of the Hall conductivity {[}solid (green) line{]} has a
finite (nonzero) value around $\omega\simeq(E_{1}+E_{0})/\hbar$.
{[}Note that the extrema of the interband Hall conductivity can be
obtained from Eq.~(\ref{eq:omega_intra_hall_exact}) by making the
replacement $\Delta_{N_{F}}\rightarrow\hbar\Delta\Omega_{N_{F}}$,
with $\Delta\Omega_{N_{F}}$ given by Eq.~(\ref{interband_gap}).{]}
The Faraday rotation given by the semiclassical model is obviously
0 {[}dashed (red) line{]} since $E_{F}=0$ {[}Eq.~(\ref{eq:sigma_xy_semiclass}){]}.
The respective Faraday rotation angle (top) is approximately proportional
to $-\sigma_{xy}(\omega)$.

At higher Fermi energies (i.e., $N_{F}>0$), two types of transitions
contribute to the Hall conductivity: in general, for $E_{N_{F}}<E_{F}<E_{N_{F}+1}$,
with $N_{F}\ge1$, the allowed transitions are (i) interband between
the hole's LLs with $n=-N_{F}$ and the electron's LLs with $n=N_{F}+1$
and (ii) intraband between LLs with $n=N_{F}$ and $n=N_{F}+1$ (Sec.~\ref{sub:FiniteMagField}).
The maximum intensity of $\sigma_{xy}^{\prime}$ falls off with the
inverse of the energy difference associated with a given electronic
transition {[}Eqs.~(\ref{eq:interband_Hall_extrema_intra}) and (\ref{eq:interband_Hall_extrema_inter}){]}.
Since, up to a good degree of approximation, the Faraday effect is
controlled by $\sigma_{xy}^{\prime}$, the latter means that the amount
of Faraday rotation induced by the interband transitions at $\omega=\Delta\Omega_{N_{F}}$
will be smaller than the Faraday rotation due to intraband processes.

The above-mentioned facts can be appreciated in Fig.~\ref{fig:Farad.Quantum.Reg},
where numerical data for $\theta_{F}$ (top), $\sigma_{xx}^{\prime}$
and $\sigma_{xy}^{\prime}$ (bottom) are shown with Fermi energy increasing
from left to right. As higher LLs in the conduction band become occupied,
the spectral weight for the interband contribution to $\sigma_{xy}^{\prime}$
shifts toward higher energies (that is, $\hbar\Delta\Omega_{N_{F}}$
increases $\rightarrow$ $\omega_{\pm}^{\textrm{inter}}$ increases).
The opposite occurs for intraband transitions, since in this case,
the relevant energy scale $\Delta_{N_{F}}$ decreases with increasing
$E_{F}$. As a result, the intraband part of $\sigma_{xy}^{\prime}$
concentrates its spectral weight at the lower edge of the plotted
spectrum, and displays a much higher amplitude than its interband
counterpart, as explained above. Similar conclusions apply to $\theta_{F}$,
as direct inspection of the bottom and top panels shows.

\begin{figure}
\begin{centering}
\includegraphics[clip,width=1\columnwidth]{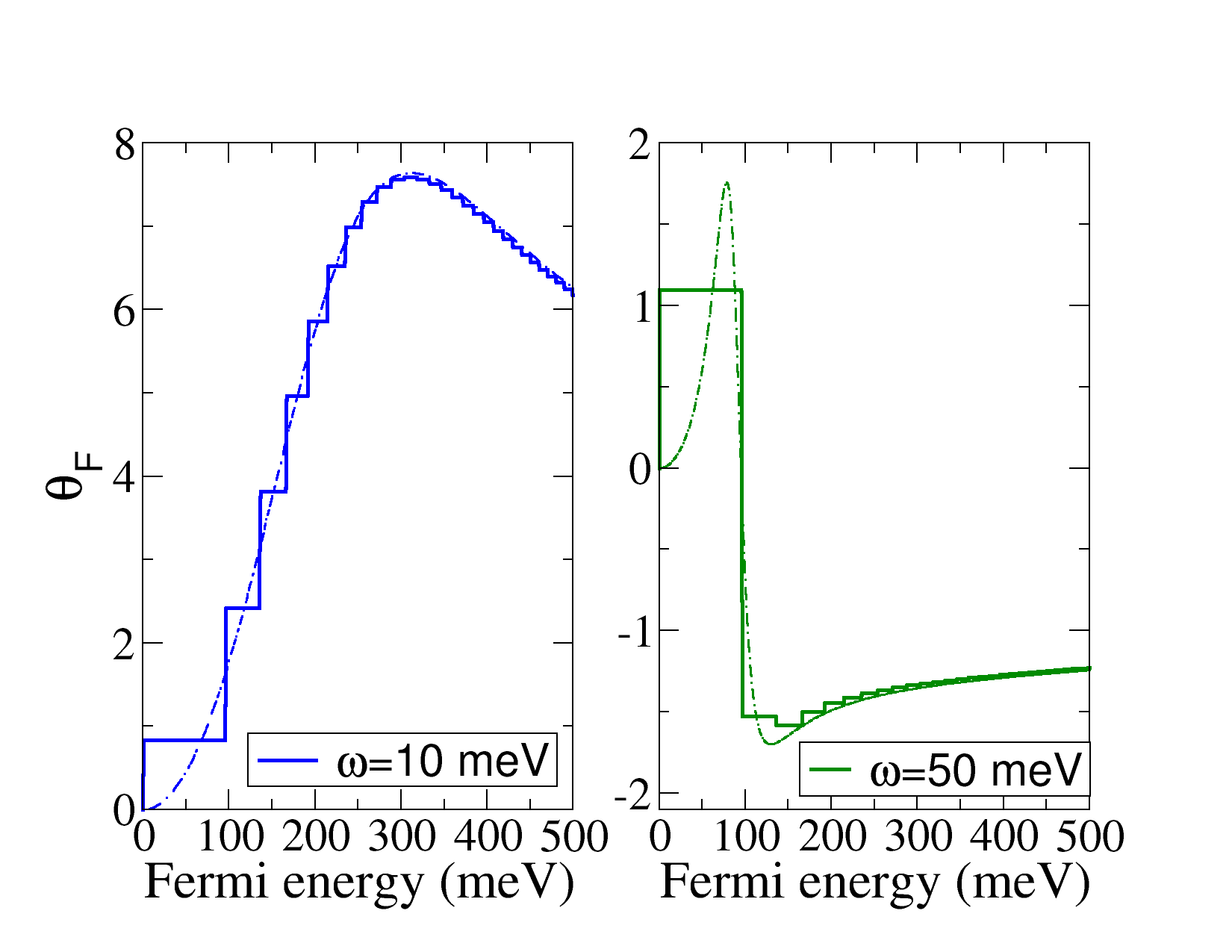} 
\par\end{centering}

\bigskip{}

\begin{centering}
\includegraphics[clip,width=1\columnwidth]{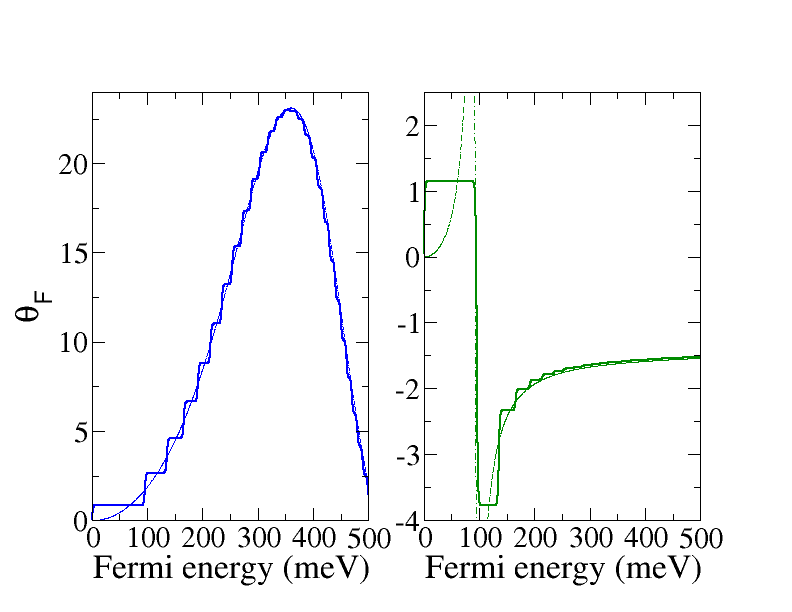} 
\par\end{centering}

\centering{}\caption{\textbf{\label{fig:Fig_ThetaF_vs_Fermi}}Quantization of the Faraday
effect in graphene. \textbf{Top}: Faraday rotation angle (in degrees)
for free-standing graphene as a function of the Fermi energy at a
magnetic field of $B=7$~T for $\hbar\omega=10$~meV (left) and
$\hbar\omega=50$~meV (right). The respective semi-classical result
is plotted by the dashed lines. Other parameters: $\Gamma=10.5$~meV
and $T=12$~K. \textbf{Bottom: }Same as the top panel but with $\Gamma=2$~meV.}
\end{figure}

When $N_{F}=1$, a significant departure from the semiclassical behavior
can be appreciated in the intraband region ($0-100$~meV). Remarkably,
though, already for $N_{F}=2$, the semiclassical Hall conductivity
approximates the quantum result well, with a significant deviation
occurring only near the interband threshold ($\approx$300~meV),
where the semiclassical approach must necessarily fail. These features
are in accordance with the general conclusions drawn in Sec.~\ref{sub:FiniteMagField}.

For comparison, the real part of the longitudinal conductivity is
also shown in the bottom panel in Fig.~\ref{fig:Farad.Quantum.Reg}.
The longitudinal current can be induced by photons which are resonant
with any interband transition allowed by the Pauli principle (i.e.,
$\hbar\omega>\hbar\Delta\Omega_{N_{F}}\simeq2E_{F}$), and hence many
absorption peaks can be observed. On the contrary, the Faraday rotation
essentially depends on $\sigma_{xy}^{\prime}$ and therefore is driven
only by two resonances.

\emph{Dependence on the Fermi energy and magnetic field}---We have
seen that when few LLs are occupied, quantum effects come into play
and the semiclassical solution no longer gives an accurate description
of the Faraday effect. The latter can even happen in the intraband
region (see, e.g., Fig.~\ref{fig:Farad.Quantum.Reg}, middle panel),
embodying the departure of the intraband gap $\Delta_{N_{F}}$ from
its semiclassical analog, the cyclotron energy, $\hbar\omega_{c}$
(see Table~\ref{tab:TableI}). Given the importance of energy quantization
for a low electronic density, we expect $\theta_{F}$ to display abrupt
behavior when the Fermi energy crosses the first few LLs. The latter
behavior should reflect directly the step structure of the optical
(or ac-) Hall conductivity $\sigma_{xy}^{\prime}(\omega)$.\cite{Morimoto2009}

Figure~\ref{fig:Fig_ThetaF_vs_Fermi} shows the Faraday rotation
angle versus $E_{F}$ for a fixed magnetic field, $B=7$~T. The heights
of the steps are not uniform since the optical Hall conductivity no
longer obeys the dc quantization rule {[}Eq.~(\ref{eq:Hall_quantification}){]}.
When the Fermi energy crosses higher LLs, the smooth semiclassical
result (dashed curves) is recovered.

Combining the approximated formula for $\theta_{F}$ {[}Eq.~(\ref{eq:theta_F_approx});
valid for $c\mu\sigma_{\mp}^{\prime}(\omega)\ll2$ and $\theta_{F}\lesssim1${]}
and the exact Hall conductivity at $T=0$ {[}Eqs.~(\ref{eq:sigma_xy_intra})
and (\ref{eq:sigma_xy_inter}){]}, explicit formulas for the step
heights can be obtained. When the Fermi energy crosses LLs with $n>1$,
the expression for $\Delta\theta_{F}$ becomes somewhat cumbersome.
Nevertheless, simple analytical expressions can be obtained in some
regimes. For instance, when the photon energy is small compared to
relevant scales, $\hbar\omega\ll\Gamma\ll E_{1}$, the steps are predicted
to be approximately uniform,
\begin{equation}
\Delta\theta_{F}^{(n\rightarrow n+1)}\simeq\frac{2c\mu e^{2}}{h}=4\alpha\simeq0.03\,\textrm{rad},\label{eq:DeltaTheta_approx}
\end{equation}
where $\alpha$ denotes the fine structure constant, $\alpha=e^{2}/(4\pi\hbar\epsilon_{0}c)$.
In Ref.~{[}\onlinecite{Morimoto2009}{]}, for estimation of the
magnitude of the effect it was assumed that the step height of $\sigma_{xy}^{\prime}$
is approximately given by $\Delta\sigma_{xy}^{\prime}(\omega)\simeq e^{2}/h$,
resulting in $\Delta\theta_{F}\simeq\alpha$. Rigorously, the step
height for the transitions $n=0\rightarrow n=1$ is about $4e^{2}/h$,
hence explaining the extra factor of 4 in our expression. In fact,
in the limit $\hbar\omega\ll\Gamma\ll E_{1}$, the steps in the Hall
conductivity will all have approximately the same height, as in the
dc case {[}see Eq.~(\ref{eq:Hall_quantification}){]}.

In Fig.~\ref{fig:Fig_ThetaF_vs_Fermi}, a decrease in the step's
height relative to the estimated value in Eq.~(\ref{eq:DeltaTheta_approx})
can be observed already for the first step. This happens because the
condition $\Gamma\ll E_{1}$ is too restrictive, and hence we relax
this condition to $\Gamma\lesssim E_{1}$, but, at the same time,
keep the low-photon-energy condition, $\hbar\omega\ll\Gamma$. Doing
so, leads to a better approximation,

\begin{equation}
\Delta\theta_{F}^{(n\rightarrow n+1)}\simeq\frac{1}{1+(6+4n+\delta_{n,0})\tilde{\gamma}^{2}}\frac{1}{1+4n\tilde{\gamma}^{2}}\times4\alpha\,,\label{eq:Delta_better_approx}
\end{equation}
 where we have defined the dimensionless parameter $\tilde{\gamma}=\Gamma/E_{1}$.
Using this parameter, the validity condition of Eq.~(\ref{eq:Delta_better_approx})
reads, $\tilde{\gamma}\lesssim1$ and $\hbar\omega\ll\Gamma$.

Two physical scenarios where the Faraday steps are not uniform are
shown in Fig.~\ref{fig:Fig_ThetaF_vs_Fermi}. In the bottom panel,
the transitions $n=0\rightarrow n=1$ ($E_{F}\simeq100$~meV) come
with a variation of $\theta_{F}$ of roughly $1.8^{\circ}$ ($\simeq0.031$~rad)
for $\hbar\omega=10$~meV, versus $-5.1^{\circ}$ ($\simeq-0.089$~rad)
for $\hbar\omega=$50~meV, which does not agree either with the rough
uniform estimative or with Eq.~(\ref{eq:Delta_better_approx}). The
reason for this discrepancy is that the condition $\hbar\omega\ll\Gamma$
is not fulfilled for the photon frequencies considered in Fig.~\ref{fig:Fig_ThetaF_vs_Fermi}.
Recall that in graphene, $\Gamma$ is about about 10 meV, and thus
infrared photons have $\hbar\omega\gtrsim\Gamma$. It is therefore
useful to derive approximate formulas for $\Delta\theta_{F}$ that
are valid in the regime $\hbar\omega\gg\Gamma$. Defining $\tilde{\omega}=E_{1}/(\hbar\omega)$,
we arrive at
\begin{equation}
\Delta\theta_{F}^{(n\rightarrow n+1)}\simeq\frac{4\alpha}{1-2(1+2n)\tilde{\omega}^{2}+\tilde{\omega}^{4}}\frac{1-\tilde{\omega}^{4}}{1-2(3+2n)\tilde{\omega}^{2}+\tilde{\omega}^{4}}\,.\label{eq:Delta_Faraday_Infrared}
\end{equation}
 Substituting for the respective values of $\tilde{\omega}$, we obtain
$\Delta\theta_{F}^{(0\rightarrow1)}=1.8^{\circ}$ and $\Delta\theta_{F}^{(0\rightarrow1)}=-5.3^{\circ}$,
for $\hbar\omega=10$~eV and $\hbar\omega=50$~meV, respectively,
which agrees well with the numerical results reported in Fig.~\ref{fig:Fig_ThetaF_vs_Fermi}
for $\Gamma=2$~meV. As for the steps observed in the top panel in
Fig.~\ref{fig:Fig_ThetaF_vs_Fermi}, they cannot be explained accurately
with Eq.~(\ref{eq:Delta_Faraday_Infrared}) since in that case we
have $\hbar\omega\approx\mathcal{O}(\Gamma)$.\emph{ }We stress that
Eqs.~(\ref{eq:Delta_better_approx}) and (\ref{eq:Delta_Faraday_Infrared})
are only accurate when the statement, Eq.~(\ref{eq:theta_F_approx}),
provides a good description of the Faraday effect in graphene, which
in practice means very high photon energies $\hbar\omega$ (see also
Fig.~\ref{fig:Faraday_Various}). For the parameters used in Fig.~\ref{fig:Fig_ThetaF_vs_Fermi},
where the photon energies are not very high, our analytical expressions
for $\Delta\theta_{F}$ are accurate only for the first few steps.

\begin{figure}
\begin{centering}
\includegraphics[clip,width=1\columnwidth]{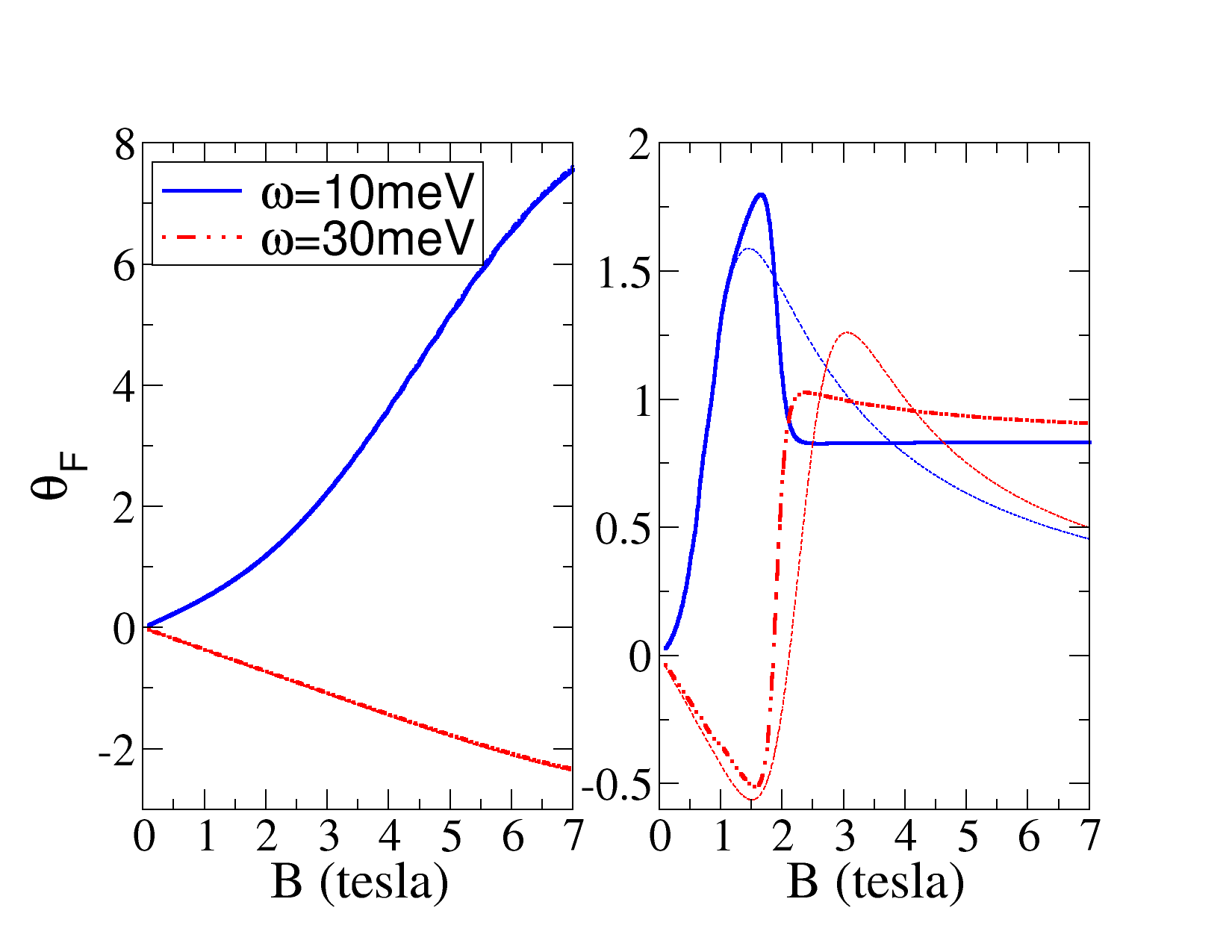} 
\par\end{centering}

\caption{\label{fig:Fig_ThetaF_vs_B}Faraday rotation angle (in degrees) as
a function of the magnetic field for $E_{F}=0.30$~eV (left) and
$E_{F}=0.05$~eV (right). In each panel two photon energies are represented---$\hbar\omega=10$~meV
{[}solid (blue) line{]} and $\hbar\omega=30$~meV {[}dashed-double-dotted
(red) line{]}---with the respective semiclassical counterparts shown
by dashed lines. Other parameters as in the top panel in Fig.~\ref{fig:Fig_ThetaF_vs_Fermi}.}
\end{figure}

Figure~\ref{fig:Fig_ThetaF_vs_B} shows the variation of $\theta_{F}$
with the magnetic field for two cases; (i) low doping ($E_{F}=0.05$~eV)
and (ii) high doping ($E_{F}=0.3$~eV). In the latter case, we are
well inside the semiclassical regime even for the maximum intensity
of the magnetic field considered ($B=$7~T), and thus no distinction
can be made between the curves computed using the semiclassical conductivity
tensor or the EOM formulas. In this regime, the Faraday effect increases
monotonously with the magnetic field.

For a low electronic density, on the other hand, the agreement between
the Boltzmann and the EOM formalisms takes place only for low magnetic
fields. For increasing values of the magnetic fields, such agreement
ceases to occur as soon as the intraband gap does not match the cyclotron
energy $\hbar\omega_{c}$. Then, energy level quantization becomes
important and the EOM expressions must be considered (i.e., $N_{F}$
is small; see Sec.~\ref{sub:SemiClassSolution}): this explains the
departure from the semiclassical value for $\theta_{F}$ observed
in the right panel in $B\thickapprox1$~T for $\hbar\omega=10$~meV
($B\thickapprox0.5$~T for $\hbar\omega=30$~meV). If the magnetic
field intensity is higher than a given value, we necessarily have
$N_{F}=0$ (for 0.05~eV this value is about 1.9~T). In this case,
the Hall conductivity, at $T=0$, is fully determined by a single
type of interband transition, and, assuming $E_{1}(B)\gg\hbar\omega,\Gamma$,
we obtain {[}see Eq.~(\ref{eq:sigma_xy_inter}){]},
\begin{equation}
\sigma_{xy}^{\prime}\underset{\textrm{large }B}{\simeq}-\frac{2e^{2}}{h}\Rightarrow\theta_{F}\simeq2\alpha\simeq3\times10^{-4}\,^{\circ}\:.\label{eq:sigma_xy_imag_approx}
\end{equation}
 The latter considerations explain the plateau formed at $B\thickapprox2$~T
{[}solid (blue) line{]} in the right panel in Fig.~\ref{fig:Fig_ThetaF_vs_B}.
The dashed-double-dotted (red) line corresponds to photons with a
higher energy, shifting the formation of the plateau toward higher
fields. Equation~(\ref{eq:sigma_xy_imag_approx}) is indeed the high-magnetic-field
limit {[}$E_{1}(B)\gg$~energy scales{]} of the Faraday rotation
induced by single-layer graphene.

Although the measured Faraday rotation angle is remarkably large given
that it comes from a single graphene layer, in both low and high doping
regimes (see Fig.~\ref{fig:Fig_ThetaF_vs_B}), it is the needed magnetic
field, $B\gtrsim1$~T. The goal is then to obtain large Faraday rotation
angles using graphene and modest fields at the same time. A simple
idea that uses the nonreciprocity of the Faraday effect is to enclose
graphene between two mirrors. We discuss this possibility in the following
section.

\subsection{Enhancement of Faraday rotation in a cavity geometry\label{sub:Enhancement-of-Faraday}}

We have seen that the existence of intraband and interband transitions
in graphene permits the generation of finite (nonzero) Faraday rotations
in different ranges of the electromagnetic spectrum. In doped graphene
($N_{F}\ge1$), e.g., the intraband gap is bounded from above by 
\begin{equation}
E^{\textrm{intra}}\le\Delta_{1}=E_{2}-E_{1}\simeq15\sqrt{B}\,\textrm{meV}\cdot\textrm{T}^{-1/2}\,,\label{eq:upper_bound_intraband}
\end{equation}
implying that, by using magnetic field intensities $\sim1$~T, graphene
can be exploited for magneto-optical applications from the microwave
up to the far-infrared regimes $f=E/h\lesssim3.6$~THz (an example
of terahertz Faraday rotation driven by intraband transitions is given
in Fig.~\ref{fig:Fig_Theta_Exp}). Another possibility is to make
use of transitions connecting the valence and conduction Dirac cones,
whose interband gaps are bounded from below, 
\begin{equation}
E^{\textrm{inter}}\ge\hbar\Delta\Omega_{1}=E_{1}\simeq36\sqrt{B}\,\textrm{meV}\cdot\textrm{T}^{-1/2}\,,\label{eq:lower_bound_interband}
\end{equation}
 thus obtaining far-infrared up to visible light light frequencies
(an example of mid-infrared Faraday rotation driven by interband transitions
is shown in Fig.~\ref{fig:Farad.Quantum.Reg}). We recall that increasing
the electronic density in order to obtain even larger interband gaps
($\hbar\Delta\Omega_{n}$ with $n>1$), and thus shifting the magneto-optical
response of graphene above the midinfrared, $\hbar\omega\sim\hbar\Delta\Omega_{N_{F}}\simeq2E_{F}$,
creates optical Hall conductivity peaks with a low intensity. As a
consequence, very small Faraday rotations are produced already in
the near-infrared regime. A good estimate for the maximum achievable
interband-induced Faraday rotation can be obtained from Eqs.~(\ref{eq:interband_Hall_extrema_inter})
and (\ref{eq:theta_F_approx}), 
\begin{equation}
\textrm{max}\,|\theta_{F}|\simeq\left(\frac{eBv_{F}^{2}}{2\omega\Gamma}\right)\times\alpha\,,\label{eq:estimate_max_infrared_rot}
\end{equation}
 which, for example, taking $B=7$~T , $\Gamma=10$~meV, and $\hbar\omega=1$~eV,
leads to $\textrm{max}\,|\theta_{F}|\simeq10^{-3}$. Although the
amounts of terahertz Faraday rotation, $\hbar\omega\simeq O\,\textrm{meV}$,
reported in our figures are well within state-of-the-art capabilities
{[}the resolution for Faraday measurements in terahertz time-domain
spectroscopy is presently limited to 1~mrad\cite{Shimano2008} ($\sim$0.06~degrees){]},
high magnetic fields, $\sim1$~T, are still needed which can be a
disadvantage for specific applications; moreover, according to Eq.~(\ref{eq:estimate_max_infrared_rot})
the needed magnetic field increases as higher photon frequencies are
to be probed.

The situation is very different in other 2D electron gases, for which
$\theta_{F}$ is proportional to the sample's thickness (as the light
travels farther through the material, more Faraday rotation accumulates).
Single-layer graphene, on the other hand, being one atom thick and
hence truly 2D, requires the use of high magnetic fields in order
to detect Faraday rotations. It is therefore natural to ask whether
it is possible to conceive a setup leading to accumulation of Faraday
effect; ideally, such setup would avoid the use of several samples
and, at the same time, take advantage of the broad magneto-optical
response of single-layer graphene.

In what follows, we discuss a graphene-based system that can enhance
the intrinsic graphene's Faraday rotation at any frequency and thus
can cope with the difficulty mentioned above. The idea consists in
enclosing graphene in an optical cavity: due to intracavity interference,
photons undergo several round trips within the cavity before leaking
out. Loosely speaking, due to nonreciprocity of the Faraday effect,
accumulation of $\theta_{F}$ then takes place each time a photon
passes through graphene; a sketch of the experimental apparatus is
shown in Fig.~\ref{fig:Graphene_Cav}.

Explicit calculations (see below) show that giant Faraday rotations
are achieved even when the optical \emph{finesse} of the cavity is
modest. The optical \emph{finesse} can be easily tuned by changing
the reflectivity of the end mirrors: the higher the latter quantity,
the larger is the number of round trips of photons inside the cavity,
and hence further Faraday accumulation occurs. Indeed, the cavity
geometry gives a straightforward solution to mimic the effect of a
sample's thickness (absent in single-layer graphene).

\begin{figure}
\begin{centering}
\includegraphics[clip,width=1\columnwidth]{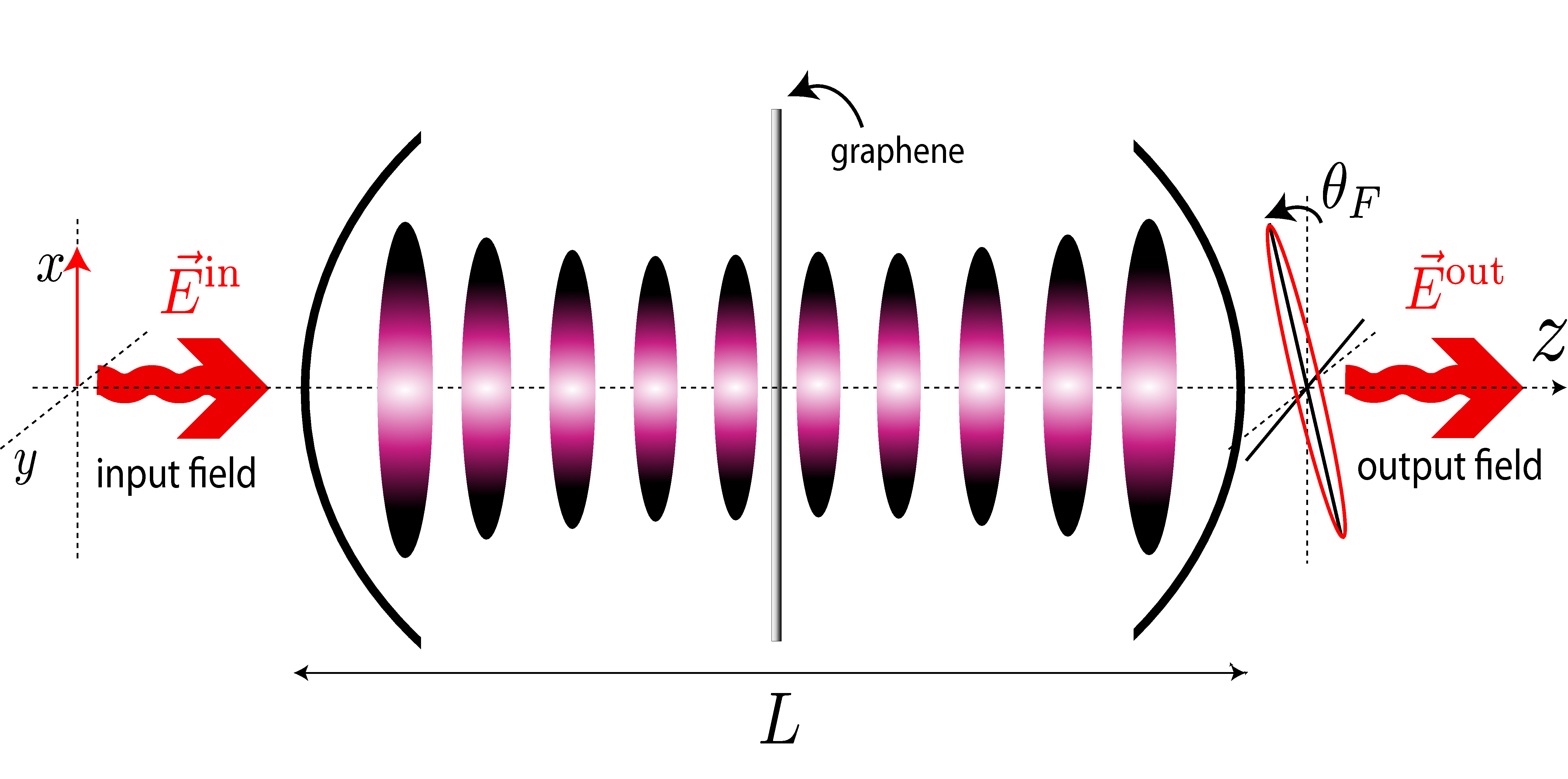} 
\par\end{centering}

\caption{\label{fig:Graphene_Cav}Schematic of the graphene-optical cavity
system: linearly polarized light shines into an optical cavity with
graphene placed at the center. The field inside the cavity perceives
graphene as an extra boundary and hence the two halves of the cavity
operate as independent cavities of effective size $L/2$. Matching
the light frequency $\hbar\omega$ with a resonant frequency of the
cavity $\omega=n\pi c/L$ ($n\in\mathbb{N}$ ) traps photons inside
the cavity for several round trips. As a consequence, Faraday rotation
accumulates due to multiple passages through graphene, leading to
an output field with a large Faraday rotation. }
\end{figure}

Following the steps in Sec.~\ref{sub:Faraday_Rot}, we write the
boundary conditions of the electromagnetic field in terms of circularly
polarized waves. Employing similar notation, we define the input and
output circular vector amplitudes, 
\begin{align}
\boldsymbol{\mathcal{E}}_{\pm}^{\textrm{in}}= & \left(E_{\pm}^{\textrm{in}}\,,E_{\pm}^{\textrm{r}}\right)^{T}\,,\label{eq:amp_vec_in}\\
\boldsymbol{\mathcal{E}}_{\pm}^{\textrm{out}}= & \left(E_{\text{\ensuremath{\pm}}}^{\textrm{t}}\,,0\right)^{T}\,,\label{eq:amp_vec_out}
\end{align}
 respectively (see also Fig.~\ref{fig:Graphene_Cav}), where $E_{\pm}^{\textrm{in}}=E_{x}^{\textrm{in}}\pm iE_{y}^{\textrm{in}}$
(the reflected $E_{\pm}^{\textrm{r}}$ and transmitted waves $E_{\text{\ensuremath{\pm}}}^{\textrm{t}}$
having analogous definitions). The first (second) component of the
vectors, Eq.~(\ref{eq:amp_vec_in}) and (\ref{eq:amp_vec_out}),
refers to the complex amplitude of light traveling in the positive
(negative) $z$ direction.

The output field, $\boldsymbol{\mathcal{E}}_{\pm}^{\textrm{out}}$,
and thus the total Faraday rotation angle, can be more conveniently
computed using the transfer matrix formalism. The method is explained
in detail in Appendix~\ref{AppendixA}. Here, we just state the basic
results: the $T$~matrix, by definition, connects the input and output
vector amplitudes, according to 
\begin{equation}
\boldsymbol{\mathcal{E}}_{\pm}^{\textrm{in}}=T_{\pm}^{\textrm{in}\rightarrow\textrm{out}}\boldsymbol{\mathcal{E}}_{\pm}^{\textrm{out}}\,,\label{eq:tm}
\end{equation}
 where $T_{\pm}^{\textrm{in}\rightarrow\textrm{out}}$ is a product
of individual transfer matrices for each boundary (optical component,
metallic surface, etc.). Its inverse permits us to compute $\boldsymbol{\mathcal{E}}_{\pm}^{\textrm{out}}$,
given the input field $\boldsymbol{\mathcal{E}}_{\pm}^{\textrm{in}}$,
and hence the optical characteristics of the cavity-graphene system.
In particular, the circular transmitted amplitudes, $t_{\pm}=E_{\text{\ensuremath{\pm}}}^{\textrm{t}}/E_{\pm}^{\textrm{in}}$,
are given by $t_{\pm}=1/[T_{\pm}^{\textrm{in}\rightarrow\textrm{out}}]_{1,1}$.

\begin{figure}
\begin{centering}
\includegraphics[clip,width=0.96\columnwidth]{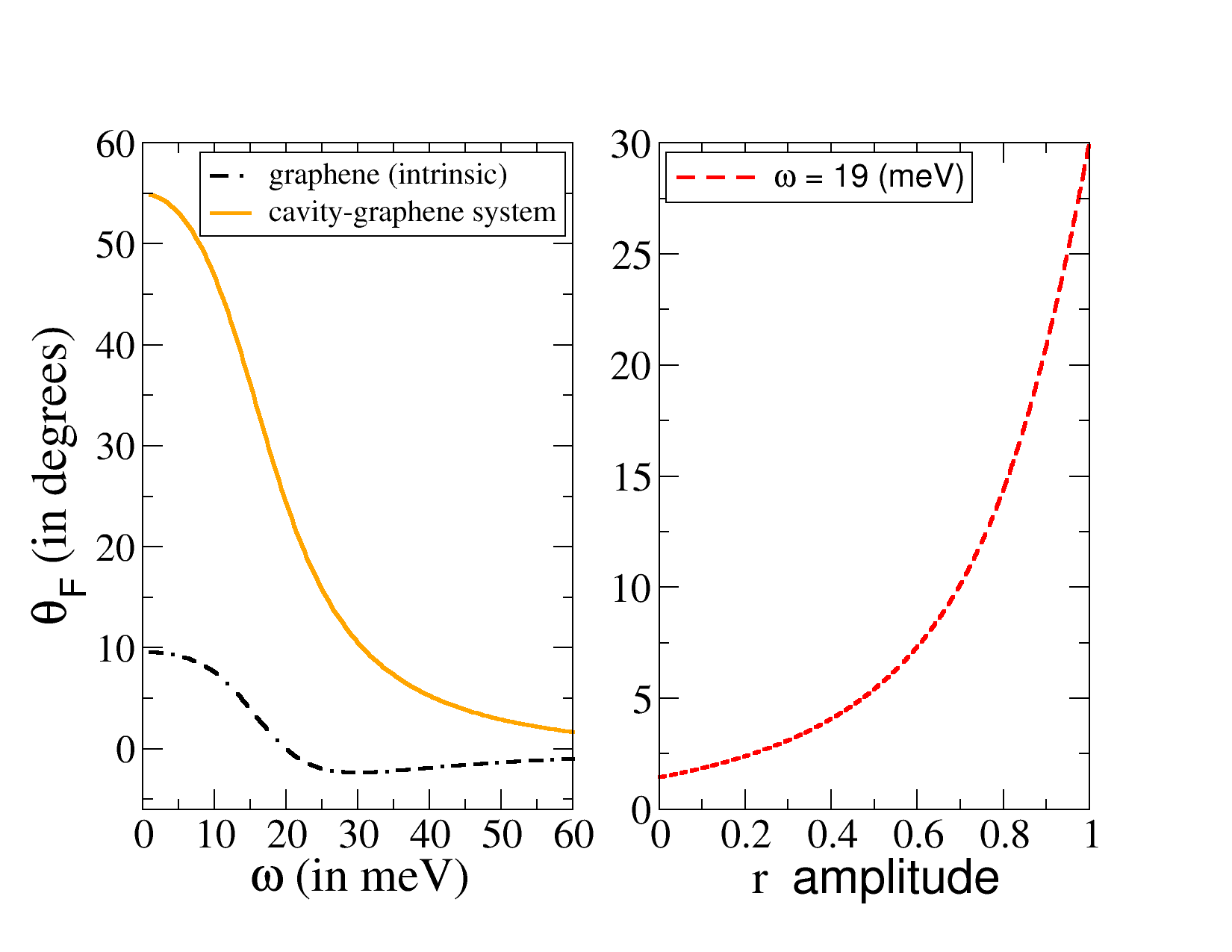} 
\par\end{centering}

\bigskip{}

\centering{}\includegraphics[clip,width=1\columnwidth]{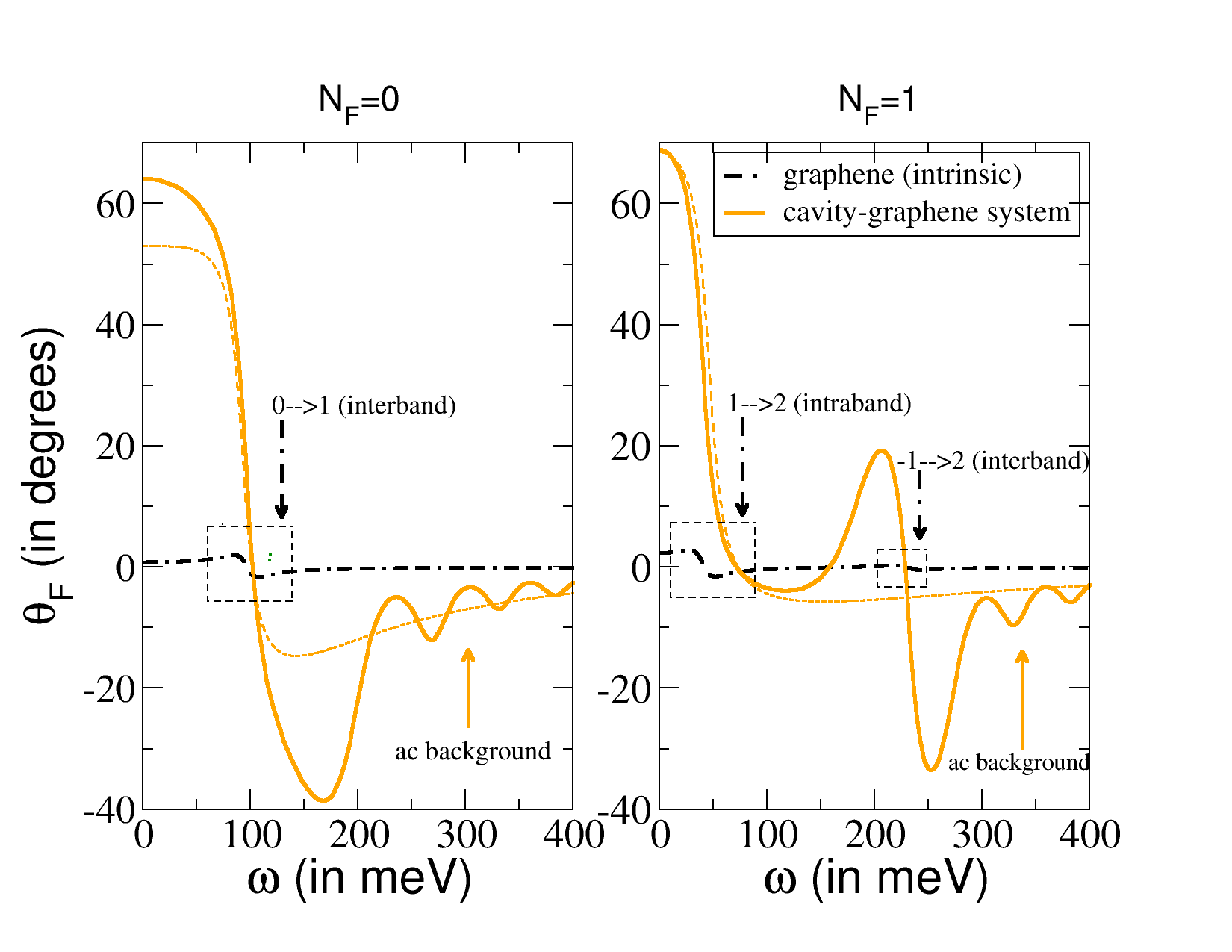}\caption{\label{fig:FaradayRotCavGraph}Faraday rotation angle of a cavity-graphene
system in the semiclassical and quantum regimes.\textbf{ Top} \textbf{left}:$\theta_{F}$
as a function of the photon energy for a cavity-graphene system in
a magnetic field of $7$~T . The Fermi energy reads $E_{F}=0.3$~eV
and the cavity mirrors have $r=0.99$. Other parameters: $\Gamma=10.5$~meV
and $T=12$~K. \textbf{Top right}: $\theta_{F}$ versus the reflection
amplitude $r$ for $\hbar\omega=19$~meV. \textbf{Bottom}: In the
left (right) panel, the Fermi energy reads $E_{F}=0.05$~eV ($E_{F}=0.1$~eV),
which corresponds to an LL occupation of $N_{F}=0$ ($N_{F}=1$).
The orange dashed line shows $\theta_{F}$ as obtained with the semiclassical
conductivity tensor. }
\end{figure}

For the geometry posed in Fig.~\ref{fig:Graphene_Cav}, the input-output
$T$~matrix reads 
\begin{align}
T_{\pm}^{\textrm{in}\rightarrow\textrm{out}} & = & T_{\textrm{m}}\cdot\left[\begin{array}{cc}
e^{-i\omega L/2c} & 0\\
0 & e^{i\omega L/2c}
\end{array}\right]\cdot T_{\pm}^{\textrm{g}}\nonumber \\
 &  & \cdot\left[\begin{array}{cc}
e^{-i\omega L/2c} & 0\\
0 & e^{i\omega L/2c}
\end{array}\right]\cdot T_{\textrm{m}}\,.\label{eq:tota_t_matrix}
\end{align}
 Each operator in Eq.~(\ref{eq:tota_t_matrix}) propagates the electric
field to the right until a boundary is reached. $T_{\textrm{m}}$
encodes the effect of the first interface, a mirror, and depends only
on the mirror's transmission and reflection amplitudes, $t$ and $r$,
respectively. It can be written as
\begin{equation}
T_{\textrm{m}}=\frac{1}{i|t|}\left[\begin{array}{cc}
1 & |r|\\
-|r| & -1
\end{array}\right]\,.\label{eq:t_mirror}
\end{equation}
 (For a derivation, see, e.g., Ref.~\onlinecite{wavepropagation}.)
After interaction with the left-end mirror, photons can enter into
the cavity and propagate for a distance of $L/2$ before the next
interaction. This means that another $T$~matrix is needed; free
propagation merely adds a phase to the electric field {[}see Eq.~(\ref{eq:free_propagation})
and text thereafter{]} and thus is represented by a diagonal matrix,
which is the second operator in Eq.~(\ref{eq:tota_t_matrix}). At
$z=L/2$, photons arrive at the air-graphene-air interface, whose
$T$~matrix we denote by $T_{\pm}^{\textrm{g}}$. (More involved
types of interfaces could be considered: for example, air-substrate-graphene-air.
The present choice has the advantage of keeping the mathematical expressions
elegant; generalization to other configurations using the present
formalism is straightforward.) The graphene's $T$~matrix depends
on the magnetic field intensity, electronic density, temperature,
and LL broadening, via the complex optical conductivity of graphene
$\sigma_{\pm}(\omega)$; its explicit form is 
\begin{equation}
T_{\pm}^{\textrm{g}}=\frac{1}{2}\left[\begin{array}{cc}
2+Z_{0}\sigma_{\mp}(\omega) & Z_{0}\sigma_{\mp}(\omega)\\
-Z_{0}\sigma_{\mp}(\omega) & 2-Z_{0}\sigma_{\mp}(\omega)
\end{array}\right]\,,\label{eq:t_matrix_graphene}
\end{equation}
 where $Z_{0}=\mu_{0}c$ denotes the vacuum impedance; see Appendix~\ref{AppendixA}
for a detailed derivation. Finally, the second line of Eq.~(\ref{eq:tota_t_matrix})
propagates the field in free space for a distance of $L/2$ and adds
the right-end mirror.

The Faraday rotation angle is obtained from $\theta_{F}=(1/2)\textrm{arg}(t_{+}/t_{-})$,
with the circular amplitude ratio $t_{+}/t_{-}$ given by $[T_{-}^{\textrm{in}\rightarrow\textrm{out}}]_{1,1}/[T_{+}^{\textrm{in}\rightarrow\textrm{out}}]_{1,1}$
{[}Eq.~(\ref{eq:transmitted_vs_incident}){]}. After some algebra,
we arrive at 
\begin{equation}
\frac{t_{+}}{t_{-}}=\frac{2+Z_{0}\sigma_{+}(\omega)-|r|\left[Z_{0}\sigma_{+}(\omega)-2\right]e^{i\omega L/c}}{2+Z_{0}\sigma_{-}(\omega)-|r|\left[Z_{0}\sigma_{-}(\omega)-2\right]e^{i\omega L/c}}\,,\label{eq:tp_over_tm_cav_graph}
\end{equation}
 from which $\theta_{F}$ can be immediately deduced. Setting $r=0$
in the latter expression leads to the previous result in the absence
of a cavity {[}compare with $t_{\pm}$ as obtained from Eq.~(\ref{eq:t_p_m})
with $\epsilon_{r}=1${]}.

When $r>0$, interference takes place and photons can make several
round trips before being transmitted through the cavity. On an intuitive
basis, we then expect that the Faraday rotation angle can be enhanced
due to multiple passages of photons through graphene, which indeed
is the case, as shown in Fig.~\ref{fig:FaradayRotCavGraph}. Hereafter,
the size of the cavity is set to $L=n\pi c/\omega$, with $n$ odd.
The solid line shows $\theta_{F}$ for the cavity-graphene system
and the dashed-dotted line shows $\theta_{F}$ for free-standing graphene
for the same parameters: clearly, in the range of frequencies considered,
the Faraday effect is greatly enhanced. For example, for a low frequency,
$\hbar\omega\approx10$ meV, $\theta_{F}$ has increased by a factor
of about $5$, reaching a value of 55~degrees, whereas for $\hbar\omega\approx19$~meV,
$\theta_{F}$ increases by a factor of about $20$, reaching a value
of approximately 25~degrees.

Direct inspection of Eq.~(\ref{eq:tp_over_tm_cav_graph}) discloses
the observed boost of the Faraday effect: when $r\rightarrow1$ and
the phase factor $\textrm{exp}(i\omega L/c)=-1$, the constant factor
of $2$ cancels in both the denominator and the numerator, leading
to, 
\begin{equation}
\left|\frac{t_{+}}{t_{-}}\right|e^{2i\theta_{F}}\underset{\underset{r\simeq1}{n\:\textrm{odd}}}{\simeq}\frac{\sigma_{+}(\omega)}{\sigma_{-}(\omega)},\label{eq:approx}
\end{equation}
 which can present large arguments, $2\theta_{F}$. The opposite limit,
$r\rightarrow0$, in which the isolated graphene system is recovered,
leads to much smaller arguments, since generally $2\gg Z_{0}\textrm{Im}\,\sigma_{\pm}$,
which implies that the real part of Eq.~(\ref{eq:t_p_m}) is predominant.
Choosing a cavity mode with $n$ odd and $r\simeq1$ is fully equivalent
to taking a large number of equally prepared graphene sheets placed
in a row (Appendix~\ref{Appendix_B}). The cavity geometry therefore
permits us to take advantage of large Faraday rotation accumulation
using a single graphene sheet.

In a cavity geometry, the Faraday rotation is no longer dominated
by the behavior of $\sigma_{xy}^{\prime}(\omega)$ {[}see Eq.~(\ref{eq:theta_F_approx}){]},
for $\theta_{F}$ now depends on the full conductivity tensor {[}Eq.~(\ref{eq:approx}){]}.
The most visible consequence of the latter fact is that photons with
$\hbar\omega\approx20$~meV undergo considerable Faraday rotation
angles in a cavity geometry, whereas, in a single passage through
graphene, photons with such energy do not produce Faraday rotation
at all (Fig.~\ref{fig:FaradayRotCavGraph}). This apparently counter-intuitive
result is due to induced ellipticity in single passages and is explained
in Appendix~\ref{Appendix_B}.

\begin{figure}
\begin{centering}
\includegraphics[clip,width=0.9\columnwidth]{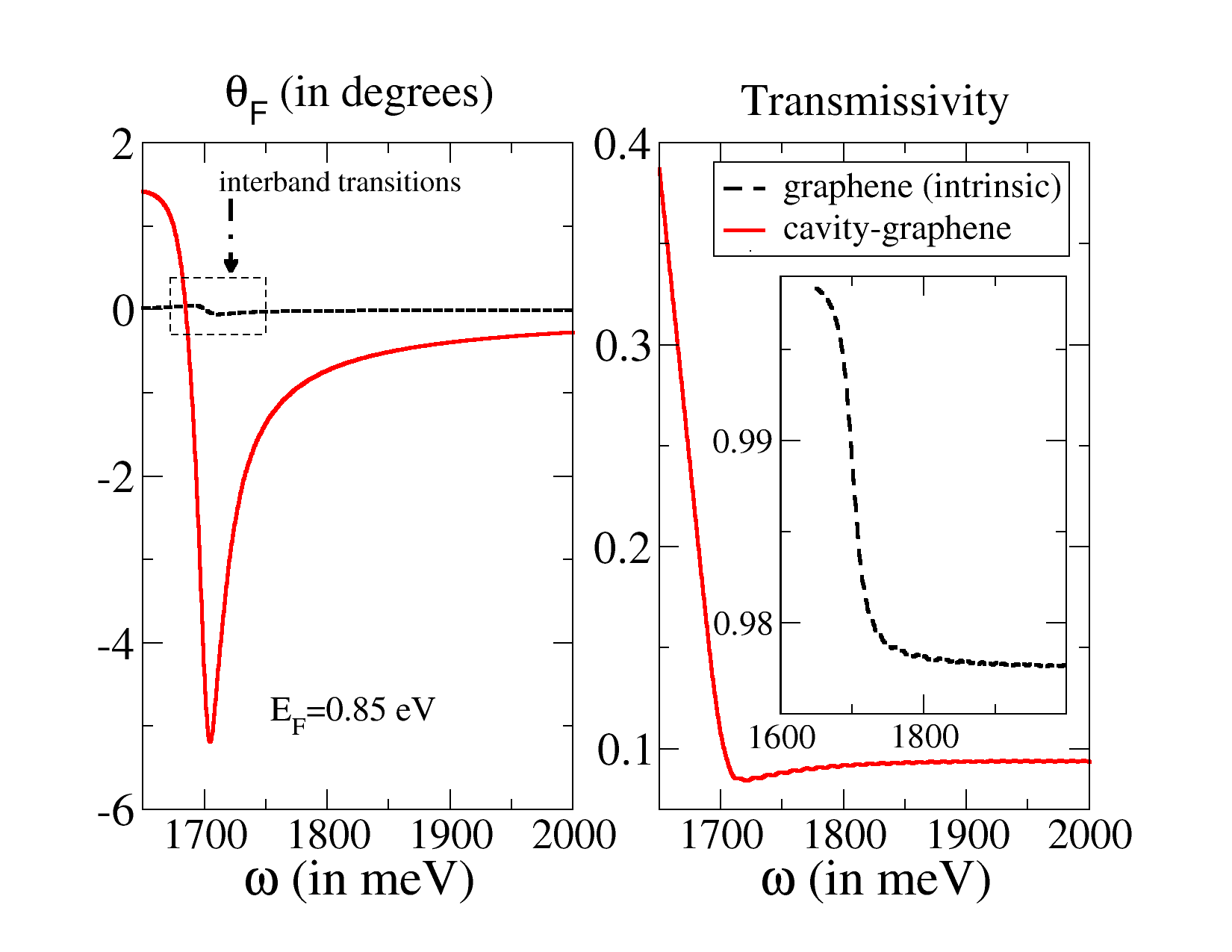} 
\par\end{centering}

\bigskip{}

\begin{centering}
\includegraphics[clip,width=0.9\columnwidth]{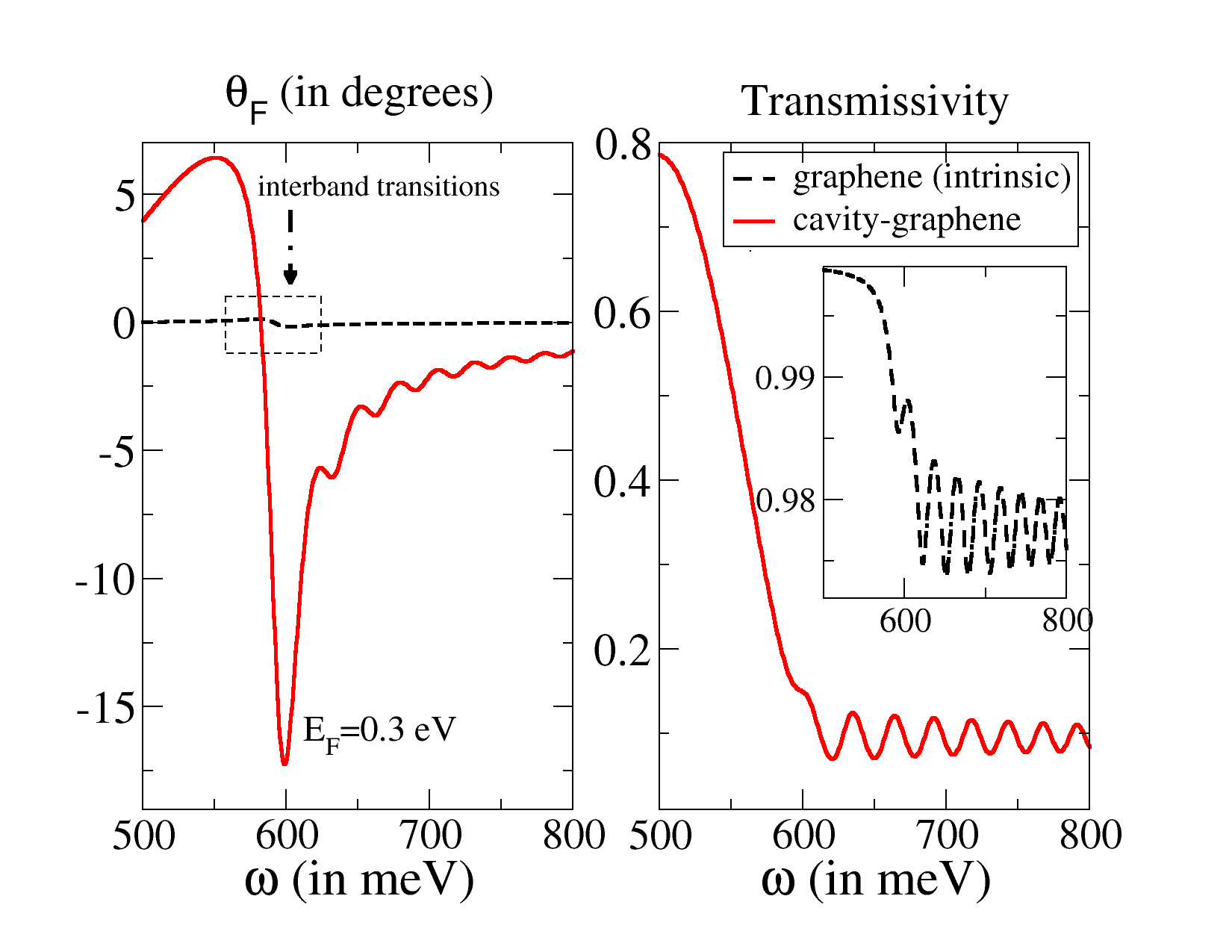} 
\par\end{centering}

\centering{}\caption{\label{fig:Faraday_visible}Faraday rotation boost in the infrared
and visible ranges.\textbf{ Left}: Faraday rotation angle versus photon
energy of a cavity-graphene system with $E_{F}=0.85$~eV (top) and
$E_{F}=0.3$~eV (bottom). \textbf{Right}: Transmissivity of a cavity-graphene
system for the same parameters considered at the left panels. Inset:
Transmissivity of intrinsic graphene for the same parameters. Other
parameters as in Fig.~\ref{fig:FaradayRotCavGraph}. In order to
obtain a measurable Faraday rotation at $\hbar\omega\approx1.7$~eV
{[}solid (red) line{]} {[}$\hbar\omega\approx0.5$~eV (infrared){]},
it is necessary to tune the intraband resonance according to $E_{F}\simeq\hbar\omega/2$.}
\end{figure}

\emph{a. Semi-classical versus quantum regimes in a cavity geometry}---Figure~\ref{fig:FaradayRotCavGraph}
(top) considers the case of $E_{F}=0.3$~eV and $B=7$~T, well inside
the semiclassical regime, for which the ac conductivity is dominated
by intraband contributions over a wide range of frequencies (Sec.~\ref{Sec2 (Magneto-Optical Transport)});
the corresponding intraband Faraday rotation is seen to be greatly
enhanced in the cavity geometry.

The low-electronic-density regime of the cavity-graphene system is
shown in the bottom panel in Fig.~\ref{fig:FaradayRotCavGraph}.
Remarkably, for energies above the interband threshold, namely, $\hbar\omega\gtrsim E_{1}\simeq95$~meV
for $N_{F}=0$ (left) and $\hbar\omega\gtrsim E_{1}+E_{2}\simeq230$~meV
for $N_{F}=1$ (right), $\theta_{F}(\omega)$ presents a behavior
qualitative different from that of an isolated graphene sheet (black
dot double-dashed curve): oscillations do emerge. These oscillations
are hindered in single-photon passages through graphene (see also
Fig.~\ref{fig:Farad.Quantum.Reg}), but for multiple-photon passages,
in the high-frequency limit, Shubnikov\textendash{}de Haas oscillations
in the longitudinal conductivity $\sigma_{xx}(\omega)$ (Fig.~\ref{fig:Cond_xx_vs_freq})
are critical in defining the orientation of light polarization axes.
These oscillations are obviously absent in the semiclassical Boltzmann
calculation {[}dashed (orange) curve{]}. In the top panel, where $E_{F}=0.3$~eV,
such oscillations are not present because the represented photon energies
are well below the threshold for interband transitions $\hbar\omega\simeq2E_{F}$.

\emph{b. Near-infrared and visible-range Faraday rotation}---We finish
this section by mentioning an important application of the cavity-graphene
system: interband-induced Faraday rotations in the near-infrared and
visible regimes. Figure~\ref{fig:Faraday_visible} shows that energetic
photons can attain $\theta_{F}\gtrsim1$ by tuning the Fermi energy
to sufficiently high values. In this regard, the top panel shows numerical
data for graphene with $E_{F}=0.85$ eV; such a high doping level
of graphene samples is feasible using chemically synthesized graphene
with ferroelectric substrates (instead of the conventional SiO$_{2}$).\cite{Wafer-scale_Graphene}

Given the mirror reflection amplitude considered, $r=0.99$, photons
are trapped for a large number of round trips. This means that it
is highly probable that photons get absorbed by graphene before leaking
into the cavity. This explains why the transmissivity of the cavity-graphene
system, as shown in the right panel in Fig.~\ref{fig:Faraday_visible},
is well below 1 (but still large enough that the effect can be measured).
One way of increasing the transmissivity of a cavity-graphene system
is to decrease the quality of the mirrors, at the expense of decreasing
the maximum achievable $\theta_{F}$.

We, finally, remark that the nonlinearity associated with next-neighbor
hopping $t^{\prime}$ in a honeycomb graphene lattice can play a role
for photons with $\hbar\omega\gtrsim1$~eV, and hence corrections
to the Dirac cone approximation (Sec.~\ref{sub:Graphene_Outline}),
and thus to the EOM solutions, may exist; such corrections are expected
to be very small, however.\cite{Stauber_VisibleReg}

\section{Conclusion and outlook}

In the first part of this work, the EOM method has been adapted to
the study of magneto-optical transport of electronic systems. To illustrate
the method, the magneto-optical conductivity tensor of single-layer
graphene in the Dirac cone approximation has been derived, accounting
for both intraband (semi-classical) transitions and interband transitions
between the valence and the conduction bands.

The general regularization procedure to obtain the regular conductivity
tensor from the solutions of the EOM for the current operator has
been established; this procedure is shown to lead to the correct formulas
without the need for evaluation of the Kubo formula. To the best of
the authors' knowledge, such a procedure has not been discussed in
the literature so far. In addition, quantitative comparisons between
the quantum EOM solutions and the semiclassical Boltzmann formulas,
in the full optical spectrum, and in both low- and high-doped graphene
samples, have been given throughout.

In the second part, the Faraday rotation effect in single-layer graphene
has been studied in detail; in particular, simple formulas for the
step heights in the quantum Hall regime have been derived. Our results
have been shown to account well for available experimental data in
the semiclassical regime.

Finally, we have proposed a simple experimental apparatus based on
an optical cavity that leads to an enhancement of the Faraday rotation
effect of graphene by orders of magnitude, thus allowing to obtain
giant Faraday rotation angles in the infrared region and modest Faraday
rotation angles in the visible region.

We hope that the present work further stimulates the research on magneto-optical
properties of ultrathin 2D gases and graphene-based solid-state devices.

\section*{Acknowledgments}

A.F. acknowledges FCT Grant No. SFRH/BPD/65600/2009. N.M.R.P. acknowledges
Fundos FEDER, through the Programa Operacional Factores de Competitividade
- COMPETE and by FCT under Project No. Past-C/FIS/UI0607/2011. A.H.C.N.
acknowledges support from the DOE Grant No. DE-FG02-08ER46512 and
the ONR Grant No. MURI N00014-09-1-1063.

\appendix
%dummy comment inserted by tex2lyx to ensure that this paragraph is not empty

\section{Transfer matrix formalism\label{AppendixA}}

The transfer matrix ($T$~matrix) approach is a widely used method
in optics and related fields and provides an efficient means of calculating
the amplitude and phase of transmitted electric fields through an
arbitrary number of interfaces. In this appendix, we give a self-contained
review of the method and derive explicitly the $T$~matrix for a
general 2D conducting media.

\subsection{General formalism}

\begin{figure}
\begin{centering}
\includegraphics[clip,width=0.9\columnwidth]{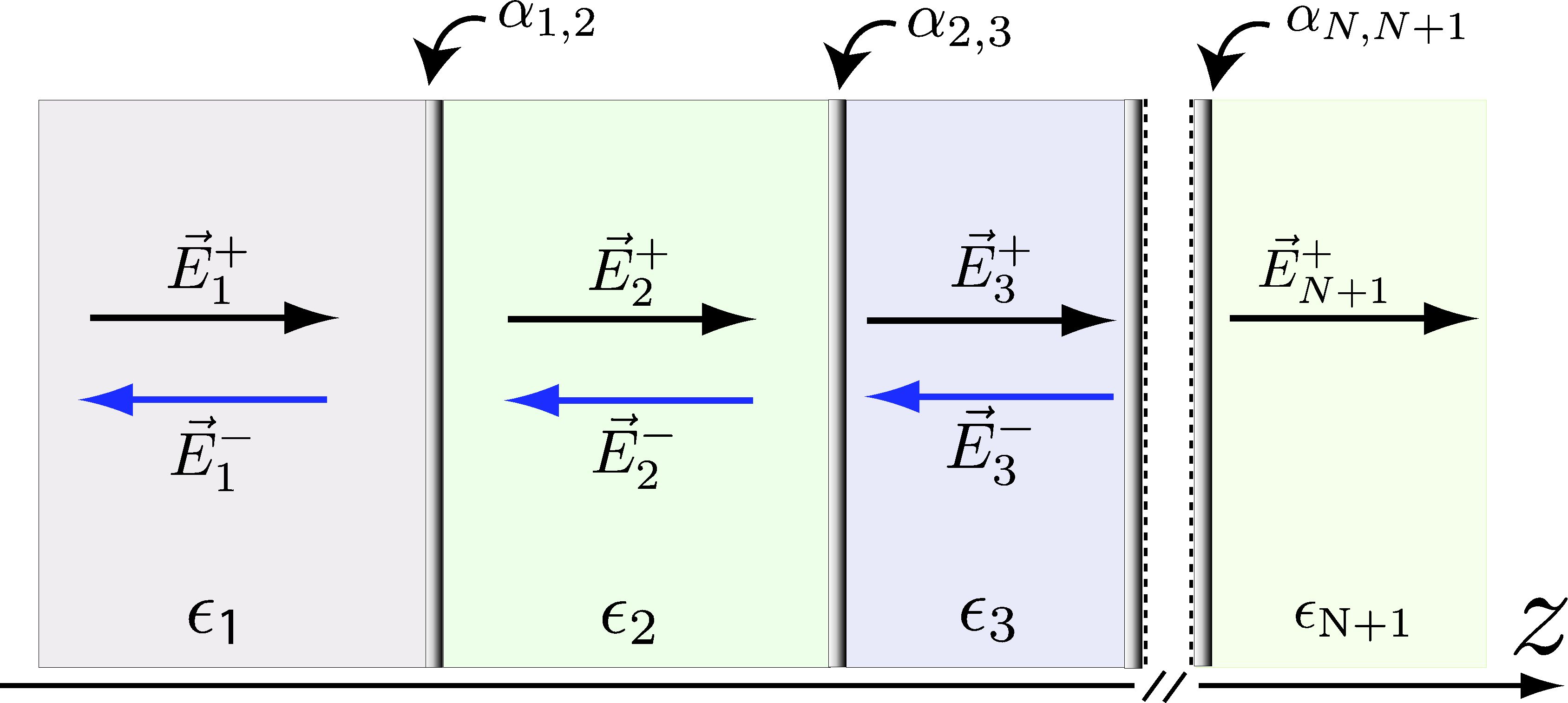} 
\par\end{centering}

\caption{\label{fig:Optical_Interfaces}Schematic of an optical system consisting
of an array of interfaces separated by different types of dielectric
media. An electromagnetic wave, $\boldsymbol{E}^{\textrm{in}}=\boldsymbol{E}_{1}^{+}$,
coming from a medium with dielectric permittivity $\epsilon_{1}$
interacts with an interface $\alpha_{12}$. As a result, it is partially
reflected and partially transmitted into the medium $\epsilon_{2}$.
Equivalent events take place at the remaining interfaces. The vectors
with superscript $+$($-$) denote the component of the electric field
traveling in the positive (negative) direction of $z$. A uniform
static magnetic field $\mathbf{B}=B\mathbf{e_{y}}$ is assumed.}
\end{figure}

For concreteness, we assume that an incident electromagnetic wave
of frequency $\omega$, travels in the $z$ direction through a set
of $N$ metallic interfaces, placed normal to the direction of propagation,
with labels $\alpha_{n,n+1}$, and located at positions $z=z_{n}$
($n\in1,2,...N$). These interfaces are separated by dielectric mediums
--- Fig.~\ref{fig:Optical_Interfaces} shows the configuration we
have in mind.

The electric field is separated according to the direction of propagation:
$\boldsymbol{E}_{n}^{+}(z)$ represents the part the electric field
traveling in the positive direction of $z$, within the region $n$,
whereas $\boldsymbol{E}_{n}^{-}(z)$ represents the part traveling
in the opposite direction.

As shown below, the calculation of transmitted and reflected amplitudes
becomes easier by writing the boundary conditions in terms of circularly
polarized waves (see also Sec.~\ref{sub:Faraday_Rot}). Therefore,
we focus on the circular amplitudes, 
\begin{equation}
E_{n,\tau}^{\pm}(z)=E_{n,x}^{\pm}(z)+i\tau E_{n,y}^{\pm}\,,\label{eq:notation_electric_field}
\end{equation}
 where $\tau$ is the polarization index: $\tau=\pm1$ {[}$+1$($-1$)
means right-handed (left-handed) circular polarization{]}. Indeed,
in a given region $i$, the total (complex) electric field is the
sum of both components,
\begin{equation}
\boldsymbol{E}_{n}(z,t)=\boldsymbol{E}_{n}^{+}(z)e^{-i\omega t}+\boldsymbol{E}_{n}^{-}(z)e^{-i\omega t}\,.\label{eq:electric field}
\end{equation}
 The physical electric field is obtained by taking the real part of
the latter expression. We omit the time-dependence in the remainder
of the appendix.

The $T$~matrix connects the amplitude of the electric field to the
left and to the right of a given boundary (interface). Take for instance,
the interface labeled $\alpha_{1,2}$ in Fig.~\ref{fig:Optical_Interfaces}.
The respective $T$~matrix, $\hat{T}^{1,2}$, is defined as,
\begin{equation}
\left(\begin{array}{c}
E_{2,\tau}^{+}(z_{1}^{+})\\
E_{2,\tau}^{-}(z_{1}^{+})
\end{array}\right)=\hat{T}_{\tau}^{1,2}\left(\begin{array}{c}
E_{1,\tau}^{+}(z_{1}^{-})\\
E_{1,\tau}^{-}(z_{1}^{-})
\end{array}\right)\,,\label{eq:t_matrix_definition}
\end{equation}
 where $z_{1}^{\pm}$ denote the position where the electric field
is to be evaluated: right after ($+$) or before ($-$) the interface
located at $z=z_{1}$. For ease of notation, we define the vector
of amplitudes,
\begin{equation}
\mathcal{\boldsymbol{E}}_{n,\tau}(z)=\left(\begin{array}{c}
E_{n,\tau}^{+}(z)\\
E_{n,\tau}^{-}(z)
\end{array}\right)\,,\label{eq:vector_amplitudes}
\end{equation}
 and drop the superscripts in the coordinates $z_{n}$.

If more than one interface is present, the light propagates a given
distance before interacting with the next component. Propagation of
light through a dielectric medium merely adds a phase to each $\tau$
circular component of the electric field. Indeed, its action can be
represented by a diagonal matrix, 
\begin{equation}
\mathcal{\boldsymbol{E}}_{n,\tau}(z_{i})=\left[\begin{array}{cc}
e^{-ik_{n}\Delta z_{n}} & 0\\
0 & e^{ik_{n}\Delta z_{n}}
\end{array}\right]\mathcal{\boldsymbol{E}}_{n,\tau}(z_{i+1})\,,\label{eq:free_propagation}
\end{equation}
 where the index $n$ just takes the values for which there is intermediate
light propagation, i.e., $n=1,...,N-1$, the wave vector depends on
the dielectric medium according to $k_{n}=\omega\sqrt{\epsilon_{n}}/c$
and $\Delta z_{n}=z_{n+1}-z_{n}$ is the width of the region $n$.
Note that Eq.~(\ref{eq:free_propagation}) defines a particular case
of a $T$~matrix, which we denote by $\hat{F}_{n}$.

The problem of finding how the output electric field, of definite
polarization $\tau$, immediately after leaving the last interface,
$E_{N+1,\tau}^{+}(z_{N})$, relates to the incoming electric field,
with the same polarization $\tau$, $E_{1,\tau}^{+}(z_{1})$, then
amounts to take the product of the individual transfer matrices, 
\begin{equation}
\mathcal{\boldsymbol{E}}_{1,\tau}(z_{1})=\underset{\hat{T}_{\tau}^{\textrm{in}\rightarrow\textrm{out}}}{\underbrace{\hat{T}_{\tau}^{1,2}\hat{F}_{2}\hat{T}_{\tau}^{2,3}...\hat{T}_{\tau}^{N-1,N}\hat{F}_{N}\hat{T}_{\tau}^{N,N+1}}}\mathcal{\boldsymbol{E}}_{N+1,\tau}(z_{N})\,.\label{eq:prod_t_matrices}
\end{equation}
 The total $T$~matrix $\hat{T}_{\tau}^{\textrm{in}\rightarrow\textrm{out}}$
has the desired information:
\begin{equation}
\frac{E_{N+1,\tau}^{+}}{E_{1,\tau}^{+}}=1/\left[\hat{T}_{\tau}^{\textrm{in}\rightarrow\textrm{out}}\right]_{1,1}\,.\label{eq:transmitted_vs_incident}
\end{equation}
 As for the relation between the output field and the reflected field
at the first boundary, $E_{1,\tau}^{-}(z_{1}),$ we obtain,
\begin{equation}
\frac{E_{N+1,\tau}^{+}}{E_{1,\tau}^{-}}=1/\left[\hat{T}^{\textrm{in}\rightarrow\textrm{out}}\right]_{2,1}\,.\label{eq:transmitted_vs_reflected}
\end{equation}

In what follows, we show how to construct the $T$~matrix for a general
2D conducting medium. Knowledge of the transfer matrices allows us
to determine the characteristics of transmitted and reflected light
through a general set of conducting 2D thin films, mirrors, etc.,
by employing Eq.~(\ref{eq:prod_t_matrices}).

\subsection{$T$ matrix for a general conducting 2D interface}

We restrict the present derivation to non-magnetic media, and assume
the standard constitutive relations hold,
\begin{eqnarray}
\boldsymbol{D}(\mathbf{r},\omega) & = & \epsilon(\mathbf{r},\omega)\boldsymbol{E}(\mathbf{r},\omega)\,,\label{eq:rel1-1}\\
\boldsymbol{J}(\mathbf{r},\omega) & = & \hat{\sigma}(\mathbf{r},\omega)\boldsymbol{E}(\mathbf{r},\omega)\,,\label{eq:rel2-1}
\end{eqnarray}
 where $\boldsymbol{D}$, $\epsilon$, and $\sigma$ denote the displacement
field, permittivity and conductivity, respectively. Also, and without
prejudice, we take the 2D conducting interface to be located at $z=0$.
The $T$~matrix is defined as
\begin{equation}
\left(\begin{array}{c}
E_{a,\tau}^{+}(z=0^{-})\\
E_{a,\tau}^{-}(z=0^{-})
\end{array}\right)=\hat{T}_{\tau}^{(ab)}\left(\begin{array}{c}
E_{b,\tau}^{+}(z=0^{+})\\
E_{b,\tau}^{-}(z=0^{+})
\end{array}\right)\,,\label{eq:tmatrix_definition}
\end{equation}
 where $a$ ($b$) is the bookkeeping index for the medium at the
left (right) of the interface.

Various constraints emerge due to continuity of $\boldsymbol{E}$
(and its derivative) at the 2D conducting interface. Indeed, Maxwell
equations imply that, 
\begin{eqnarray}
\boldsymbol{E}_{a}(0) & = & \boldsymbol{E}_{b}(0)\,,\label{eq:Cont_1-1}\\
\left(\frac{\partial\boldsymbol{E}_{a}}{\partial z}\right)_{z=0}-\left(\frac{\partial\boldsymbol{E}_{b}}{\partial z}\right)_{z=0} & = & i\omega\mu_{0}\hat{\sigma}\boldsymbol{E}_{b}(0)\,,\label{eq:Cont_2-1}
\end{eqnarray}
 where the conductivity tensor reads 
\begin{equation}
\hat{\sigma}=\sigma_{ij}(\omega)\delta(z)\,.\label{eq:cond_tensor}
\end{equation}
 The conductivity depends on the light frequency $\omega$, and, generally,
also on other quantities (Fermi energy of the interface, temperature,
etc.). In the latter expression, the subscripts $i,j=x,y$ are Cartesian
coordinates. In terms of circularly polarized fields, Eq.~(\ref{eq:Cont_2-1})
reads,
\begin{align}
k_{a}(E_{a,\tau}^{+}-E_{a,\tau}^{-})-k_{b}(E_{b,\tau}^{+}-E_{b,\tau}^{-}) & =\omega\mu_{0}\times\nonumber \\
 & \times(E_{b,\tau}^{+}+E_{b,\tau}^{-})\sigma_{-\tau}(\omega)\,,\label{eq:Cont_Circular}
\end{align}
 where we have admitted an isotropic medium, $\sigma_{xx}=\sigma_{yy}$,
and have defined
\begin{eqnarray}
\sigma_{\pm}(\omega) & = & \sigma_{xx}(\omega)\pm i\sigma_{xy}(\omega)\,.\label{eq:def_comp_cond}
\end{eqnarray}
 The statement, Eq.~(\ref{eq:Cont_Circular}), shows that the two
circularly polarizations are decoupled, even in the presence of a
complex conductivity $\sigma_{\pm}(\omega)$. This is why it is advantageous
to write the boundary conditions in terms of circularly polarized
fields (Sec.~\ref{sub:Faraday_Rot}).

According to the definition of the $T$~matrix {[}Eq.~(\ref{eq:tmatrix_definition}){]},
we need to relate $E_{a,\tau}^{+}$ with $E_{b,\tau}^{\pm}$ and $E_{a,\tau}^{-}$
with $E_{b,\tau}^{\pm}$, separately. To do so, we make use of the
continuity condition, Eq.~(\ref{eq:Cont_1-1}), written in circular
waves, $E_{a,\tau}^{+}+E_{a,\tau}^{-}=E_{b,\tau}^{+}+E_{b,\tau}^{-}$,
in order to arrive at,
\begin{align}
\pm2k_{a}E_{a,\tau}^{\pm} & =k_{b}(E_{b,\tau}^{+}-E_{b,\tau}^{-})\nonumber \\
 & +[\omega\mu_{0}\sigma_{-\tau}(\omega)\pm k_{a}](E_{b,\tau}^{+}+E_{b,\tau}^{-})\,.\label{eq:aux-1}
\end{align}

Combining Eq.~(\ref{eq:tmatrix_definition}) and the latter expression,
we arrive at the desired result,
\begin{equation}
\hat{T}_{\tau}^{(ab)}=\frac{1}{2k_{a}}\left[\begin{array}{cc}
\Lambda_{\tau,++}^{ab} & \Lambda_{\tau,-+}^{ab}\\
\Lambda_{\tau,--}^{ab} & \Lambda_{\tau,+-}^{ab}
\end{array}\right]\,,\label{eq:TMatrix_Interface2D}
\end{equation}
 where
\begin{equation}
\Lambda_{\tau,\pm\pm}^{ab}=k_{a}\pm k_{b}\pm\omega\mu_{0}\sigma_{-\tau}(\omega)\,.\label{eq:def_c}
\end{equation}

\subsection{Example: $T$ matrix of suspended graphene}

The $T$~matrix of suspended graphene can be obtained immediately
from Eq.~(\ref{eq:def_comp_cond}). Admitting that the mediums at
the left and right of the single-layer graphene sheet are air, we
obtain
\begin{equation}
\hat{T}_{\tau}^{\textrm{graph}}=\frac{1}{2}\left[\begin{array}{cc}
2+Z_{0}\sigma_{-\tau}^{\textrm{graph}}(\omega)\qquad & Z_{0}\sigma_{-\tau}^{\textrm{graph}}(\omega)\\
-Z_{0}\sigma_{-\tau}^{\textrm{graph}}(\omega)\qquad\quad & 2-Z_{0}\sigma_{-\tau}^{\textrm{graph}}(\omega)
\end{array}\right]\,,\label{eq:TMatrix_suspended_graphene}
\end{equation}
 where $Z_{0}=\mu_{0}c$ is the vacuum impedance.

\section{Faraday Effect\label{Appendix_B}}

In the present appendix, we derive the exact analytical conditions
for the existence of Faraday rotation and discuss their modification
when graphene is enclosed in an optical cavity. Despite the focus
on graphene, most of the conclusions drawn here apply generally for
systems possessing in-plane symmetry. Once again, for simplicity,
we consider the case of suspended graphene; generalization to the
case of graphene on top of a substrate is straightforward using the
general formulas given in Appendix~\ref{AppendixA}.

\subsection{Conditions for Faraday effect in free space}

We consider a target graphene sheet, placed on the $xy$ plane, subjected
to a normally incident electromagnetic wave, linearly polarized along
the $x$ axis, $E_{x}e^{-i\omega t}$. The magneto-optical Faraday
effect takes place when a magnetic field $\boldsymbol{B}=Be_{z}$
is applied. Then, Lorentz force acts on free carriers, producing a
Hall electronic ac current, which, under specific conditions (see
below), will produce out-of-phase radiation polarized transversely
to the impinging field, $E_{y}e^{-i\omega t}e^{i\phi}$. As a consequence,
the resulting electromagnetic wave sees its polarization plane rotated.

Without loss of generality, consider the graphene sheet to be placed
at $z=0$. In the circular basis, $\boldsymbol{e}_{\tau}=(1/2)(\boldsymbol{e}_{x}+\tau i\boldsymbol{e}_{y})$,
the electromagnetic field at $z=0^{-}$, reads
\begin{equation}
\boldsymbol{E}(0^{-})=E_{0}e^{-i\omega t}(\boldsymbol{e}_{+}+\boldsymbol{e}_{-})\,.\label{eq:impinging_field}
\end{equation}

Note that the actual electric field is given by the real part of the
latter equation. After interaction with graphene, each of the circular
components $\tau=\pm1$ changes according to Eq.~(\ref{eq:TMatrix_suspended_graphene}).
The field right after the graphene plane is given by 
\begin{equation}
\boldsymbol{E}(0^{+})=E_{0}e^{-i\omega t}\left[\frac{1}{1+\beta_{-}}\boldsymbol{e}_{+}+\frac{1}{1+\beta_{+}}\boldsymbol{e}_{-}\right]\,,\label{eq:field_after_graph}
\end{equation}
 with $\beta_{\pm}=Z_{0}\sigma_{\pm}(\omega)/2$. To determine whether
the plane of polarization has rotated, we write the latter equation
in the Cartesian basis,
\begin{align}
\boldsymbol{E}(0^{+}) & =\frac{E_{0}e^{-i\omega t}}{2(1+\beta_{+})(1+\beta_{-})}\times\nonumber \\
 & \times\left[\left(2+Z_{0}\sigma_{xx}\right)\boldsymbol{e}_{x}-Z_{0}\sigma_{xy}\boldsymbol{e}_{y}\right]\,,\label{eq:cartesian_basis}
\end{align}
 where we have used the definition of $\beta_{\pm}$ to simplify the
term inside brackets. Obviously, no Faraday rotation takes place when
$\sigma_{xy}(\omega)=0$. On the other hand, having $\sigma_{xy}(\omega)\neq0$
does not suffice to rotate the polarization plane; linear polarization
can change to elliptic polarization with main axes along $x$ and
$y$ (this is the case for $B=5$~T and $\hbar\omega\approx15$~meV,
as shown in the top panel in Fig.~\ref{fig:Faraday_Various}: elliptic
polarized light leaves the graphene sheet, $\delta\approx0.15$, but
still $\theta_{F}=0$). For this reason, the actual condition for
the existence of Faraday rotation is 
\begin{equation}
|\sigma_{xy}|>0\,\wedge\,\textrm{Arg}\left(\frac{2+Z_{0}\sigma_{xx}}{Z_{0}\sigma_{xy}}\right)\neq\pm(2m+1)\frac{\pi}{2}\,,m\in\mathbb{N}_{0}\,.\label{eq:condition_faraday_rotation}
\end{equation}
 The amount of Faraday rotation is given by Eq.~(\ref{eq:theta_faraday-1})
and thus can be obtained directly from Eq.~(\ref{eq:field_after_graph}),
reading,
\begin{equation}
\theta_{F}=\frac{1}{2}\textrm{Arg}\left(\frac{2+Z_{0}\sigma_{-}}{2+Z_{0}\sigma_{+}}\right)\,.\label{eq:theta_F}
\end{equation}
 In many situations (e.g. high photon energies and high electronic
density), the longitudinal conductivity obeys $Z_{0}\sigma_{xx}^{\prime\prime}\ll2+Z_{0}\sigma_{xx}^{\prime}$,
thus leading to the approximate condition, $|\sigma_{xy}^{\prime}|>0\Rightarrow\theta_{F}>0$.
This is consistent with the approximated formula derived for the Faraday
rotation angle {[}Eq.~(\ref{eq:theta_F_approx}){]}, which states
that $\theta_{F}$ is proportional to $\sigma_{xy}^{\prime}$ (see
also Fig.~\ref{fig:Faraday_Various}).

\subsection{Conditions for the Faraday effect in an optical cavity}

In Sec.~\ref{sub:Enhancement-of-Faraday}, we have seen that large
Faraday rotations $\theta_{F}$ can be achieved in a cavity-graphene
system, even for such photon energies that do not cause Faraday rotation
in free space. An example is given in Fig.~\ref{fig:FaradayRotCavGraph}:
in free space, impinging light with $\hbar\omega\thickapprox20$~meV
does not change its polarization direction, $\theta_{F}=0$, whereas
$\theta_{F}$ can be as large as $25{}^{\circ}$ for graphene mounted
on a cavity geometry.

In order to explain the above-described phenomenon, it is sufficient
to consider the simplified situation where a normally incident photon
interacts with graphene twice in a row. For concreteness, we take
two graphene samples, equally prepared, separated by a given distance
$W$. Let the photon frequency $\bar{\omega}$ be such that no Faraday
rotation is produced in the passage through the first graphene sample,
that is, 
\begin{equation}
\textrm{Arg}\left[\frac{2+Z_{0}\sigma_{xx}(\bar{\omega})}{Z_{0}\sigma_{xy}(\bar{\omega})}\right]=\pm(2m+1)\frac{\pi}{2}\,,\label{eq:condition_zero}
\end{equation}
 for some $m\in\mathbb{N}_{0}$ {[}see Eq.~(\ref{eq:condition_faraday_rotation}){]}.
In the latter expression, it is assumed that $\sigma_{xy}(\bar{\omega})\neq0$
which is the case when a magnetic field is present. In these conditions,
after the first passage, the electric field {[}Eq.~(\ref{eq:cartesian_basis}){]},
can be written as
\begin{align}
\boldsymbol{E}_{1} & =\frac{E_{0}e^{-i\bar{\omega}t}}{2[1+\beta_{+}(\bar{\omega})][1+\beta_{-}(\bar{\omega})]}e^{i\phi}\times\nonumber \\
 & \times\left[\left|2+Z_{0}\sigma_{xx}(\bar{\omega})\right|\boldsymbol{e}_{x}\pm i\left|Z_{0}\sigma_{xy}(\bar{\omega})\right|\boldsymbol{e}_{y}\right]\,,\label{eq:cartesian_basis-1}
\end{align}
 where $\phi=\textrm{Arg}[2+Z_{0}\sigma_{xx}(\bar{\omega})]$ and
the sign $\pm$ depends on the actual argument of $\sigma_{xy}(\bar{\omega})$.
The latter equation describes a field elliptically polarized with
main axes along $x$ and $y$ (i.e., $\theta_{F}=0$). We thus see
that although no Faraday rotation occurs when Eq.~(\ref{eq:condition_zero})
is fulfilled, the polarization changes from linear to elliptic, an
unavoidable consequence for Lorentz force enforces some radiation
to be emitted that is polarized along the $y$ axis.

In order to determine the field after the second passage, and hence
demonstrate our point, i.e., that some Faraday rotation must necessarily
be produced in multiple passages through graphene (such as in a cavity
geometry), we make use of the transfer matrix formalism. Indeed, we
approximate the total $T$~matrix by $\hat{T}_{\tau}^{\textrm{graph}}\cdot\hat{T}_{\tau}^{\textrm{graph}}$
(this approximation is exact when the phase for free propagation between
the graphene sheets, $\omega W/c$, equals $2m\pi$). Employing Eq.~(\ref{eq:transmitted_vs_incident}),
we obtain

\begin{align}
\boldsymbol{E}_{2} & =\frac{E_{0}e^{-i\bar{\omega}t}}{[1+2\beta_{+}(\bar{\omega})][1+2\beta_{-}(\bar{\omega})]}\times\nonumber \\
 & \times\left\{ [1+Z_{0}\sigma_{xx}(\bar{\omega})]\boldsymbol{e}_{x}-Z_{0}\sigma_{xy}(\bar{\omega})\boldsymbol{e}_{y}\right\} \,.\label{eq:cartesian_basis-1-1}
\end{align}
 This time, the condition for zero Faraday rotation, 
\begin{equation}
\textrm{Arg}\left[\frac{1+Z_{0}\sigma_{xx}(\bar{\omega})}{Z_{0}\sigma_{xy}(\bar{\omega})}\right]=\pm(2m+1)\frac{\pi}{2}\,,\label{eq:condition_zero_2}
\end{equation}
 cannot be fulfilled because Eq.~(\ref{eq:condition_zero}) fixes
the photon frequency in this example. Then a finite (nonzero) Faraday
rotation is produced in the second passage.

The case of graphene in a cavity geometry is more involved because
intracavity interference takes place. Nevertheless, the physics behind
the boost of Faraday rotation is analogous: if, for graphene subjected
to a transverse magnetic field, it turns out that the first photon
passage yields $\theta_{F}=0$, then, in the following passages it
must be that $\theta_{F}>0$. See, for instance, Eq.~(\ref{eq:approx}),
valid for an optical cavity made of mirrors with very high reflection
amplitudes: because $|\sigma_{xy}(\omega)|>0$, for $B>0$, then $\theta_{F}>0$
for all light frequencies.

\subsection{Row of graphene sheets}

Taking a number $N$ of graphene sheets separated by $W$, such that
$\omega W/c=2m\pi$, leads to the following electric field, right
after the last graphene plane: 
\begin{equation}
\boldsymbol{E}_{N}=E_{0}e^{-i\omega t}\left[\frac{1}{1+N\beta_{-}}\boldsymbol{e}_{+}+\frac{1}{1+N\beta_{+}}\boldsymbol{e}_{-}\right]\,,\label{eq:field_after_graph-1}
\end{equation}
 and hence in the limit $N\gg1$ we obtain,
\begin{equation}
\frac{t_{+}}{t_{-}}\simeq\frac{\sigma_{+}(\omega)}{\sigma_{-}(\omega)},\label{eq:aux-2}
\end{equation}
 which coincides with the result obtained for the cavity-graphene
system, given by Eq.~(\ref{eq:approx}).

\section{Regularization of the EOM optical conductivity\label{Appendix_C}}

The EOM approach consists in extracting the optical conductivity from
the average of the current operator $\boldsymbol{J}(t)$ (obtained
through the corresponding Heisenberg equation).

This method avoids the calculation of current correlations, and hence
short-circuits the calculation of $\sigma_{ij}(\omega)$. The crucial
point of the EOM approach is the regularization of the following expression,
\begin{equation}
\psi_{ij}(\omega)=\frac{\tilde{J}_{i}(\omega)}{\tilde{E}_{j}(\omega)}\,,\label{eq:ohm}
\end{equation}
 where $\tilde{O}(\omega)$ ($O=\boldsymbol{J},\boldsymbol{E}$) is
defined via
\begin{equation}
O(t)=\tilde{O}(\omega)e^{-i\omega t}+\textrm{c.c.}\,.\label{eq:def_O(t)}
\end{equation}
 Equation.~(\ref{eq:def_O(t)}) is valid for a monochromatic electromagnetic
field, $\mathbf{A}=\boldsymbol{A}_{0}e^{i\omega t}+\textrm{c.c.}$,
and for EOM solutions $\tilde{J}_{i}(\omega)$ in first order in $\boldsymbol{A}_{0}$.
For convenience, we write the external electric field as $\mathbf{E}(t)=\boldsymbol{E}_{+}(t)+\boldsymbol{E}_{-}(t)$,
with $\boldsymbol{E}_{\pm}(t)=\pm i\omega\boldsymbol{A}_{0}e^{\mp i\omega t}$.

Despite the resemblance of Eq.~(\ref{eq:ohm}) to the Ohm's law,
$\psi_{ij}(\omega)$ is not the optical conductivity: in the linear
response regime, the EOM solution can be put into the form 
\begin{equation}
\boldsymbol{J}(t)=\hat{\psi}(\omega)\boldsymbol{E}_{+}(t)+\textrm{c.c.}\,,\label{eq:J_t}
\end{equation}
 with $\hat{\psi}(\omega)$ as defined in Eq.~(\ref{eq:ohm}). On
the other hand, the conductivity, $\hat{\sigma}(t)$, is defined via
the relation 
\begin{equation}
\boldsymbol{J}(t)=\int_{-\infty}^{\infty}d\tau\hat{\sigma}(t-\tau)\mathbf{E}(\tau)\,.\label{eq:kubo}
\end{equation}
 The Fourier transform of Eq.~(\ref{eq:kubo}) is nothing more than
Ohm's law, $\boldsymbol{J}(\omega)=\hat{\sigma}(\omega)\mathbf{E}(\omega)$,
with $\hat{\sigma}(\omega)=\int_{-\infty}^{\infty}dte^{i(\omega+i0^{+})t}\sigma(t)$.
The function $\hat{\sigma}(\omega)$ is analytic in the upper complex
plane and therefore satisfies Kramers-Kronig causality relations.

From Eqs.~(\ref{eq:J_t}) and (\ref{eq:kubo}), we immediately conclude
that, $\hat{\psi}(\omega)\neq\hat{\sigma}(\omega)$. The bottom line
of the EOM approach is that the tensor $\hat{\psi}(\omega)$ can be
exactly transformed into $\hat{\sigma}(\omega)$ via a simple regularization
procedure, as we show in what follows.

Without loss of generality let $\boldsymbol{A}_{0}=A_{0}\mathbf{e_{x}}$,
with $A_{0}\in\mathbb{R}$, and consider that no current flows in
the absence of external perturbations, $\langle\mathcal{J}_{j}(t)\rangle=0$.
Since we are interested in the regular part of the optical response,
we also take $\mathcal{J}_{j}(t)=J_{j}^{P}(t)\equiv J_{j}(t)$; then,
in first order in $A_{0}$,

\begin{align}
\langle J_{i}(t)\rangle_{H}=- & \frac{i}{\hbar}\int_{-\infty}^{t}d\tau A(\tau)\langle\left[J_{x}^{I}(\tau),J_{i}^{I}(t)\right]\rangle_{\beta},\label{eq:int_0}
\end{align}
 with $i=x,y$. Using the Lehman representation, and similar notation
as employed above, the latter expression can be written as
\begin{align}
\langle J_{i}(t)\rangle_{H} & =-\frac{i}{\mathcal{Z}\hbar}\sum_{n\neq m}\int_{-\infty}^{t}d\tau A(\tau)\langle m|J_{x}|n\rangle\langle n|J_{i}|m\rangle\times\nonumber \\
 & \times e^{i\omega_{mn}(\tau-t)}\left(e^{-\beta E_{m}}-e^{-\beta E_{n}}\right)\,.\label{eq:int_1}
\end{align}
 Since we wish to find the explicit form of $\hat{\psi}(\omega)$,
we perform the integration over the variable $\tau$. We obtain
\begin{align}
\langle J_{i}(t)\rangle_{H}= & \frac{1}{\mathcal{Z}\hbar}\sum_{n\neq m}\frac{1}{\omega+\omega_{nm}+i0^{+}}\langle m|J_{x}|n\rangle\langle n|J_{i}|m\rangle\times\nonumber \\
 & \times\left(e^{-\beta E_{m}}-e^{-\beta E_{n}}\right)A_{0}e^{-i\omega t}+\textrm{c.c.}\,.\label{eq:int_2}
\end{align}
 where a small imaginary part has been added to ensure convergence.
Making use of the definition, Eq.~(\ref{eq:J_t}), we arrive at the
desired result,
\begin{align}
\psi_{ij}(\omega) & =-\frac{1}{\mathcal{Z}\hbar}\frac{1}{i\omega}\sum_{n\neq m}\frac{1}{\omega+\omega_{nm}+i0^{+}}\times\nonumber \\
 & \times\langle m|J_{j}|n\rangle\langle n|J_{i}|m\rangle\left(e^{-\beta E_{n}}-e^{-\beta E_{m}}\right)\,,\label{eq:Psi_(Omega)}
\end{align}
 where $i=x$. We also have $\psi_{xy}(\omega)=-\psi_{yx}(\omega)$.

On the other hand, the frequency-dependent conductivity is obtained
from the Fourier transform of $\sigma(t)$, leading to the well-known
Kubo formula: 
\begin{align}
\sigma_{ij}(\omega) & =\frac{1}{\mathcal{Z}\hbar}\sum_{n\neq m}\frac{1}{i\omega_{nm}}\frac{1}{\omega+\omega_{nm}+i0^{+}}\times\nonumber \\
 & \times\langle m|J_{j}|n\rangle\langle n|J_{i}|m\rangle\left(e^{-\beta E_{n}}-e^{-\beta E_{m}}\right)\,.\label{eq:Sigma_ij}
\end{align}
 Comparison of Eq.~(\ref{eq:Psi_(Omega)}) with Eq.~(\ref{eq:Sigma_ij})
yields the general regularization procedure:
\begin{equation}
\sum_{n\neq m}\frac{e^{-\beta E_{n}}-e^{-\beta E_{m}}}{\omega}[...]\rightarrow\sum_{n\neq m}-\frac{e^{-\beta E_{n}}-e^{-\beta E_{m}}}{\omega_{nm}}[...]\,.\label{eq:regularization}
\end{equation}
 In a single-electron representation, the Gibbs factors $\mathcal{Z}^{-1}e^{-\beta E_{n}}$
are substituted for the Fermi occupation numbers $n_{F}(E_{n})$.
This procedure was used in Sec.~\ref{sub:FiniteMagField} to regularize
the EOM solutions of graphene in the presence of a magnetic field.

In Sec.~\ref{sub:ZeroMagField}, no regularization was employed to
derive the interband universal conductivity of graphene in zero field,
$\textrm{Re}\,\sigma_{xx}(\omega)$; see Eqs.~(\ref{eq:Cond_Inter_Graph_ZeroField})
and (\ref{eq:Univ_Cond_InterBand}). The reason is that the $\frac{1}{\omega}$
pre-factor {[}coming from the electric field $\tilde{E}_{x}(\omega)=i\omega A_{0}${]}
is canceled by numerator in Eq.~(\ref{eq:ohm}) in this particular
example, since in zero field, $\tilde{J}_{x}(\omega)\sim\omega$.
It is straightforward to show that applying the regularization, Eq.~(\ref{eq:regularization}),
to Eq.~(\ref{eq:Cond_Inter_Graph_ZeroField}) yields exactly Eq.~(\ref{eq:Univ_Cond_InterBand}).
As for the imaginary part of the conductivity, the regularization,
Eq.~(\ref{eq:regularization}), is compulsory in order to obtain
a consistent result; the imaginary part of Eq.~(\ref{eq:Cond_Inter_Graph_ZeroField}),
as it stands, diverges.

The regularization prescription, Eq.~(\ref{eq:regularization}),
is general and makes the link between the solutions of the EOM $\hat{\psi}(\omega)$
{[}Eq.~(\ref{eq:J_t}){]} and the exact regular optical conductivity
$\hat{\sigma}(\omega)$ of electronic systems.


\begin{thebibliography}{References}
\bibitem{Regularizaton_Procedure}Note that the oscillator strength
$\tilde{E}_{j}(\omega)$ of an ac monochromatic external field polarized
along $j$ is proportional to the radiation frequency $\omega$ and
thus Eq.~(\ref{eq:Ohm's Law}) diverges in the dc limit $\omega\rightarrow0$.
This ill-behavior is not accidental and stems from the fact that the
actual conductivity $\hat{\sigma}(t)$ relates the current at a time
$t$ to the external electric field according to $\boldsymbol{J}(t)=\int d\tau\hat{\sigma}(t-\tau)\mathbf{E}(\tau)$,
whereas the EOM solution for $\boldsymbol{J}(t)$ rather defines the
``EOM conductivity'', $\hat{\psi}(\omega)$, satisfying $\boldsymbol{J}(t)=\hat{\psi}(\omega)\mathbf{E}_{+}(t)+\textrm{c.c.}$,
where $\mathbf{E}_{+}(t)$ is the component of the electric field
proportional to $e^{-i\omega t}$. As shown in Appendix~\ref{Appendix_C},
the EOM solution can be regularized to yield the actual conductivity
$\hat{\psi}(\omega)\rightarrow\hat{\sigma}(\omega)$ according to
a general prescription.
\end{thebibliography}
\end{document}